\documentclass[fleqn,usenatbib, onecolumn]{mnras}

\usepackage{newtxtext,newtxmath}
\usepackage[T1]{fontenc}

\DeclareRobustCommand{\VAN}[3]{#2}
\let\VANthebibliography\thebibliography
\def\thebibliography{\DeclareRobustCommand{\VAN}[3]{##3}\VANthebibliography}

\usepackage{graphicx}	% Including figure files
\usepackage{amsmath}	% Advanced maths commands
\usepackage{mathtools}
\usepackage{orcidlink}
\usepackage[caption=false,lofdepth,lotdepth]{subfig} 
\graphicspath{{Figures/}}

\title[Radiative GRBs]{Synchrotron self-Compton in a radiative-adiabatic fireball scenario:\\ Modelling the multiwavelength observations in some {\itshape Fermi}/LAT bursts}

\author[N. Fraija et al.]{
Nissim Fraija\orcidlink{0000-0002-0173-6453},$^{1}$\thanks{E-mail: nifraija@astro.unam.mx}
P. Veres\orcidlink{0000-0002-2149-9846},$^{2}$
B. Betancourt Kamenetskaia\orcidlink{0000-0002-2516-5739}$^{3,4}$
A. Galvan-Gamez\orcidlink{0000-0001-5193-3693}$^{1}$
M.G. Dainotti\orcidlink{0000-0003-4442-8546}$^{5,6,7,8}$
\newauthor
Simone Dichiara\,\orcidlink{0000-0001-6849-1270}$^{9}$,
and R.~L. Becerra\orcidlink{0000-0002-0216-3415}$^{10}$
\\
$^{1}$Instituto de Astronom\' ia, Universidad Nacional Aut\'onoma de M\'exico, Circuito Exterior, C.U., A. Postal 70-264, 04510 Cd. de M\'exico, M\'exico.\\
$^{2}$Center for Space Plasma and Aeronomic Research (CSPAR), University of Alabama in Huntsville, Huntsville, AL 35899, USA\\
$^{3}$Technical University of Munich, TUM School of Natural Sciences, Physics Department, James-Franck-Str 1, 85748 Garching, Germany\\
$^{4}$Max-Planck-Institut f\"ur Physik (Werner-Heisenberg-Institut), F\"ohringer Ring 6, 80805 Munich, Germany\\
$^{5}$Division of Science, National Astronomical Observatory of Japan, 2-21-1 Osawa, Mitaka, Tokyo 181-8588, Japan \\
$^{6}$The Graduate University for Advanced Studies (SOKENDAI),
2-21-1 Osawa, Mitaka, Tokyo 181-8588, Japan \\
$^{7}$Space Science Institute, 4750 Walnut Street, Boulder, CO
80301, USA \\
$^{8}$SLAC National Accelerator Laboratory, 2575 Sand Hill Road, Menlo Park, CA 94025, USA \\
$^{9}$Department of Astronomy and Astrophysics, The Pennsylvania State University, 525 Davey Lab, University Park, PA 16802, USA\\
$^{10}$ Department of Physics, University of Rome - Tor Vergata, via della Ricerca Scientifica 1, 00100 Rome, IT\\
}
\date{Accepted XXX. Received YYY; in original form ZZZ}

\pubyear{2015}

\begin{document}
\label{firstpage}
\pagerange{\pageref{firstpage}--\pageref{lastpage}}
\maketitle

\begin{abstract}
Energetic GeV photons expected from the closest and the most energetic Gamma-ray bursts (GRBs) provide an unique opportunity to study the very-high-energy emission as well as the possible correlations with lower energy bands in realistic GRB afterglow models. In the standard GRB afterglow model, the relativistic homogeneous shock is usually considered to be fully adiabatic, however, it could be partially radiative. Based on the external forward-shock scenario in both stellar wind and constant-density medium. We present a radiative-adiabatic analytical model of the synchrotron self-Compton (SSC) and synchrotron processes considering an electron energy distribution with a power-law index of $1<p<2$ and $ 2\leq p $. We show that the SSC scenario plays a relevant role in the radiative parameter $\epsilon$, leading to a prolonged evolution during the slow cooling regime. In a particular case, we derive the {\itshape Fermi}/LAT light curves together with the photons with energies $\geq 100$~ MeV in a sample of nine bursts from the second {\itshape Fermi}/LAT GRB catalog that exhibited temporal and spectral indices with $\gtrsim 1.5$ and $\approx 2$, respectively. These events can hardly be described with closure relations of the standard synchrotron afterglow model, and also exhibit energetic photons above the synchrotron limit. We have modeled the multi-wavelength observations of our sample to constrain the microphysical parameters, the circumburst density, the bulk Lorentz factor and the mechanism responsible for explaining the energetic GeV photons.
\end{abstract}


%%%%%%%%%%%%%%%%%%%%%%%%%%%%%%%%%%%%%%%%%%%%%%%%%%

%%%%%%%%%%%%%%%%% BODY OF PAPER %%%%%%%%%%%%%%%%%%

\section{Introduction}
Gamma-ray bursts (GRBs) releasing $\sim 10^{51} - 10^{55}$~erg of isotropic equivalent gamma-ray energy are the most energetic transient sources in the Universe. These transient events can last from a few milliseconds to several hours \citep{1999PhR...314..575P, 2015PhR...561....1K}, and the duration of this prompt episode classifies GRBs into either short or long. The prompt emission is usually detected in the~keV-MeV energy range \citep{1993ApJ...413L.101K} and is described by an empirical Band function \citep{1993ApJ...413..281B}. The late and long-lasting emission, named ``afterglow", detected from radio to TeV gamma-rays, is usually interpreted within the fireball scenario \citep[e.g. see][]{1978MNRAS.183..359C}. The fireball model predicts an external shock when a relativistic jet transfers a large part of its energy to the circumburst medium \citep{1999PhR...314..575P}. A fraction of the total energy is constantly transferred during the shock to accelerate electrons ($\varepsilon_e$) amplifying the magnetic field ($\varepsilon_B$). These electrons are cooled down mainly by synchrotron radiation emitting photons from radio to gamma-rays \citep{1998ApJ...497L..17S,2009MNRAS.400L..75K, 2010MNRAS.409..226K,2013ApJ...763...71A, 2016ApJ...818..190F} and by very-high-energy (VHE $\geq$ 100~GeV) photons, which are radiated through the synchrotron self-Compton (SSC) mechanism \citep{2019ApJ...885...29F, 2019ApJ...883..162F, 2019arXiv191109862Z}.\\
In the standard GRB afterglow scenario, the relativistic forward shock (FS) is usually considered to be fully adiabatic, although it can be partially or fully radiative \citep{1998MNRAS.298...87D, 1998ApJ...497L..17S, 2000ApJ...532..281B, 2010MNRAS.403..926G}. The shock-accelerated electrons are in the fast-cooling regime when the dynamical timescale is larger than the cooling timescale. When accelerated electrons lie in the fast-cooling regime and also the microphysical parameter $\varepsilon_e$ is much greater than $0.1$, the afterglow phase should be in the radiative regime instead of the adiabatic one \citep{2000ApJ...532..281B,2002MNRAS.330..955L, 2019ApJ...886..106P}. Once the afterglow enters the slow-cooling regime, the hydrodynamic evolution can be approximated by the adiabatic case \citep[e.g., see][]{2000ApJ...529..151M}. The deceleration of the GRB fireball by the circumburst medium is faster when the fireball is radiative than when it is adiabatic. Therefore, the temporal afterglow evolution of synchrotron light curves and the shock energetics are modified \citep{2000ApJ...532..281B, 2005ApJ...619..968W}. \cite{2010MNRAS.403..926G} studied the high-energy ($>$ 100~MeV) emission of 11 GRBs detected by the {\itshape Fermi}/LAT (Large Area Telescope) until 2009 October. They found that the temporal decay indices were consistent with synchrotron afterglow flux from a GRB fireball in the fully radiative regime. They proposed that a radiatively efficient fireball could explain the efficiency problem observed during the early afterglow \citep{2007ApJ...655..989Z}.\\
\cite{Ajello_2019} reported the second {\itshape Fermi}/LAT GRB catalog (2FLGC), covering 10 years (from 2008 to 2018) with a total of 169 bursts with high-energy photons above 100~MeV and 29 bursts above $10$~GeV. In this sample, half of them (86/169) of LAT-detected bursts exhibited a temporally-extended component whereas 21/169 GRBs displayed a temporal break spanning from a few dozen to thousands of seconds. While the persistent emission is often attributed to standard synchrotron radiation originating from external FSs \citep{2009MNRAS.400L..75K,2010MNRAS.409..226K}, it is worth noting that not all of the light curves detected by the LAT instrument adhere to the predicted closure relations between power-law (PL) temporal ($\alpha_{\rm L}$) and spectral ($\Gamma_{\rm L}$) indices \cite[e.g., see][]{2019ApJ...883..134T, 2022ApJ...934..188F}. This is the case of those GRBs that exhibit PL temporal and spectral indices of $\alpha_{\rm L}\gtrsim 1.5$ and $\Gamma_{\rm L}\approx 2$, respectively.\\

\cite{2019ApJ...883..162F} derived the light curves and spectra of the SSC model for a wind and homogeneous medium for the adiabatic regime and an electron spectral index larger than 2 ($2< p$). The authors showed that the SSC framework could explain photons beyond the synchrotron limit in GRB~190114C. In this paper, we consider the deceleration phase of a relativistic outflow in a stellar wind and homogeneous medium and derive the expected light curves and spectra of the synchrotron and SSC FS model in the radiative regime for an electron spectral index in the range of $1<p<2$ and $ 2\leq p$. The closure relations of the synchrotron and the SSC FS model as a function of the radiative parameter ($\epsilon$) are presented.  We apply the proposed model to a representative sample of GRBs reported in the 2FLGC with values of temporal and spectral indices with $\alpha_{\rm L}\gtrsim 1.5$ and $\Gamma_{\rm L}\approx 2$, which can hardly be described with closure relations of the standard (non-radiative) synchrotron afterglow model, and also exhibit energetic photons above the synchrotron limit \cite[e.g., see][]{2010MNRAS.403..926G, 2019ApJ...883..134T, 2022ApJ...934..188F}. The paper is arranged as follows. Section~\ref{sec:model} presents the SSC FS model as a function of the radiative parameter when the afterglow model evolves in a stellar wind and in a homogeneous medium. In Section~\ref{sec:app}, we apply our current model to a representative sample of GRBs reported in the 2FLGC and discuss the results, and finally, in Section~\ref{sec:summary}, we summarise our results and present our conclusions. The convention $Q_{\rm x}=Q/10^{\rm x}$ in cgs units and the universal constants $c=\hbar=$1 in natural units will be adopted throughout this paper.

\section{Synchrotron self-Compton afterglow model}\label{sec:model}
The long-lived afterglow emission is generated when a relativistic GRB outflow decelerates and drives a FS into the circumstellar medium. The outflow transfers a large amount of its energy to this surrounding external medium during the deceleration phase. Here, we extend the SSC afterglow model introduced in \cite{2019ApJ...883..162F}, considering that the dynamics of the afterglow emission evolves in the radiative regime, and the electron distribution is described with a hard spectral index in the range of $1<p<2$. Additionally, we show the synchrotron light curves in the radiative regime for $1<p<2$ and $2\leq p$. We note that the self-absorption frequency does not affect the VHE emission.

%, and assume the value of $C_R=4/5$ \citep{2013MNRAS.433.2107N}
%($\gamma_{\rm c}=\frac{6\pi m_e c}{\sigma_T}(1+Y)^{-1}\Gamma^{-1}B'^{-2}t^{-1}$)
\subsection{Dynamics and afterglow emission in a stellar-wind medium}
The dynamics of the afterglow emission for an outflow propagating into a stellar wind medium with a density profile $\rho(r)\propto r^{-2}$ has been widely discussed in the particular case when it evolves in the adiabatic regime, and the electron distribution is described with $2<p$ \citep[e.g., see][]{2000ApJ...536..195C, 1998ApJ...501..772P}. Relativistic electrons are accelerated in the FS and efficiently cooled by synchrotron and SSC processes. In general, with the isotropic-equivalent kinetic energy\footnote{The isotropic gamma-ray energy $E_{\rm \gamma, iso}$ defines the isotropic-equivalent kinetic energy through kinetic efficiency $\eta_K=E_{\rm \gamma, iso}/(E_{\rm K}+E_{\rm \gamma, iso})$.} $E_{\rm K}= E \left( \frac{\Gamma}{\Gamma_0}\right)^\epsilon$ \citep{2010MNRAS.403..926G, 2023ApJ...958..126F} and at a considerable distance from the progenitor $R=2\,\Gamma^2\,t/(1+z)$, the evolution of the bulk Lorentz factor in the adiabatic and radiative regime considering the Blandford-McKee solution ($E_{\rm K}\propto (2-\epsilon) \rho(r) r^{3} \Gamma^2$) \citep{1976PhFl...19.1130B} is $\Gamma=92.6 \left(\frac{1+z}{1.1}\right)^{\frac{1}{4-\epsilon}}\, A_{\rm W,-1}^{-\frac{1}{4-\epsilon}}E_{53}^{\frac{1}{4-\epsilon}}\Gamma_{0,2}^{-\frac{\epsilon}{4-\epsilon}}t_{4.7}^{-\frac{1}{4-\epsilon}}$. The term $A_{\rm W}$ is the parameter of wind density \citep{1998MNRAS.298...87D, 2000ApJ...536..195C, 2005A&A...442..587V}, and $\Gamma_0$ is the initial Lorentz factor. The parameter $\epsilon$ takes into account the hydrodynamic evolution of the FS in the fully adiabatic ($\epsilon=0$), fully radiative ($\epsilon=1$) or partially radiative or adiabatic ($0 <\epsilon < 1$) regimes \citep{2000ApJ...532..281B, 2005ApJ...619..968W,2010MNRAS.403..926G}. The Lorentz factors of the lowest-energy electrons are $\gamma_{\rm m}=\left[\tilde{g}\frac{m_p}{m_e} \varepsilon_{e}\Gamma \gamma_{\rm M}\right]^{\frac{1}{p-1}}$ with $\tilde{g}=\frac{2-p}{p-1}$ for $1<p<2$ and $\gamma_{\rm m}=g\frac{m_p}{m_e}\varepsilon_{e}\Gamma$ for $2\leq p$, In this last case, $g={\rm ln\left( \frac{\gamma_M}{\gamma_m} \right)}$ for $p=2$ and $g=\frac{p-2}{p-1}$ for $2 < p$ \citep{2001ApJ...558L.109D}. The term  $\gamma_{\rm M}$ corresponds to the maximum electron Lorentz factor, and $m_{\rm p}$ and $m_{\rm e}$ to the proton and electron mass, respectively.  The Lorentz factor above which the electrons cool efficiently is $\gamma_{\rm c}=\frac{6\pi m_e c}{\sigma_T}(1+Y)^{-1}\Gamma^{-1}B'^{-2}t^{-1}$ where $B'$ is the comoving magnetic field in the blastwave, $c$ is the speed of light, $\sigma_T$ is the Thompson cross section and $Y$ is the Compton parameter \citep[e.g., see][]{2023ApJ...958..126F}.  When SSC emission becomes the dominant one in electron cooling \citep[as compact radio sources; see ][]{2007A&A...463..145T}
, the electron population preferentially decreases the amount of energy available on $U_{\rm SSC, 1st}$ rather than $U_{\rm syn}$, where $U_{\rm syn}$ and $U_{\rm SSC, 1st}$  are the energy density of synchrotron emission and the first-order scattering, respectively \cite[e.g., see][]{2008MNRAS.391L..19K, 2007ApJ...655..391K, 2015MNRAS.452.3226P, 2018ApJ...859...70F}. In this case, it is needed to take into account the second-order Compton scatterings and therefore, the Lorentz factor above which the electrons cool efficiently is reduced by $(1 + Y + Y^2)$, where the Compton parameter of second-order
scattering is $Y^2\equiv \frac{U_{\rm IC, 1st}}{U_{\rm B}}=\frac{U_{\rm SSC, 1st}}{U_{\rm syn}}\frac{U_{\rm syn}}{U_{\rm B}}$  with $U_{\rm B}$ the energy density of magnetic field \citep{2007ApJ...655..391K, 2018ApJ...859...70F}.  It is worth noting that the Klein-Nishina (KN) effect allows us to ignore higher order (three or more) scattering. For a particular analysis and discussion of the range of parameter space considering this effect, see \cite{2015MNRAS.452.3226P, 2018ApJ...859...70F}.
We have adopted the unprimed and primed quantities to denote them in the observer and comoving frames, respectively.  The Lorentz factors of the lowest-energy electrons and the Lorentz factor of the higher-energy electrons that cool efficiently by synchrotron process are given by  
\begin{small} %*******************
\begin{equation}
\label{ele_win}
\gamma_{\rm m}= \begin{cases}
4.9\times 10^2\,\tilde{g}^{\frac{1}{p-1}}\left(\frac{1+z}{1.1}\right)^{\frac{8+p(\epsilon-3)-2\epsilon}{2(4-\epsilon)(p-1)}}\, \varepsilon^{\frac{1}{p-1}}_{e,-1} \varepsilon^{\frac{-(p-2)}{4(p-1)}}_{B,-4} A_{\rm W,-1}^{\frac{8+p(\epsilon-6)-2\epsilon }{4(p-1)(4-\epsilon)}}E_{53}^{\frac{p}{2(p-1)(4-\epsilon)}}\, \Gamma^{-\frac{p\epsilon}{2(4-\epsilon)(p-1)}}_{0,2} t_{4.7}^{\frac{2\epsilon+ p ( 3 - \epsilon) - 8}{2(4-\epsilon)(p-1)}} \hspace{0.5cm} {\rm for} \hspace{0.2cm} { 1<p<2 }\cr
%&& \hspace{0.1cm}\,  \hspace{0.8cm} \cr
1.6\times 10^3\,g \left(\frac{1+z}{1.1}\right)^{\frac{1}{4-\epsilon}}\, \varepsilon_{e,-1} \, A_{\rm W,-1}^{-\frac{1}{4-\epsilon}}\, E_{53}^{\frac{1}{4-\epsilon}}\, \Gamma^{-\frac{\epsilon}{4-\epsilon}}_{0,2} t_{4.7}^{-\frac{1}{4-\epsilon}} \hspace{6.05cm} {\rm for} \hspace{0.2cm} {2\leq p }
\end{cases}
\end{equation}
\end{small}%***************************
\begin{small}
\begin{equation*}
\gamma_{\rm c} = 3.3\times 10^3\,(1+Y(\gamma_{\rm c}))^{-1} \left(\frac{1+z}{1.1}\right)^{\frac{\epsilon-3}{4-\epsilon}}\,\varepsilon^{-1}_{B,-4} A_{\rm W,-1}^{\frac{\epsilon-5}{4-\epsilon}}E_{53}^{\frac{1}{4-\epsilon}}\,\Gamma^{-\frac{\epsilon}{4-\epsilon}}_{0,2} t_{4.7}^{\frac{3-\epsilon}{4-\epsilon}}.\hspace{7cm}
\end{equation*}
\end{small}

%8\sqrt{2}/3 %$\epsilon=\varepsilon_{\rm e}\, {t'}^{-1}_{\rm syn}/({t'}^{-1}_{\rm syn} + {t'}^{-1}_{\rm ex})$
%respectively, where $\tilde{g}=\frac{2-p}{p-1}$ and the term $g$ is ${\rm ln\left( \frac{\gamma_M}{\gamma_m} \right)}$ for $p=2$ and $\frac{p-2}{p-1}$ for $2 < p$, with $m_{\rm p}$ and $m_{\rm e}$ the proton and electron mass, respectively.  The term $\gamma_{\rm M}$ corresponds to the maximum electron Lorentz factor.

Hereafter, we use the spectral index $p=1.9$ ($\tilde{g}\simeq 0.1$) for $1<p<2$ and $p=2.1$ ($g\simeq 0.09$) for $2 < p$ to estimate the proportionality constant in each quantity. The terms $\epsilon$ and the Compton parameter ($Y$) are defined in subsections~\ref{subsec:eps} and \ref{subsec:kn}, respectively. The transition time from the fast- to the slow-cooling regime is \citep[i.e., see][]{2000ApJ...532..281B, 2005ApJ...619..968W}

%{\small
%\begin{eqnarray}
%\label{t_cm}
%t_{\rm cm} = \cases{ 
%8.4\times 10^3\,{\rm s}\, \tilde{g}^{\frac{2(4-\epsilon)}{S_1}}\chi_{\rm e}^{-\frac{2(p-1)(4-\epsilon)}{S_1}}\left(\frac{1+z}{1.1}\right)\,(1+Y(\gamma_{\rm c}))^{\frac{2(p-1)(4 - \epsilon)}{S_1}} \varepsilon^{\frac{2(4 - \epsilon)}{S_1}}_{e,-1} \varepsilon^{\frac{(3p-2)(4 - \epsilon)}{2S_1}}_{B,-4} A_{W,-1}^{\frac{14p - 12 + \epsilon(2 - 3p)}{2S_1}} E_{53}^{-\frac{p-2}{S_1}}\, \Gamma^{\frac{\epsilon(p - 2)}{S_1}}_{0,2} \hspace{0.1cm} {\rm for}\, \hspace{0.cm} { 1<p<2 }\hspace{0.3cm}\cr
%2.4\times 10^4\,{\rm s}\,g\,\chi_{\rm e}^{-1}\,\left(\frac{1+z}{1.1}\right)\,(1+Y(\gamma_{\rm c}))\, \varepsilon_{e,-1}\, \varepsilon_{B,-4}\, A_{W,-1}\, \hspace{7.85cm} {\rm for} \hspace{0.2cm} {2 \leq p }\,,\cr
%}
%\end{eqnarray}
%}

\begin{small}
\begin{equation*}
\label{t_cm}
t_{\rm cm} = 
\begin{cases} 
8.4\times 10^3\,{\rm s}\, \tilde{g}^{\frac{2(4-\epsilon)}{S_1}}\left(\frac{1+z}{1.1}\right)\,(1+Y(\gamma_{\rm c}))^{\frac{2(p-1)(4 - \epsilon)}{S_1}} \varepsilon^{\frac{2(4 - \epsilon)}{S_1}}_{e,-1} \varepsilon^{\frac{(3p-2)(4 - \epsilon)}{2S_1}}_{B,-4} A_{W,-1}^{\frac{14p - 12 + \epsilon(2 - 3p)}{2S_1}} E_{53}^{-\frac{p-2}{S_1}}\, \Gamma^{\frac{\epsilon(p - 2)}{S_1}}_{0,2} \hspace{0.1cm} {\rm for}\, \hspace{0.cm} { 1<p<2 }\hspace{0.3cm}\cr
2.4\times 10^4\,{\rm s}\,g\,\left(\frac{1+z}{1.1}\right)\,(1+Y(\gamma_{\rm c}))\, \varepsilon_{e,-1}\, \varepsilon_{B,-4}\, A_{W,-1}\, \hspace{6.0cm} {\rm for} \hspace{0.4cm} {2 \leq p }\,,\cr
\end{cases}
\end{equation*}
\end{small}

where $S_1=3p + 2 - p\epsilon$.  Given the total number of emitting electrons ($N_e$), and the synchrotron radiation power \citep[$P_{\nu}$;][]{1998ApJ...501..772P, 1998ApJ...497L..17S,  2000ApJ...536..195C}, the spectral breaks ($\nu^{\rm syn}_{\rm m,c}=\frac{q_e}{2\pi m_ec}(1+z)^{-1}\Gamma\gamma^{2}_{\rm m,c}B'$ with $q_e$ the electron charge) and the maximum flux ($F^{\rm  syn}_{\rm max}=\frac{(1+z)^2}{4\pi D_z^2}N_eP_{\nu}$) in the synchrotron scenario can be written as
%spectral breaks $\nu^{\rm syn}_{\rm m,c}=\frac{\Gamma}{(1+z)} {\nu'}_{\rm m,c}^{\rm syn}$, 
%Here, it is relevant to define the radiation efficiency in both cases; for $1<p< 2$ and $2\leq p$.

\begin{small}
\begin{equation*}
\label{break_win}
\nu^{\rm syn}_{\rm m}= \begin{cases} 
4.2\times 10^{11}\,{\rm Hz}\,\tilde{g}^{\frac{2}{p-1}}\left(\frac{1+z}{1.1}\right)^{\frac{8-3p-2\epsilon + p\epsilon}{(4-\epsilon)(p-1)}}\,\varepsilon^{\frac{2}{p-1}}_{e,-1} \varepsilon^{\frac{1}{2(p-1)}}_{B,-4} A_{\rm W,-1}^{\frac{4-2p-\epsilon}{2(p-1)(4-\epsilon)}} \,E_{53}^{\frac{p}{(p-1)(4-\epsilon)}}\, \Gamma^{-\frac{p\epsilon}{(4-\epsilon)(p-1)}}_{0,2} t_{4.7}^{\frac{\epsilon - p - 4}{(4-\epsilon)(p-1)}} \hspace{0.5cm} {\rm for} \hspace{0.2cm} { 1<p<2 }\cr
4.2\times 10^{12}\,{\rm Hz}\,g^2 \left(\frac{1+z}{1.1}\right)^{\frac{2}{4-\epsilon}}\, \varepsilon^{\frac12}_{B,-4} \varepsilon^{2}_{e,-1}\, A_{\rm W,-1}^{-\frac{\epsilon}{2(4-\epsilon)}}\, E_{53}^{\frac{2}{4-\epsilon}}\, \Gamma^{-\frac{2\epsilon}{4-\epsilon}}_{0,2} t_{4.7}^{\frac{\epsilon-6}{4-\epsilon}} \hspace{4.6cm} {\rm for} \hspace{0.2cm} {2\leq p }\cr
\end{cases}
\end{equation*}
\end{small}
{\small
\label{syn_br_win}
\begin{equation*}
 \begin{multlined}
\nu^{\rm syn}_{\rm c} = 6.8\times 10^{12}\,{\rm Hz}(1+Y(\gamma_{\rm c}))^{-2} \left(\frac{1+z}{1.1}\right)^{\frac{2(\epsilon-3)}{4-\epsilon}}\,\varepsilon^{-\frac32}_{B,-4} A_{\rm W,-1}^{\frac{3\epsilon-16}{2(4-\epsilon)}} E_{53}^{\frac{2}{4-\epsilon}}\, \Gamma^{-\frac{2\epsilon}{4-\epsilon}}_{0,2} t_{4.7}^{\frac{2-\epsilon}{4-\epsilon}}\\
F^{\rm syn}_{\rm max} = 9.1\times 10^3\,{\rm mJy}\, \left(\frac{1+z}{1.1}\right)^{\frac{2(5-\epsilon)}{4-\epsilon}}\, \varepsilon^{\frac12}_{B,-4} D_{\rm z,27}^{-2} A_{\rm W,-1}^{\frac{8-3\epsilon}{2(4-\epsilon)}}\, E_{53}^{\frac{2}{4-\epsilon}}\, \Gamma^{-\frac{2\epsilon}{4-\epsilon}}_{0,2} t_{4.7}^{-\frac{2}{4-\epsilon}}\,,\hspace{5.5cm}
 \end{multlined}
\end{equation*}
}

where the term $D_{\rm z}$ corresponds to the luminosity distance, which is estimated using the cosmological parameters reported in \cite{2018arXiv180706209P}.  Given the maximum Lorentz factor of the electron distribution $\gamma_{\rm max}=\left( \frac{3q_e}{\xi\sigma_T}{B'}_{\rm }^{-1}\right)^{\frac12}$ with  $\xi$ the Bohm parameter\footnote{The value of this parameter becomes $\xi\sim 1$ in the Bohm limit.} \citep{2010ApJ...718L..63P}, the evolution of the maximum energy photon radiated by the synchrotron process in the stellar-wind medium ($h\nu^{\rm syn}_{\rm max}= \frac{3q^2_e}{2\pi \sigma_T m_e c}(1+z)^{-1}\Gamma$) can be written as \citep{2024MNRAS.527.1884F}

\begin{equation}
\label{Emax_win}
h\nu^{\rm syn}_{\rm max} = 0.6\,{\rm GeV}\,\left(\frac{1+z}{1.1}\right)^{\frac{\epsilon-3}{4-\epsilon}}\, A_{\rm W,-1}^{-\frac{1}{4-\epsilon}}E_{53}^{\frac{1}{4-\epsilon}}\Gamma_{0,2}^{-\frac{\epsilon}{4-\epsilon}}t_{4.7}^{-\frac{1}{4-\epsilon}}\,.
\end{equation}

Using Eqs.~(\ref{syn_br_win}) and the synchrotron spectra for the fast- and slow-cooling regimes \citep{1998ApJ...497L..17S}, we find that the synchrotron light curves at an observed frequency $\nu$ and a given time $t$ for $1<p<2$ and $2\leq p$ evolve as

\begin{small}
\begin{equation*}
\label{syn_esp_win}
F^{\rm syn}_{\nu}\propto 
\begin{cases}
\hspace{0.6cm}(1<p<2)\hspace{3.5cm} (2 \leq p) \cr
\{ t^{\frac{\epsilon-8}{3(4-\epsilon)}}, t^{-\frac{5p+\epsilon-10}{3(p-1)(4-\epsilon)}}\}\nu^{\frac13} \hspace{1.cm} \{ t^{\frac{\epsilon-8}{3(4-\epsilon)}}, t^{-\frac{\epsilon}{3(4-\epsilon)}}\} \nu^{\frac13},\hspace{0.4cm} {\rm for} \hspace{0.1cm} \nu < \{ \nu^{\rm syn}_{\rm c}, \nu^{\rm syn}_{\rm m}\}, \cr
t^{-\frac{\epsilon+2}{2(4-\epsilon)}}  \nu^{-\frac12}, \hspace{2.8cm} t^{-\frac{\epsilon+2}{2(4-\epsilon)}}  \nu^{-\frac12} \hspace{1.6cm} {\rm for} \hspace{0.19cm}\nu^{\rm syn}_{\rm c}<\nu <\nu^{\rm syn}_{\rm m} ,\hspace{.2cm} \cr
t^{\frac{\epsilon-p-8}{2(4-\epsilon)}}\nu^{-\frac{p-1}{2}}, \hspace{2.7cm} t^{\frac{2-6p-\epsilon+p\epsilon}{2(4-\epsilon)}}\nu^{-\frac{p-1}{2}} \hspace{0.95cm} {\rm for} \hspace{0.2cm} \nu^{\rm syn}_{\rm m}<\nu <\nu^{\rm syn}_{\rm c},\hspace{.2cm}\cr
t^{-\frac{p+6}{2(4-\epsilon)}}\,\nu^{-\frac{p}{2}} \hspace{2.8cm} t^{\frac{4+p(\epsilon-6)-2\epsilon}{2(4-\epsilon)}}\,\nu^{-\frac{p}{2}} ,\hspace{0.95cm} {\rm for} \hspace{0.2cm} \{\nu^{\rm syn}_{\rm m},\nu^{\rm syn}_{\rm c}\}<\nu\,.\cr
\end{cases}
\end{equation*}
\end{small}

In order to show the evolution of the rest of parameters as a function of $\epsilon$, e.g. for $\{\nu^{\rm syn}_{\rm m},\nu^{\rm syn}_{\rm c}\}<\nu$ the synchrotron flux yields 

%for $\{\nu^{\rm syn\epsilon$}_{\rm m},\nu^{\rm syn}_{\rm c}\}<\nu$, the predicted synchrotron flux  as a function of the rest of parameters can be written as 

\begin{small}
\begin{equation*}
\begin{multlined}
F^{\rm syn}_{1<p<2} \propto g(p) (1+z)^{\frac{22+p(\epsilon-3)-4\epsilon}{2(4-\epsilon)}}(1+Y(\gamma_{\rm c}))^{-1}A_{\rm W}^{\frac{2(1-\epsilon)-p}{2(4-\epsilon)}}D_{z}^{-2}\epsilon_{e}E^{\frac{p+6}{2(4-\epsilon)}}\Gamma_0^{-\frac{\epsilon(p+6)}{2(4-\epsilon)}},\\
F^{\rm syn}_{2\leq p} \propto g(p)^{p-1} (1+z)^{\frac{6+p-\epsilon}{4-\epsilon}}(1+Y(\gamma_{\rm c}))^{-1}A_{\rm W}^{-\frac{\epsilon(p+2)}{4(4-\epsilon)}}D_{z}^{-2}\epsilon_{B}^{\frac{p-2}{4}}\epsilon_{e}^{p-1}E^{\frac{p+2}{4-\epsilon}}\Gamma_0^{-\frac{\epsilon(p+2)}{4-\epsilon}}.\,\,\,
\end{multlined}
\end{equation*}
\end{small} 

 It is worth noting that for $p\simeq 2$, synchrotron fluxes for $1<p<2$ and $2\leq p$ are equal (i.e., $F^{\rm syn}_{1<p<2}\simeq F^{\rm syn}_{2\leq p}$).\\

The SSC process occurs when the same electron population that radiates synchrotron photons up-scatters them to higher energies as $h\nu^{\rm ssc}_{\rm k}\sim \gamma^2_{\rm k} h\nu^{\rm syn}_{\rm k}$. Here, the notation ${\rm k=m, c}$ refers to the minimum and cooling frequencies and $h$ stands for the Planck constant \citep[e.g., see][]{2001ApJ...548..787S}. The maximum flux that the SSC process can reach $F^{\rm ssc}_{\rm max}\sim \tau F^{\rm syn}_{\rm max}$ depends on the maximum synchrotron flux given in Eqs.~(\ref{syn_br_win}) and the optical depth $\tau\propto \frac13\,A_{\rm W}\,R^{-1}$. Therefore, the spectral breaks and the maximum flux in the SSC scenario for $1<p<2$ and $2\leq p$ are \citep[e.g., see][]{2019ApJ...883..162F, 2022ApJ...934..188F}

%{\small
%\begin{eqnarray}\nonumber
%\label{break_ssc_win}
%h\nu^{\rm ssc}_{\rm m}= \cases{ 
%4.2\times 10^2\,{\rm eV}\, \tilde{g}^{\frac{4}{p-1}}\chi_{\rm e}^{-4}\left(\frac{1+z}{1.1}\right)^{\frac{2[8+p(\epsilon-3) - 2 \epsilon]}{(4-\epsilon)(p-1)}} \varepsilon^{\frac{4}{p-1}}_{e,-1} \varepsilon^{\frac{3-p}{2(p-1)}}_{B,-4} A_{\rm W,-1}^{\frac{p(\epsilon - 8) + 3(4 - \epsilon) }{2(p-1)(4-\epsilon)}}\, E_{53}^{\frac{2p}{(p-1)(4-\epsilon)}} \Gamma^{-\frac{2p\epsilon}{(4-\epsilon)(p-1)}}_{0,2} t_{4.7}^{\frac{3(\epsilon - 4) -p (\epsilon - 2)}{(4-\epsilon)(p-1)}} \hspace{0.6cm} {\rm for} \hspace{0.2cm} { 1<p<2 }\cr
%4.1\times10^4\,{\rm eV}\,g^{4}\chi_{\rm e}^{-4}\left(\frac{1+z}{1.1}\right)^{\frac{4}{4-\epsilon}}\, \varepsilon^{\frac12}_{B,-4} \varepsilon^{4}_{e,-1}\, A_{\rm W,-1}^{-\frac{4+\epsilon}{2(4-\epsilon)}}\, E_{53}^{\frac{4}{4-\epsilon}}\, \Gamma^{-\frac{4\epsilon}{4-\epsilon}}_{0,2} t_{4.7}^{\frac{\epsilon-8}{4-\epsilon}} \hspace{5.95cm} {\rm for} \hspace{0.1cm} {2\leq p}\cr
%}
%\end{eqnarray}
%}
%
\begin{small}
\begin{equation}\nonumber
\label{break_ssc_win}
h\nu^{\rm ssc}_{\rm m} = \begin{cases} 
4.2\times 10^2\,{\rm eV}\, \tilde{g}^{\frac{4}{p-1}} \left(\frac{1+z}{1.1}\right)^{\frac{2[8+p(\epsilon-3) - 2 \epsilon]}{(4-\epsilon)(p-1)}} \varepsilon^{\frac{4}{p-1}}_{e,-1} \varepsilon^{\frac{3-p}{2(p-1)}}_{B,-4} A_{\rm W,-1}^{\frac{p(\epsilon - 8) + 3(4 - \epsilon) }{2(p-1)(4-\epsilon)}}\, E_{53}^{\frac{2p}{(p-1)(4-\epsilon)}} \Gamma^{-\frac{2p\epsilon}{(4-\epsilon)(p-1)}}_{0,2} t_{4.7}^{\frac{3(\epsilon - 4) -p (\epsilon - 2)}{(4-\epsilon)(p-1)}} \hspace{0.6cm} {\rm for} \hspace{0.2cm} { 1<p<2 }\cr
4.1\times10^4\,{\rm eV}\,g^{4} \left(\frac{1+z}{1.1}\right)^{\frac{4}{4-\epsilon}}\, \varepsilon^{\frac12}_{B,-4} \varepsilon^{4}_{e,-1}\, A_{\rm W,-1}^{-\frac{4+\epsilon}{2(4-\epsilon)}}\, E_{53}^{\frac{4}{4-\epsilon}}\, \Gamma^{-\frac{4\epsilon}{4-\epsilon}}_{0,2} t_{4.7}^{\frac{\epsilon-8}{4-\epsilon}} \hspace{5.55cm} {\rm for} \hspace{0.1cm} {2\leq p}\cr
\end{cases}
\end{equation}
\end{small}
\begin{small}
 \begin{equation}\label{ssc_br_win}
 \begin{multlined}
h\nu^{\rm ssc}_{\rm c} = 8.57\times 10^{-3}\,{\rm GeV} (1+Y(\gamma_{\rm c}))^{-4} \left(\frac{1+z}{1.1}\right)^{\frac{4(\epsilon-3)}{4-\epsilon}}\, \varepsilon^{-\frac72}_{B,-4} A_{\rm W,-1}^{\frac{7\epsilon-36}{2(4-\epsilon)}}\, E_{53}^{\frac{4}{4-\epsilon}}\, \Gamma^{-\frac{4\epsilon}{4-\epsilon}}_{0,2} t_{4.7}^{\frac{8-3\epsilon}{4-\epsilon}} \\
F^{\rm ssc}_{\rm max} = 4.6\times 10^{-5}\,{\rm mJy}\,g^{-1} \left(\frac{1+z}{1.1}\right)^3\, \varepsilon^{\frac12}_{B,-4}\, D_{\rm z,27}^{-2} A_{\rm W,-1}^{\frac{5}{2}}\, t_{4.7}^{-1}\,.\hspace{9cm}
 \end{multlined}
\end{equation}
\end{small}

Given the evolution of the spectral breaks and the maximum flux (Eqs.~\ref{ssc_br_win}), it is possible to write the SSC light curves at an observed frequency $\nu$ and a given time $t$ as \citep[e.g., see][]{2019ApJ...883..162F, 2022ApJ...934..188F}

\begin{small}
\begin{equation}
\label{ssc_wind}
F^{\rm ssc}_{\nu}\propto \begin{cases}
\hspace{0.6cm}(1<p<2)\hspace{4cm} (2\leq p) \cr
\{ t^{\frac{2(3\epsilon - 10)}{3(4-\epsilon)}}, t^{\frac{2[12-7p+\epsilon(2p-3)]}{3(p-1)(4-\epsilon)}} \} \nu^{\frac13},\hspace{1.cm}
\{ t^{\frac{2(3\epsilon - 10)}{3(4-\epsilon)}}, t^{\frac{2(\epsilon-2)}{3(4-\epsilon)}} \} \nu^{\frac13},
\hspace{0.4cm} {\rm for} \hspace{0.2cm} \nu < \{ \nu^{\rm ssc}_{\rm c}, \nu^{\rm ssc}_{\rm m}\}, \cr
t^{-\frac{\epsilon}{2(4-\epsilon)}} \nu^{-\frac12},\hspace{3.6cm}
t^{-\frac{\epsilon}{2(4-\epsilon)}}  \nu^{-\frac12}
\hspace{1.7cm} {\rm for} \hspace{0.2cm}\nu^{\rm ssc}_{\rm c} < \nu<\nu^{\rm ssc}_{\rm m} ,\hspace{.3cm}\cr
t^{\frac{2p+5\epsilon - 20 - p\epsilon }{2(4-\epsilon)}} \nu^{-\frac{p-1}{2}},\hspace{2.55cm}
t^{\frac{p(\epsilon - 8) + \epsilon}{2(4-\epsilon)}} \nu^{-\frac{p-1}{2}}
\hspace{1.4cm} {\rm for} \hspace{0.2cm} \nu^{\rm ssc}_{\rm m}<\nu< \nu^{\rm ssc}_{\rm c},\hspace{.3cm} \cr
t^{\frac{2(\epsilon-6)-p(\epsilon-2)}{2(4-\epsilon)}}\,\nu^{-\frac{p}{2}},\hspace{2.7cm}
t^{\frac{8+p(\epsilon-8)-2\epsilon}{2(4-\epsilon)}}\,\nu^{-\frac{p}{2}}
\hspace{1.2cm} {\rm for} \hspace{0.2cm} \{\nu^{\rm ssc}_{\rm m},\nu^{\rm ssc}_{\rm c}\}<\nu\,. \cr
\end{cases}
\end{equation}
\end{small}

In order to show the evolution of the rest of parameters as a function of $\epsilon$, e.g. for $\{\nu^{\rm ssc}_{\rm m},\nu^{\rm ssc}_{\rm c}\}<\nu$, the SSC flux yields 
 
\begin{small}
\begin{equation*}
\begin{multlined}
 F^{\rm ssc}_{1<p<2} \propto g(p) (1+z)^{\frac{14+p(\epsilon-3)+3\epsilon}{4-\epsilon}}(1+Y(\gamma_{\rm c}))^{-2}A_{\rm W}^{\frac{8(2-p)+\epsilon(p-6)}{4(4-\epsilon)}}D_{z}^{-2}\epsilon_{B}^{-\frac{p+2}{4}}\epsilon_{e}^{2}E^{\frac{p+2}{4-\epsilon}}\Gamma_0^{-\frac{(p+2)\epsilon}{4-\epsilon}}\\
 F^{\rm ssc}_{2\leq p} \propto g(p)^{2p-3} (1+z)^{\frac{4+2p-\epsilon}{4-\epsilon}}(1+Y(\gamma_{\rm c}))^{-2}A_{\rm W}^{\frac{4(2-p)-\epsilon(p+2)}{4(4-\epsilon)}}D_{z}^{-2}\epsilon_{B}^{\frac{p-6}{4}}\epsilon_{e}^{2(p-1)}E^{\frac{2p}{4-\epsilon}}\Gamma_0^{-\frac{2\epsilon p}{4-\epsilon}}\,.
\end{multlined}
\end{equation*}
\end{small} 

We can see that SSC fluxes for $1<p<2$ and $2\leq p$ are equal (i.e., $F^{\rm ssc}_{1<p<2}\simeq F^{\rm ssc}_{2\leq p}$) when $p\approx 2$. \\

%Another spectral break at VHEs could be due to the internal attenuation by pair production through interaction with low-energy photons.

It is worth noting that very energetic photons of energy $h\nu_{\rm h}\approx 1\,{\rm TeV}$ interacting with low-energy photons $h\nu_{\rm l}=60.1\,{\rm eV}\left(\frac{\Gamma_1}{1+z}\right)^2 \frac{1}{\left(h\nu_{\rm h}/{\rm 1\,TeV}\right)}$ are absorbed to produce pairs \cite[e.g., see][]{2019ApJ...885...29F}. The optical depth of attenuation of this interaction process is $\tau_{\gamma\gamma}=1.8\times 10^{-8} \frac{R_{17}}{\Gamma_{1}} n_{\gamma,5.3}$, where $n_\gamma\approx 2.4\times 10^3\,{\rm cm^{-3}} \frac{L_{\gamma,44}}{R^2_{17}\Gamma_1 \left(h\nu_{\rm l}/{\rm 60.1\,eV}\right)}$ and $L_{\gamma, 44}$ are the photon density and luminosity of the seed photons, respectively \cite[e.g., see][]{1999PhR...314..575P}.

\subsection{Dynamics and afterglow emission in a uniform-density medium}

Once the outflow begins to be decelerated at a significant distance from the progenitor by a uniform-density medium ($\rho=n$), the evolution of the bulk Lorentz factor in the adiabatic and radiative regimes considering the Blandford-McKee solution ($E_{\rm K}=\frac{2\pi}{3}(2-\epsilon) m_pc^2 n r^{3} \Gamma^2$) \citep{1976PhFl...19.1130B} becomes $\Gamma=17.0 \left(\frac{1+z}{1.1}\right)^{\frac{3}{8-\epsilon}}\, n^{-\frac{1}{8-\epsilon}}E_{53}^{\frac{1}{8-\epsilon}}\Gamma_{0,2}^{-\frac{\epsilon}{8-\epsilon}}t_{4.7}^{-\frac{3}{8-\epsilon}}$ \citep{2010MNRAS.403..926G, 2023ApJ...958..126F} . In the case of a uniform-density medium, the Lorentz factors of both, the lowest-energy electrons and higher-energy electrons that cool efficiently by synchrotron process evolve as \citep[i.e., see][]{2001ApJ...558L.109D, 1998ApJ...497L..17S, 2023ApJ...958..126F}

\begin{small}
\begin{equation}
\label{ele_Lorent_ism}
\gamma_{\rm m}= \begin{cases}
9.3\times 10\,\tilde{g}^{\frac{1}{p-1}} \left(\frac{1+z}{1.1}\right)^{\frac{3(4-p)}{2(8-\epsilon)(p-1)}}\, \varepsilon^{\frac{1}{p-1}}_{e,-1} \varepsilon^{\frac{-(p-2)}{4(p-1)}}_{B,-4} n^{\frac{8-6p+p\epsilon-2\epsilon }{4(p-1)(8-\epsilon)}} E_{53}^{\frac{4-p}{2(p-1)(8-\epsilon)}}\, \Gamma^{-\frac{\epsilon(4-p)}{2(8-\epsilon)(p-1)}}_{0,2} t_{4.7}^{-\frac{3(4-p)}{2(8-\epsilon)(p-1)}} \hspace{0.75cm} {\rm for} \hspace{0.2cm} { 1<p<2 }\cr
3.4\times 10^2\,g \left(\frac{1+z}{1.1}\right)^{\frac{3}{8-\epsilon}}\, \varepsilon_{e,-1} n^{-\frac{1}{8-\epsilon}}\, E_{53}^{\frac{1}{8-\epsilon}}\, \Gamma^{-\frac{\epsilon}{8-\epsilon}}_{0,2} t_{4.7}^{-\frac{3}{8-\epsilon}} \hspace{6.2cm} {\rm for} \hspace{0.2cm} {2\leq p }\cr
\end{cases}
\end{equation}
\end{small}
\begin{small}
\begin{equation}
\gamma_{\rm c} = 1.6\times 10^{3}\,(1+Y(\gamma_{\rm c}))^{-1} \left(\frac{1+z}{1.1}\right)^{-\frac{1+\epsilon}{8-\epsilon}}\, \varepsilon^{-1}_{B,-4} n^{\frac{\epsilon-5}{8-\epsilon}}\, E_{53}^{-\frac{3}{8-\epsilon}}\, \Gamma^{\frac{3\epsilon}{8-\epsilon}}_{0,2} t_{4.7}^{\frac{1+\epsilon}{8-\epsilon}}\,.
\end{equation}
\end{small}

The terms $\epsilon$ and $Y$ are defined in subsections \ref{subsec:kn} and \ref{subsec:eps}, respectively. In this afterglow model, the transition time from the fast- to the slow-cooling regime takes place at \citep[i.e., see][]{2000ApJ...532..281B, 2005ApJ...619..968W}

\begin{small}
\begin{equation*}
\label{t_cm_ism}
t_{\rm cm} = \begin{cases} 
4.8\times 10^2\,{\rm s}\, \tilde{g}^{\frac{2(8-\epsilon)}{S_2}} \left(\frac{1+z}{1.1}\right) \,(1+Y(\gamma_{\rm c}))^{\frac{2(p-1)(8 - \epsilon)}{S_2}} \varepsilon^{\frac{2(8 - \epsilon)}{S_2}}_{e,-1} \varepsilon^{\frac{(3p-2)(8 - \epsilon)}{2S_2}}_{B,-4} n^{\frac{14p - 12 + \epsilon(2 - 3p)}{2S_2}} E_{53}^{\frac{5p-2}{S_2}}\, \Gamma^{\frac{\epsilon(2 - 5p)}{S_2}}_{0,2} \hspace{0.2cm} {\rm for} \hspace{0.2cm} { 1<p<2 }\cr
3.0\times 10^3\,{\rm s}\,g^{\frac{8-\epsilon}{4+\epsilon}} \left(\frac{1+z}{1.1}\right)\,(1+Y(\gamma_{\rm c}))^{\frac{8-\epsilon}{4+\epsilon}} \, \varepsilon^{\frac{8-\epsilon}{4+\epsilon}}_{e,-1}\, \varepsilon^{\frac{8-\epsilon}{4+\epsilon}}_{B,-4} n^{\frac{4-\epsilon}{4+\epsilon}} E_{53}^\frac{4}{4+\epsilon} \Gamma_{0,2}^{-\frac{4\epsilon}{4+\epsilon}} \hspace{4.0cm} {\rm for} \hspace{0.2cm} {2\leq p },\cr
\end{cases}
\end{equation*}
\end{small}

where $S_2=10 - p + 2\epsilon(p - 1)$. In this case, given the electron Lorentz factors (Eq.~\ref{ele_Lorent_ism}), the synchrotron spectral breaks ($\nu^{\rm syn}_{\rm m,c}\propto (1+z)^{-1}\Gamma\gamma^{2}_{\rm m,c}B'$) and the maximum flux ($F^{\rm  syn}_{\rm max}\propto (1+z)^2 D_z^{-2}N_eP_{\nu}$) become

\begin{small}
\begin{equation*}
\label{syn_esp_ims}
\nu^{\rm syn}_{\rm m} = \begin{cases} 
9.1\times 10^{9}\,{\rm Hz}\,\tilde{g}^{\frac{2}{p-1}} \left(\frac{1+z}{1.1}\right)^{\frac{14-5p+\epsilon(p-1)}{(8-\epsilon)(p-1)}}\, \varepsilon^{\frac{2}{p-1}}_{e,-1} \varepsilon^{\frac{1}{2(p-1)}}_{B,-4} n^{\frac{4-2p-\epsilon}{2(p-1)(8-\epsilon)}} E_{53}^{\frac{p+2}{(p-1)(8-\epsilon)}}\, \Gamma^{-\frac{\epsilon(p+2)}{(8-\epsilon)(p-1)}}_{0,2} t_{4.7}^{-\frac{3(p+2)}{(8-\epsilon)(p-1)}} \hspace{0.2cm} {\rm for} \hspace{0.2cm} { 1<p<2 }\cr
3.5\times 10^{11}\,{\rm Hz}\,g^2 \left(\frac{1+z}{1.1}\right)^{\frac{4+\epsilon}{8-\epsilon}}\, \varepsilon^2_{e,-1} \varepsilon^{\frac12}_{B,-4} n^{-\frac{\epsilon}{2(8-\epsilon)}}\, E_{53}^{\frac{4}{8-\epsilon}}\, \Gamma^{-\frac{4\epsilon}{8-\epsilon}}_{0,2} t_{4.7}^{-\frac{12}{8-\epsilon}} \hspace{4.3cm} {\rm for} \hspace{0.2cm} {2\leq p }\cr
\end{cases}
\end{equation*}
\end{small}
\begin{small}
\begin{equation} \label{syn_br_hom}
\begin{multlined}
\nu^{\rm syn}_{\rm c} = 2.2\times 10^{12}\,{\rm Hz} (1+Y(\gamma_{\rm c}))^{-2} \left(\frac{1+z}{1.1}\right)^{-\frac{(4+\epsilon)}{8-\epsilon}}\, \varepsilon^{-\frac32}_{B,-4} n^{\frac{3\epsilon-16}{2(8-\epsilon)}}\, E_{53}^{-\frac{4}{8-\epsilon}}\, \Gamma^{\frac{4\epsilon}{8-\epsilon}}_{0,2} t_{4.7}^{\frac{2(\epsilon-2)}{8-\epsilon}}\cr
F^{\rm syn}_{\rm max} = 8.1\times 10\,{\rm mJy}\, \left(\frac{1+z}{1.1}\right)^{\frac{\epsilon+16}{8-\epsilon}}\, \varepsilon^\frac12_{B,-4}\,D_{\rm z,27}^{-2}\, n^{\frac{8-3\epsilon}{2(8-\epsilon)}}\, E_{53}^{\frac{8}{8-\epsilon}}\, \Gamma^{-\frac{8\epsilon}{8-\epsilon}}_{0,2} t_{4.7}^{-\frac{3\epsilon}{8-\epsilon}}\,.
\end{multlined}
\end{equation}
\end{small}
The evolution of the maximum energy photon radiated by the synchrotron process in a homogeneous medium ($h\nu^{\rm syn}_{\rm max} \propto (1+z)^{-1}\Gamma$) can be written as \citep{2024MNRAS.527.1884F} 

\begin{equation}
\label{Emax_ism}
h\nu^{\rm syn}_{\rm max} = 0.2\,{\rm GeV}\,\left(\frac{1+z}{1.1}\right)^{\frac{\epsilon-5}{8-\epsilon}}\, n^{-\frac{1}{8-\epsilon}}E_{53}^{\frac{1}{8-\epsilon}}\Gamma_{0,2}^{-\frac{\epsilon}{8-\epsilon}}t_{4.7}^{-\frac{3}{8-\epsilon}}\,. 
\end{equation}

Using Eqs.~\ref{syn_br_hom}, the synchrotron light curves at a specific time $t$ and frequency $\nu$ for $1 < p < 2$ and $2\leq p$ can be written as \citep[i.e., see][]{2000ApJ...532..281B, 2005ApJ...619..968W}
 
\begin{small}
\begin{equation}
\label{syn_esp_hom}
F^{\rm syn}_{\nu}\propto \begin{cases}
\hspace{0.6cm}(1<p<2)\hspace{4cm} (2\leq p) \cr
\{ t^{\frac{4-11\epsilon}{3(8-\epsilon)}} , t^{\frac{(p+2) - 3\epsilon(p-1)}{(8-\epsilon)(p-1)}} \} \nu^{\frac13},\hspace{1.5cm}
\{ t^{\frac{4-11\epsilon}{3(8-\epsilon)}} , t^{\frac{4 - 3\epsilon}{8-\epsilon}} \} \nu^{\frac13}
\hspace{1.2cm} {\rm for} \hspace{0.2cm} \nu < \{ \nu^{\rm syn}_{\rm c}, \nu^{\rm syn}_{\rm m} \}, \cr
t^{-\frac{2(\epsilon+1)}{8-\epsilon}}  \nu^{-\frac12},\hspace{3.5cm}
t^{-\frac{2(\epsilon+1)}{8-\epsilon}}  \nu^{-\frac12} 
\hspace{2.cm} {\rm for} \hspace{0.2cm} \nu^{\rm syn}_{\rm c} <\nu<\nu^{\rm syn}_{\rm m} ,\hspace{.1cm} \cr
t^{-\frac{3(p+2+2\epsilon)}{2(8-\epsilon)}}\nu^{-\frac{p-1}{2}},\hspace{2.7cm}
t^{\frac{3(2-2p-\epsilon)}{8-\epsilon}}\nu^{-\frac{p-1}{2}},
\hspace{1.33cm} {\rm for} \hspace{0.2cm} \nu^{\rm syn}_{\rm m}<\nu< \nu^{\rm syn}_{\rm c},\hspace{.1cm}\cr
t^{-\frac{3p+10+4\epsilon}{2(8-\epsilon)}}\,\nu^{-\frac{p}{2}},\hspace{3.0cm}
t^{\frac{2(2-3p-\epsilon)}{8-\epsilon}}\,\nu^{-\frac{p}{2}}
\hspace{1.75cm} {\rm for} \hspace{0.2cm} \{\nu^{\rm syn}_{\rm m},\nu^{\rm syn}_{\rm c}\}<\nu\,. \cr
\end{cases}
\end{equation}
\end{small}

Given the electron Lorentz factors (Eqs.~\ref{ele_Lorent_ism}), the spectral breaks and the maximum flux of the synchrotron process (Eqs.~\ref{syn_br_hom}) with the optical depth given by $\tau\propto n\,R$, the spectral breaks and the maximum flux in the SSC scenario for $1<p<2$ and $2\leq p$ are \citep[e.g., see][]{2019ApJ...883..162F, 2022ApJ...934..188F}

\begin{small}
\begin{equation*}
\label{break_ssc_hom}
h\nu^{\rm ssc}_{\rm m} = \begin{cases} 
0.3\,{\rm eV}\,\tilde{g}^{\frac{4}{p-1}} \left(\frac{1+z}{1.1}\right)^{\frac{26+p(\epsilon-8)-\epsilon}{(8-\epsilon)(p-1)}}\, \varepsilon^{\frac{4}{p-1}}_{e,-1} \varepsilon^{\frac{3-p}{2(p-1)}}_{B,-4} n^{\frac{p(\epsilon-8)-3(\epsilon-4)}{2(p-1)(8-\epsilon)}} E_{53}^{\frac{6}{(p-1)(8-\epsilon)}}\, \Gamma^{-\frac{6\epsilon}{(8-\epsilon)(p-1)}}_{0,2} t_{4.7}^{-\frac{18}{(8-\epsilon)(p-1)}} \hspace{1.2cm} {\rm for} \hspace{0.2cm} { 1<p<2 }\cr
1.7\times 10^{2}\,{\rm eV}\,g^4 \left(\frac{1+z}{1.1}\right)^{\frac{10+\epsilon}{8-\epsilon}}\, \varepsilon^4_{e,-1} \varepsilon^{\frac12}_{B,-4} n^{-\frac{4+\epsilon}{2(8-\epsilon)}}\, E_{53}^{\frac{6}{8-\epsilon}}\, \Gamma^{-\frac{6\epsilon}{8-\epsilon}}_{0,2} t_{4.7}^{-\frac{18}{8-\epsilon}} \hspace{4.8cm} {\rm for} \hspace{0.2cm} {2\leq p }\cr
\end{cases}
\end{equation*}
\end{small}

\begin{small}
\begin{equation} \label{ssc_br_hom}
\begin{multlined}
h\nu^{\rm ssc}_{\rm c} = 2.2\times 10^{-5}\,{\rm GeV} \left(\frac{1+z}{1.1}\right)^{-\frac{3(\epsilon+2)}{8-\epsilon}} (1+Y(\gamma_{\rm c}))^{-4}\, \varepsilon^{-\frac72}_{B,-4} n^{\frac{7\epsilon-36}{2(8-\epsilon)}}\, E_{53}^{-\frac{10}{8-\epsilon}}\, \Gamma^{\frac{10\epsilon}{8-\epsilon}}_{0,2} \,t_{4.7}^{\frac{2(2\epsilon-1)}{8-\epsilon}}\\
F^{\rm ssc}_{\rm max} = 3.9\times 10^{-5}\,{\rm mJy}\, g^{-1} \left(\frac{1+z}{1.1}\right)^{\frac{2(\epsilon+7)}{8-\epsilon}}\, \varepsilon^\frac12_{B,-4} D_{\rm z,27}^{-2}\,n^{\frac{5(4-\epsilon)}{2(8-\epsilon)}}\, E_{53}^{\frac{10}{8-\epsilon}}\, \Gamma^{-\frac{10\epsilon}{8-\epsilon}}_{0,2} t_{4.7}^{\frac{2(1-2\epsilon)}{8-\epsilon}}\,,
\end{multlined}
\end{equation}
\end{small}
respectively. Given the evolution of the spectral breaks and the maximum flux (Eqs.~\ref{ssc_br_hom}), it is possible to write the SSC light curves at an observed frequency $\nu$ and a given time $t$ for $1<p<2$ and $2\leq p$ as \citep[e.g., see][]{2019ApJ...883..162F, 2022ApJ...934..188F}

\begin{small}
\begin{equation}
\label{ssc_esp_hom}
F^{\rm ssc}_{\nu}\propto \begin{cases} 
\hspace{0.6cm}(1<p<2)\hspace{4cm} (2\leq p) \cr
\{ t^{\frac{8(1-2\epsilon)}{3(8-\epsilon)}}, t^{\frac{2(p+2)-4\epsilon(p-1)}{(8-\epsilon)(p-1)}} \} \nu^{\frac13},\hspace{1.2cm}
\{ t^{\frac{8(1-2\epsilon)}{3(8-\epsilon)}} , t^{\frac{4(2-\epsilon)}{8-\epsilon}} \} \nu^{\frac13}
\hspace{1.2cm} {\rm for} \hspace{0.2cm} \nu< \{ \nu^{\rm ssc}_{\rm c}, \nu^{\rm ssc}_{\rm m} \}, \cr
t^{\frac{1-2\epsilon}{8-\epsilon}}  \nu^{-\frac12}, \hspace{3.8cm}
t^{\frac{1-2\epsilon}{8-\epsilon}}  \nu^{-\frac12}
\hspace{2.7cm} {\rm for} \hspace{0.2cm} \nu^{\rm ssc}_{\rm c}<\nu<\nu^{\rm ssc}_{\rm m},\hspace{.0cm} \cr
t^{-\frac{7+4\epsilon}{8-\epsilon}} \nu^{-\frac{p-1}{2}}, \hspace{3.3cm}
t^{\frac{11-4\epsilon-9p}{8-\epsilon}}\nu^{-\frac{p-1}{2}}
\hspace{1.8cm} {\rm for} \hspace{0.2cm}\nu^{\rm ssc}_{\rm m}<\nu< \nu^{\rm ssc}_{\rm c},\hspace{.0cm} \cr
t^{-\frac{2(4+\epsilon)}{8-\epsilon}}\,\nu^{-\frac{p}{2}}, \hspace{3.4cm}
t^{\frac{10 - 2\epsilon-9p}{8-\epsilon}}\,\nu^{-\frac{p}{2}}
\hspace{2.cm} {\rm for} \hspace{0.2cm}\{\nu^{\rm ssc}_{\rm m},\nu^{\rm ssc}_{\rm c}\}<\nu\,, \cr
\end{cases}
\end{equation}
\end{small}

respectively. As discussed in the case of the stellar wind medium, VHE photons $\sim 1\,{\rm TeV}$ could be attenuated by interactions with softer photons.

%In this case, the energy of softer photons, the optical depth, and density of softer photons are $h\nu_{\gamma,l}=40.3\,{\rm eV}$, $\tau_{\gamma\gamma}=6.0\times10^{-7}$ and $n_\gamma=1.7\times 10^2\,{\rm cm^{-3}}$, respectively.
%instead of the adiabatic one
%then the fireball evolution is the radiative regime. On the other hand, when the dissipated energy is larger than radiated energy radiative losses are small, and the fireball evolution lies in the adiabatic regime.
%
%The evolution of the radiative parameter can be estimated as \citep[for details see][]{2000ApJ...543...90H}
\subsection{Evolution of the radiative parameter}\label{subsec:eps}

The radiative parameter is defined by the ratio between the radiated and dissipated energy \citep{2014MNRAS.445.1625N, 2000ApJ...529..151M}. When the radiated and dissipated energies are similar almost most of the internal energy is dissipated as radiation, and the afterglow phase lies in the radiative regime. This parameter is defined as

\begin{equation}
\label{eps_y}
\epsilon \equiv \varepsilon_{e} \zeta\,,
\end{equation}

where $\zeta$ is the fraction of the radiated energy. It is estimated as the ratio of the power radiated during the slow- and fast-cooling regimes \citep{2014MNRAS.445.1625N}

\begin{equation}
 \zeta=\begin{cases} \label{eps_y1}
\frac{\gamma_{\rm m}}{\gamma_{\rm c}}\frac{p-2}{3-p}\left[\frac{1}{p-2} \left(\frac{\gamma_{\rm c}}{\gamma_{\rm m}} \right)^{3-p} -1 \right] \hspace{1cm}{\rm for}\hspace{1cm} \gamma_{\rm m}\leq \gamma_{\rm c}\cr
1\hspace{4.6cm}{\rm for}\hspace{1cm} \gamma_{\rm c} < \gamma_{\rm m}\,.\cr
\end{cases}
\end{equation}

Eqs.~\ref{eps_y} and \ref{eps_y1} show that the radiative parameter is constant $\epsilon=\varepsilon_e$ during the fast cooling regime. Afterwards it is expected to decrease to $0$ during the slow cooling regime for $2\leq p$. We notice that the radiative parameter decreases slowly if the value of $p$ does not deviate from $2$, and in the particular case when $p\to 2$, the radiative parameter approaches the same value $\epsilon\simeq \varepsilon_e$. At the previous point, an adiabatic break was not observed. On the other hand, if $p$ largely deviates from $2$, adiabatic breaks around the transition time are expected in the synchrotron and SSC light curves. Eqs. \ref{eps_y} and \ref{eps_y1} shows that the radiative parameter evolves during GRB.

\paragraph{Equivalence with the radiative parameter $s$.} Some authors have introduced the radiative parameter $s$ instead of $\epsilon$ through the variation explicitly of the equivalent kinetic energy as \citep{2000ApJ...532..281B, 2005ApJ...619..968W,2010MNRAS.403..926G} 
\begin{equation}
\label{Ef}
E\left(\frac{t}{t_{\rm dec}}\right)^{-s}\propto (2-\epsilon) \rho(r) \Gamma^2\,R^{3-k}\,,
\end{equation}
where $\rho(r) \propto \, r^{\rm -k}$ with ${\rm k=0}$ corresponds to the density of the constant-density medium ($\rho=n$), and ${\rm k = 2}$ to the density of the stellar wind ejected by its progenitor ($\rho(r)\propto r^{-2})$. For instance, Eq.~\ref{Ef} for $k=0$ can be obtained from $E \propto n\Gamma_0^{\epsilon} \Gamma^{8-\epsilon}\,t^3$ and $\Gamma=\Gamma_0\,\left(\frac{t}{t_{\rm dec}}\right)^{-\frac{3}{8-\epsilon}}$ with $t_{\rm dec}$ the deceleration time scale. In this scenario, the evolution of the radiative parameters ($\epsilon, s$) is related through

\begin{eqnarray}
(3+s)(8-\epsilon)-24=0,\,\,\,\,\,\,\,\,&&{\rm for\,\,\,\, k=0}\cr
(1+s)(4-\epsilon)-4=0,\,\,\,\,\,\,\,\,\,\,\,&&{\rm for\,\,\,\, k=2}\,,
\end{eqnarray}

where the parameter~s lies in the range of $0\leq~s \leq 3/7$ and $0\leq~s \leq 1/3$ for the homogeneous and stellar wind medium, respectively.

\subsection{Klein-Nishina effects}\label{subsec:kn}
%Just above the transition of the scattering cross section from the Thomson to the KN regime, direct and indirect effects can be observed.
%and for $2 \leq p$, the Compton parameter (Eq.~\ref{Yth}) is explicitly derived in \cite{2001ApJ...548..787S}.
A direct effect on the SSC spectrum due to the KN regime is the suppression of up-scattered synchrotron photons, and an indirect effect occurs when the SSC emission dominates, and at least some of the injected electrons with different Lorentz factors have enough time to cool down. The SSC spectra could have several breaks depending on the location of the spectral breaks in the KN regime; $h\nu^{\rm syn}_{\rm KN}(\gamma_m) \simeq\frac{2\Gamma}{(1+z)}\,\frac{m_e c^2}{\gamma_m}$ for $\nu^{\rm syn}_{\rm c} < \nu^{\rm syn}_{\rm m}$ and $h\nu^{\rm syn}_{\rm KN} (\gamma_c) \simeq\frac{2\Gamma}{(1+z)}\,\frac{m_e c^2}{\gamma_c}$ for $\nu^{\rm syn}_{\rm m} < \nu^{\rm syn}_{\rm c}$ \citep[see][]{2009ApJ...703..675N, 2010ApJ...712.1232W}. For example, the value of the Compton parameter in the case of the slow-cooling regime ($\nu^{\rm syn}_{\rm m} < \nu^{\rm syn}_{\rm c}$) might not be constant and be defined by
\begin{small}
\begin{equation}
\label{Yth_kn}
Y(\gamma_c)[Y(\gamma_c)+1] = \frac{\varepsilon_{e}}{\varepsilon_{B}} \left(\frac{\gamma_{\rm c}}{\gamma_{\rm m}}\right)^{2-p}\,\begin{cases} 
\left( \frac{\nu^{\rm syn}_{\rm m}}{\nu^{\rm syn}_{\rm c}} \right)^{-\frac{p-3}{2}}\,\left( \frac{\nu^{\rm syn}_{\rm KN}(\gamma_{\rm c})}{\nu^{\rm syn}_{\rm m}}\right)^\frac43 \, \hspace{1cm} {\rm for} \hspace{0.2cm} {\nu^{\rm syn}_{\rm KN}(\gamma_{\rm c}) <\nu^{\rm syn}_{\rm m} }\cr
\left(\frac{\nu^{\rm syn}_{\rm KN}(\gamma_{\rm c})}{\nu^{\rm syn}_c}\right)^{-\frac{p-3}{2}} \, \hspace{2.2cm} {\rm for} \hspace{0.2cm} { \nu^{\rm syn}_{\rm m} < \nu^{\rm syn}_{\rm KN}(\gamma_{\rm c}) < \nu^{\rm syn}_{\rm c}}\cr
1\, \hspace{4.25cm} {\rm for} \hspace{0.2cm} {\nu^{\rm syn}_{\rm c} < \nu^{\rm syn}_{\rm KN}(\gamma_{\rm c})}\,.
%&& \hspace{0.1cm}\,  \hspace{0.8cm} \cr
%\hspace{4.05cm} {\rm for} \hspace{0.2cm} {2 \leq p }\,,
\end{cases}
\end{equation}
\end{small}

The last cooling condition ($\nu^{\rm syn}_{\rm c}<\nu^{\rm syn}_{\rm KN}(\gamma_{\rm c}) $) corresponds to the Thomson regime, and then KN effects are neglected.  In this case, the value of the Compton parameter becomes

\begin{small}
\begin{equation}
\label{Yth}
Y(\gamma_c)[Y(\gamma_c)+1] = \eta \begin{cases} 
\,\frac{\varepsilon_{e}^{\frac{1}{p-1}}}{\varepsilon_{B}} \left[\frac{m_p\tilde{g}\,\Gamma}{m_e\gamma_{\rm M}} \right]^{\tilde{g}} \hspace{2.3cm} {\rm for} \hspace{0.2cm} { 1<p<2 }\cr
%&& \hspace{0.1cm}\,  \hspace{0.8cm} \cr
\frac{\varepsilon_{e}}{\varepsilon_{B}} \hspace{4.05cm} {\rm for} \hspace{0.2cm} {2 \leq p }\,,
\end{cases}
\end{equation}
\end{small}
with $\eta=1$ and $\left(\frac{\gamma_{\rm c}}{\gamma_{\rm m}}\right)^{2-p}$ for the fast- and slow-cooling regime, respectively \citep{2001ApJ...548..787S}. Due to the terms $h\nu^{\rm syn}_{\rm KN}(\gamma_m)$ and $h\nu^{\rm syn}_{\rm KN}(\gamma_c)$ and the electron Lorentz factors (Eqs. \ref{ele_win} and \ref{ele_Lorent_ism}), the SSC spectral breaks in the KN regime can be written explicitly as 

\begin{small}
\begin{equation*}\nonumber
\label{KN_win}
h\nu^{\rm ssc}_{\rm m, KN}= \begin{cases} 
4.6\times 10\,{\rm GeV}\, \tilde{g}^{\frac{1}{p-1}} \left(\frac{1+z}{1.1}\right)^{\frac{6-p-2\epsilon+p\epsilon}{2(p-1)(4-\epsilon)}}\, \varepsilon^{\frac{1}{p-1}}_{e,-1} \varepsilon^{\frac{2-p}{4(p-1)}}_{B,-4} A_{W,-1}^{\frac{12-10p + \epsilon(p -2)}{4(p-1)(4-\epsilon)}}E_{53}^{\frac{3p-2}{2(p-1)(4-\epsilon)}}\, \Gamma^{\frac{-\epsilon(3p-2)}{2(4-\epsilon)(p-1)}}_{0,2} t_{4.7}^{\frac{p-6 - \epsilon(p-2)}{2(4-\epsilon)(p-1)}} \hspace{0.2cm} {\rm for} \hspace{0.2cm} { 1<p<2 }\cr
1.5\times 10^2\,{\rm GeV}\,g \left(\frac{1+z}{1.1}\right)^{\frac{2}{4 -\epsilon}} \, \varepsilon_{e,-1} \, A_{W,-1}^{-\frac{2}{4-\epsilon}} 
E_{53}^{\frac{2}{4-\epsilon}}\, \Gamma^{-\frac{2\epsilon}{4-\epsilon}}_{0,2} t_{4.7}^{-\frac{2} {4-\epsilon}} \hspace{5.55cm} {\rm for} \hspace{0.2cm} {2\leq p }\cr
\end{cases}
\end{equation*}
\end{small}

%and
\begin{small}
\begin{equation}
h\nu ^{\rm ssc}_{\rm c, KN} = 7.1\times 10^2\,{\rm GeV}\left(\frac{1+z}{1.1}\right)^{\frac{\epsilon - 2}{4 -\epsilon}} (1+Y(\gamma_{\rm c}))^{-1} \varepsilon^{-1}_{B,-4} A_{W,-1}^{\frac{\epsilon - 6}{4-\epsilon}}
E_{53}^{\frac{2}{4-\epsilon}}\, \Gamma^{-\frac{2\epsilon}{4-\epsilon}}_{0,2} t_{4.7}^{\frac{2-\epsilon} {4-\epsilon}}\,, \hspace{5.5cm}
\end{equation}
\end{small}
for a stellar-wind environment, and as
%The value of the Compton parameter is estimated depending on spectral breaks \citep[see;][]{2010ApJ...712.1232W}. In this case, the spectral breaks in the KN regime which could modify the SSC spectra are
%
\begin{small}
\begin{equation*}
\label{KN_hom}
h\nu^{\rm ssc}_{\rm m, KN} = \begin{cases} 
1.6\,{\rm GeV}\,\tilde{g}^{\frac{1}{p-1}} \left(\frac{1+z}{1.1}\right)^{\frac{3(p+2)}{2(p-1)(8-\epsilon)}} \varepsilon^{\frac{1}{p-1}}_{e,-1} \varepsilon^{-\frac{p-2}{4(p-1)}}_{B,-4} n^{\frac{12-10p+\epsilon(p - 2)}{4(p-1)(8-\epsilon)}} E_{53}^{\frac{p+2}{2(p-1)(8-\epsilon)}}\, \Gamma^{-\frac{\epsilon(p+2)}{2(8-\epsilon)(p-1)}}_{0,2} t_{4.7}^{-\frac{3(p+2)}{2(8-\epsilon)(p-1)}} \hspace{0.2cm} {\rm for} \hspace{0.2cm} { 1<p<2 }\cr
5.9\,{\rm GeV}\,g \left(\frac{1+z}{1.1}\right)^{\frac{6}{8-\epsilon}}\, \varepsilon_{e,-1} \, n^{-\frac{2}{8-\epsilon}}\, E_{53}^{\frac{2}{8-\epsilon}}\, \Gamma^{-\frac{2\epsilon}{8-\epsilon}}_{0,2} t_{4.7}^{-\frac{6}{8-\epsilon}} \hspace{6.05cm} {\rm for} \hspace{0.2cm} {2\leq p }\cr
\end{cases}
\end{equation*}
\end{small}
\begin{small}
\begin{equation}
 h\nu^{\rm ssc}_{\rm c, KN} = 2.7\times 10\,{\rm GeV} \left(\frac{1+z}{1.1}\right)^{\frac{2-\epsilon}{8-\epsilon}} (1+Y(\gamma_{\rm c}))^{-1} \varepsilon^{-1}_{B,-4} n^{\frac{\epsilon - 6}{8-\epsilon}} E_{53}^{-\frac{2}{8-\epsilon}}\, \Gamma^{\frac{2\epsilon}{8-\epsilon}}_{0,2} t_{4.7}^{\frac{\epsilon-2} {8-\epsilon}}\,,
\end{equation}
\end{small}

for a constant-density medium. It should be noted that the total cross section in the KN regime can be approximated by $\sigma\approx 3/8 \sigma_T x^{-1}\left(\rm ln2x+1/2\right)$ with $x=hv/m_e\gg1$ \citep{1986rpa..book.....R}.

\subsection{Analysis and Discussion}
\label{analysis}

\begin{figure*}
\centering
\includegraphics[width=0.49\textwidth]{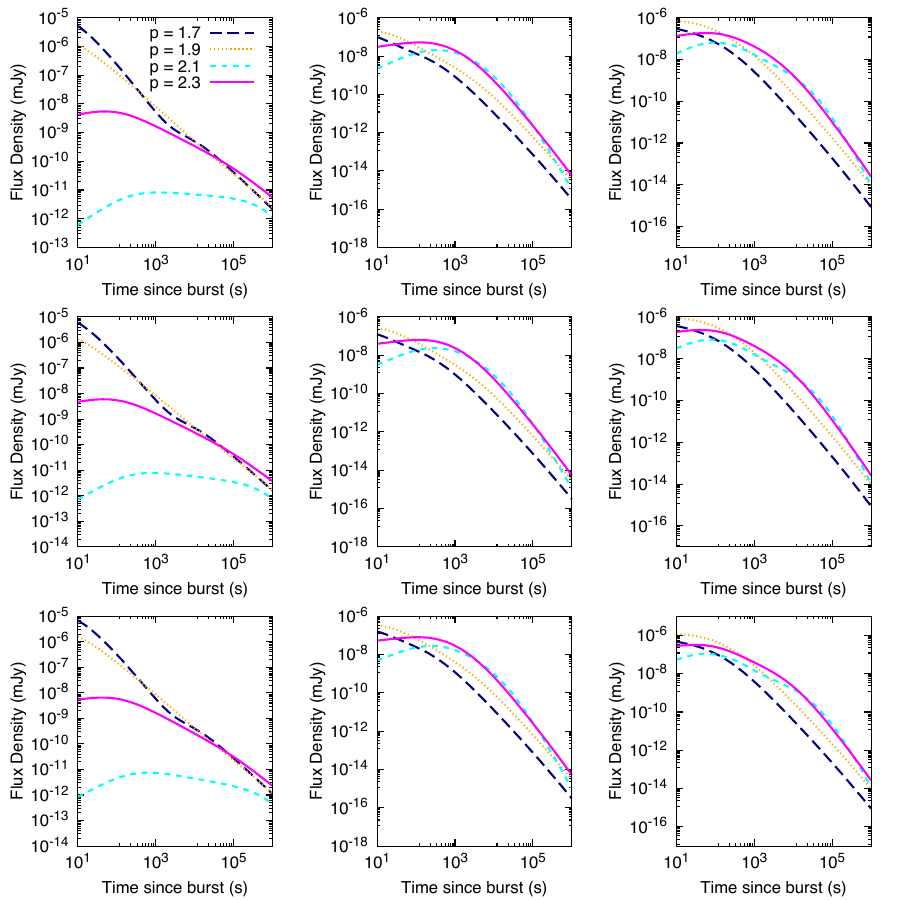}
\includegraphics[width=0.49\textwidth]{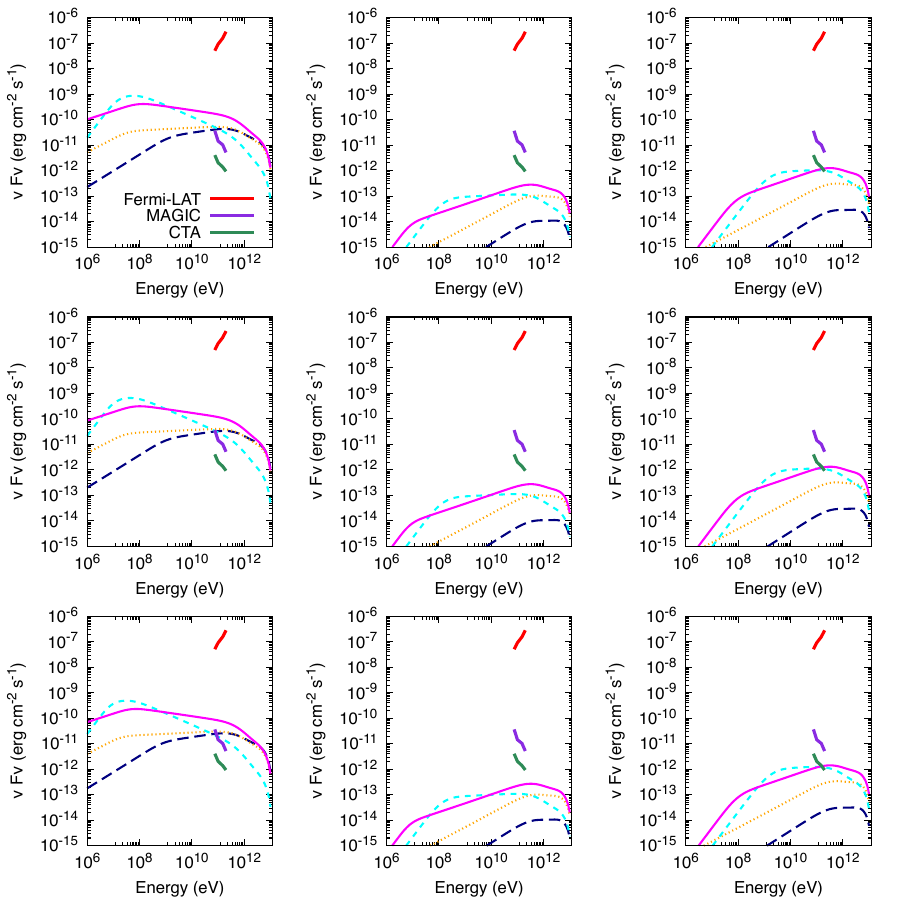}
\caption{SSC light curves (left) and spectra (right) in the stellar-wind afterglow model for $\epsilon=0$ (upper panels), $0.2$ (center panels) and $0.4$ (lower panels) and $p=1.7$, $1.9$, $2.1$ and $2.3$. The light curves and SED are shown at 1 TeV and $5\times 10^{4}\,{\rm s}$, respectively, for $\epsilon_{\rm e}=0.1$, $\epsilon_{\rm B}=10^{-4}$ and $\zeta=0.5$. The values of parameter pairs ($A_{\rm W}=0.1$ and $E=10^{52}\,{\rm erg}$), ($A_{\rm W}=10^{-3}$ and $E=10^{52}\,{\rm erg}$), and ($A_{\rm W}=10^{-3}$ and $E=10^{53}\,{\rm erg}$) are used for the left, middle and right panels. The sensitivities of CTA (Southern array, green line), MAGIC (purple line) and {\itshape Fermi}/LAT (red line) are shown between 75 and 250~GeV at $3\times 10^4\,{\rm s}$ for a zenith angle of 20$^\circ$\citep{2019ICRC...36..673F}. We have considered a hypothetical burst located at $z=0.1$ and the effect of the EBL absorption proposed in \citep{2017A&A...603A..34F}.}
\label{LC_wind}
\end{figure*}

Based on the external FS scenario, we have presented a general analytical model of the synchrotron and SSC processes in three cases: i) the fully adiabatic ($\epsilon=0$), ii) fully radiative ($\epsilon=1$), and iii) partially radiative or adiabatic ($0 <\epsilon < 1$) regimes. We plotted the expected light curves and spectra for an electron spectral index $1<p<2$ and $2\leq p$ when the outflow decelerates in a stellar wind (see Figure~\ref{LC_wind}).  Significant variations of the spectral and temporal features of the afterglow emission are introduced by radiative losses only if $\epsilon$ is large and approaches to unity.  In particular, when $\epsilon=0$, the synchrotron and SSC light curves derived in the standard FS scenario are recovered \citep{2001BASI...29..107B, 1998ApJ...497L..17S, 2001ApJ...548..787S, 2019ApJ...883..162F}. We want to highlight the synchrotron and SSC spectral breaks and light curves for $1<p<2$ and $2\leq p$ become equal when $p\to 2$.\\

We have obtained and shown in Table~\ref{ClosureRelations} the closure relations that describe the evolution of the synchrotron and SSC flux ($F_{\nu}\propto t^{-\alpha}\nu^{-\beta}$) as a function of $\epsilon$ and $p$. The cooling conditions in the constant density medium of $\nu^{\rm ssc}_{\rm m}\leq \nu \leq \nu^{\rm ssc}_{\rm c}$, and $\{\nu^{\rm ssc}_{\rm m}, \nu^{\rm ssc}_{\rm c}\}<\nu$ for $1<p<2$, are the only ones where the SSC fluxes do not depend on $p$, hence their closure relations cannot be estimated. In this case, the SSC flux evolves as $\propto t^{-\alpha}$ with $\alpha=\frac{7+4\epsilon}{8-\epsilon}$ and $\frac{2(4 + \epsilon)}{8-\epsilon}$ for the spectral indices $\beta=\frac{p-1}{2}$ and $\frac{p}{2}$, respectively. Note that adiabatic breaks are expected around the transition time between the fast- and slow-cooling regimes. The transition time refers to the temporal interval of applicability of the synchrotron and SSC light curves. 
%Eq.~\ref{eps_y} shows that the radiative parameter is constant $\epsilon=\varepsilon_e$ during the fast cooling regime, and later it is expected to decrease to $0$ during the slow cooling regime for $2\leq p$. We notice that the radiative parameter decreases slowly as the value of $p$ does not deviate from $2$, and in the particular case of $p\to 2$, the radiative parameter approaches the same value $\epsilon\simeq \varepsilon_e$. At the previous point, an adiabatic break was not observed. On the other hand, if $p$ largely deviates from $2$, adiabatic breaks around the transition time are expected in the synchrotron and SSC light curves. The evolution of the radiative parameter during the slow-cooling regime for $1<p<2$ is not applicable when we consider Eq.~\ref{eps_y}.\\ 
%
Given the closure relations (see Table \ref{ClosureRelations}), we can notice that the synchrotron and SSC fluxes will evolve similarly in time and energy if the condition $\alpha_{\rm syn}(\beta) \approx \alpha_{\rm ssc}(\beta)$ is satisfied. Considering both, the stellar wind and the homogeneous afterglow model, this condition is satisfied when the observed frequency evolves in $\{\nu^{\rm j}_{\rm m}, \nu^{\rm j}_{\rm c}\}<\nu$ (with ${\rm j=syn}$ and ${\rm ssc}$) and $\beta\to 1$ (i.e., $p\to 2$). Irrespective of whether $p$ approaches the value of 2 from $1<p<2$ (e.g. $p\simeq 1.98$) or from $2\leq p$ (e.g. $p\simeq 2.02$), the temporal decay indices become $\alpha\approx\frac{4}{4-\epsilon}$ and $\approx\frac{2(4+\epsilon)}{8-\epsilon}$ for the wind and constant-density afterglow model, respectively. In the particular case of $\epsilon\approx 0$, the expected flux obtained from the stellar-wind and constant-density afterglow model evolves with the same temporal index of $\alpha\approx 1$. The temporal and spectral similarities that could be observed during the afterglow phase in two or more different bands could be interpreted in the synchrotron and SSC FC scenario with a hard spectral index $p\approx 2$, and the parameter $\epsilon$ would be useful to discriminate between the stellar wind or homogeneous afterglow model.  One of the more relevant features to be observed in the radiative regime should occur during the afterglow transition between a stellar wind and a constant-density medium. For $\epsilon=0$, the X-ray flux evolving in the stellar wind and constant density afterglow has the same temporal evolution $t^{-\frac{3p-2}{4}}$ when $p$ lies in the range of $2<p$. Therefore, depending on the values of observable quantities and parameters, a transition between stellar-wind and homogeneous medium afterglow could be noticeable. For an electron spectral index that does not deviate from 2 (i.e., $p\sim 2$), the expected synchrotron and SSC fluxes would evolve as $F_\nu\propto t^{-\alpha}$ with $\alpha=1$ for any value of $p$. In this case, a transition between stellar-wind and constant-density afterglow could not be noticeable in the SSC and synchrotron light curves. Otherwise, this transition could be more highlighted as $p$ deviates from 2 and $\epsilon$ from 0. On the other hand, the development of the FS in the fully radiative ($\epsilon=1$) or partially radiative or adiabatic ($0 <\epsilon < 1$) regimes presents an X-ray flux that has different evolution. Regardless of the values, a transition between stellar wind and homogeneous medium afterglow must be observed. \\ 

Deviations around $\epsilon=0$ will also modify the evolution of synchrotron and SSC spectral breaks, producing distinct variations in the afterglow tails. The SSC and sychrotron spectral breaks for $1<p<2$ ($2<p$) evolve as $\nu^{\rm ssc}_{\rm m} \propto t^{\frac{3(\epsilon - 4) - p(\epsilon - 2)}{(4-\epsilon)(p-1)}}$($t^{\frac{\epsilon - 8}{4-\epsilon}}$), $\nu^{\rm syn}_{\rm m} \propto t^{\frac{\epsilon - p - 4}{(4-\epsilon)(p-1)}}$($t^{\frac{\epsilon - 6}{4-\epsilon}}$), $\nu^{\rm ssc}_{\rm c}\propto t^{\frac{8 - 3\epsilon}{4-\epsilon}}$ and $\nu^{\rm syn}_{\rm c}\propto t^{\frac{2 -\epsilon}{4-\epsilon}}$ for a stellar wind medium, and as $\nu^{\rm ssc}_{\rm m} \propto t^{-\frac{18}{(8-\epsilon)(p-1)}}$($t^{-\frac{18}{8-\epsilon}}$), $\nu^{\rm syn}_{\rm m} \propto t^{-\frac{3 (p + 2)}{(8-\epsilon)(p-1)}}$($t^{-\frac{12}{8-\epsilon}}$), $\nu^{\rm ssc}_{\rm c}\propto t^{\frac{2(2\epsilon - 1)}{8-\epsilon}}$ and $\nu^{\rm syn}_{\rm c}\propto t^{\frac{2(\epsilon - 2)}{8-\epsilon}}$ for a constant-density medium. For instance, considering $2<p$ the spectral breaks in the stellar wind (constant-density) medium evolve under the cooling condition $\nu^{\rm syn}_{\rm m}\propto t^{-[1.50-1.67]}$ ($t^{-[1.50-1.71]}$), $\nu^{\rm syn}_{\rm c}\propto t^{[0.33-0.50]}$ ($ t^{-[0.29-0.50]}$), $\nu^{\rm ssc}_{\rm m}\propto t^{-[2.0-2.33]}$ ($t^{-[2.25-2.51]}$) and
$\nu^{\rm ssc}_{\rm c}\propto t^{[1.67-2.0]}$ ($t^{[-0.25-0.29]}$) instead of the typical evolution of spectral breaks of $\nu^{\rm syn}_{\rm m}\propto t^{-\frac32}$ ($t^{-\frac32}$), $\nu^{\rm syn}_{\rm c}\propto t^{-\frac12}$ ($t^{\frac12}$), $\nu^{\rm ssc}_{\rm m}\propto t^{-2}$ ($t^{-\frac52}$) and $\nu^{\rm ssc}_{\rm c}\propto t^{-2}$ ($t^{-\frac14}$), respectively.\\ 

\begin{figure*}
\centering
\includegraphics[width=0.49\textwidth]{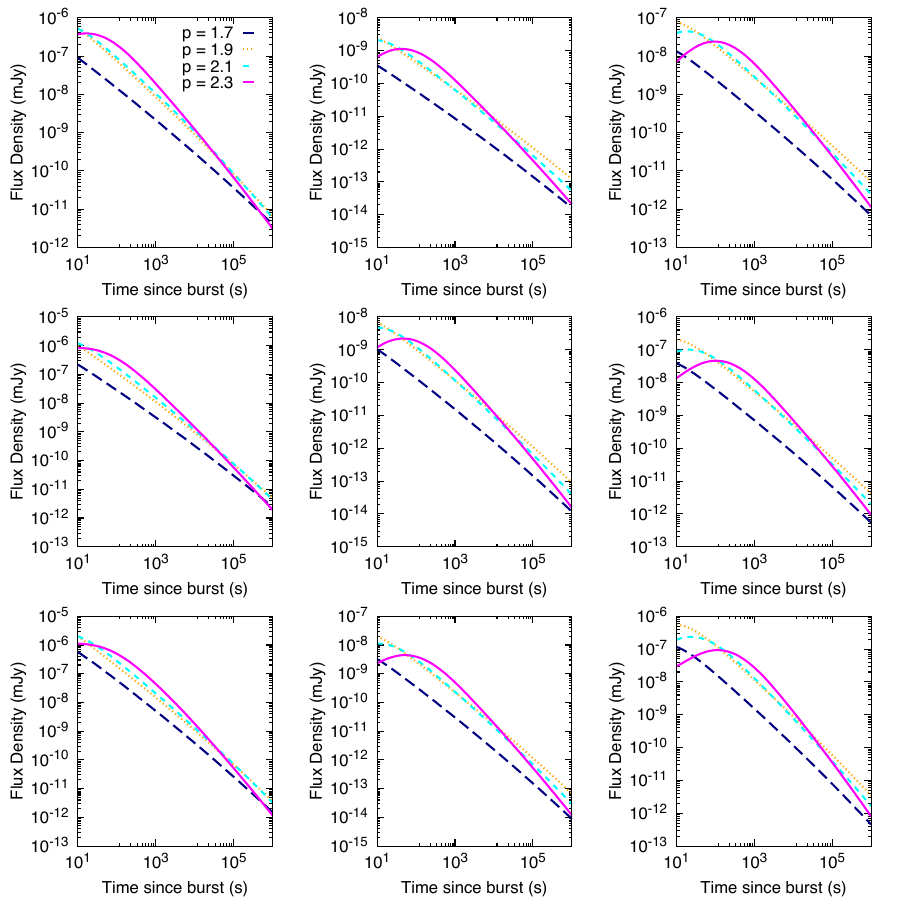}
\includegraphics[width=0.49\textwidth]{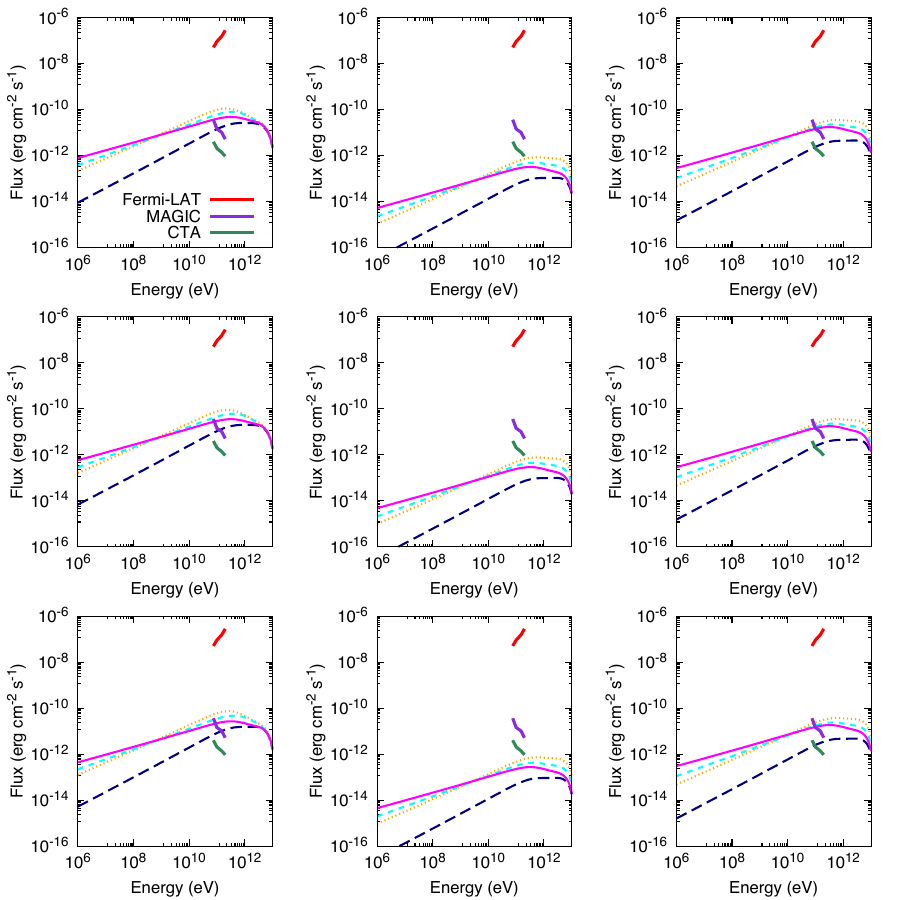}
\caption{The same as Figure \ref{LC_wind}, but for a constant-density medium. The values of parameter pairs ($n=0.1\,{\rm cm^{-3}}$ and $E=10^{52}\,{\rm erg}$), ($n=10^{-3}\,{\rm cm^{-3}}$ and $E=10^{52}\,{\rm erg}$), and ($n=10^{-3}\,{\rm cm^{-3}}$ and $E=10^{53}\,{\rm erg}$) are used for the left, middle and right panels.}
\label{LC_ism}
\end{figure*}

Figures~\ref{LC_wind} and~\ref{LC_ism} illustrate the light curves and spectra of the SSC process evolving in stellar-wind and homogeneous medium for typical GRB afterglow parameters, respectively. These light curves and spectra are shown at 1~TeV and at $5\times 10^4\,{\rm s}$, respectively, for $\epsilon_{\rm e}=\epsilon$, $\epsilon_{\rm B}=10^{-4}$ and different spectral indices ($p=1.7$, $1.9$, $2.1$ and $2.3$). These panels are shown from top to bottom for $\epsilon=0$, $0.2$ and $0.4$, and from left to right for the values of [$A_{\rm W}$ ($n$); $E$]=[$0.1$ ($0.1\,{\rm cm^{-3}}$); $10^{52}\,{\rm erg}$], [$10^{-3}$ ($10^{-3}\,{\rm cm^{-3}}$); $10^{52}\,{\rm erg}$], and [$10^{-3}$ ($10^{-3}\,{\rm cm^{-3}}$); $10^{53}\,{\rm erg}$]. We consider a hypothetical burst located at $z=0.1$ and the model of the Extragalactic Background Light (EBL) absorption proposed in \cite{2017A&A...603A..34F}.  The left-hand panels in Figure \ref{LC_wind} show that, depending on the parameter values, the expected flux will have a different behavior. For example, the light curves during the early time show a plateau phase followed by a normal decay for $2\leq p$, but only in a few cases for $1<p<2$. It is also shown that irrespective of the value of the spectral index $p$ and the parameter $\epsilon$, the expected flux increases as $E$ also increases. The panels show that the expected flux $F^{\rm ssc}_{\rm \nu}\propto t^{-\alpha}$ evolves from $1.03\leq \alpha \leq 1.13$ ($1.10\leq \alpha \leq 1.22$) to $2.03\leq \alpha \leq 2.08$ ($2.10\leq \alpha \leq 2.16$) for $p=1.9$ ($p=2.1$). This is due to the evolution of the SSC spectral breaks; $\nu^{\rm ssc}_{\rm m}\propto t^{-\alpha}$ with $1.64\leq\alpha \leq 1.70$ ($1.50\leq\alpha \leq 1.56$) for $p=1.9$ ($p=2.1$), and $\nu^{\rm ssc}_{\rm c}\propto t^{-\alpha}$ with observe that the expected flux increases in some panels as $A_{\rm w}$ increases and in other panels when $A_{\rm w}$ decreases. This result can be explained in terms of the density parameter and the cooling condition. Given the spectral index in the range $1<p<2$ ($2\leq p$), the expected flux as a function of the density parameter is $F^{\rm}_{\nu}\propto A_{\rm W}^{\frac{52-8p-13\epsilon + p\epsilon}{4(4-\epsilon)}}$ ($A_{\rm W}^{\frac{44 - 9\epsilon - p(4+\epsilon)}{4(4-\epsilon)}}$) for $\nu^{\rm ssc}_{\rm m} < \nu < \nu^{\rm ssc}_{\rm c}$ and $\propto A_{\rm W}^{\frac{16-8p-6\epsilon + p\epsilon}{4(4-\epsilon)}}\,(A_{\rm W}^{\frac{8 - 2\epsilon - p(4+\epsilon)}{4(4-\epsilon)}})$ for $\nu^{\rm ssc}_{\rm m} < \nu^{\rm ssc}_{\rm c} < \nu$. Considering the value of $p=1.9$ (2.1), the expected flux as a function of the density parameter is $F^{\rm}_{\nu}\propto A_{\rm w}^{\alpha_w}$ with $2.3 \leq\alpha_w \leq 2.25$ ($2.23 \leq\alpha_w \leq 2.16$) for $\nu^{\rm ssc}_{\rm m} < \nu < \nu^{\rm ssc}_{\rm c}$ and $-0.05 \leq -\alpha_w \leq 0.06$ ($0.03 \leq -\alpha_w \leq 0.14$) for $\nu^{\rm ssc}_{\rm m} < \nu^{\rm ssc}_{\rm c} < \nu $. Therefore, as $A_{\rm w}$ increases, the expected flux increases if it evolves in the cooling condition $\nu^{\rm ssc}_{\rm m} < \nu < \nu^{\rm ssc}_{\rm c}$ and decreases if it evolves in $\nu^{\rm ssc}_{\rm m} <\nu^{\rm ssc}_{\rm c} < \nu $.  The left-hand panels in Figure \ref{LC_ism} exhibit a similar behavior to those shown in Figure \ref{LC_wind}. For example, the evolution of the expected flux $F^{\rm ssc}_{\rm \nu}\propto t^{-\alpha}$ from $0.88\leq \alpha \leq 1.13$ ($0.99\leq \alpha \leq 1.25$) to $1.0\leq \alpha \leq 1.16$ ($1.11\leq \alpha \leq 1.28$) for $p=1.9$ ($2.1$) can be interpreted in terms of SSC spectral breaks; $\nu^{\rm ssc}_{\rm m}\propto t^{-\alpha}$ with $1.63\leq\alpha \leq 1.71$ ($1.50\leq\alpha \leq 1.58$) for $p=1.9$ ($2\leq p$), and $\nu^{\rm ssc}_{\rm c}\propto t^{-\alpha}$ with $0.42\leq\alpha \leq 0.50$. Similarly, the variation of the expected flux as a function of the density can be explained in terms of the cooling condition. In the case of $1<p<2$ ($2\leq p$), the expected flux is $F^{\rm}_{\nu}\propto n^{\frac{52-8p-13\epsilon + p\epsilon}{4(8-\epsilon)}}$ ($n^{\frac{44 - 9\epsilon - p(4+\epsilon)}{4(8-\epsilon)}}$) for $\nu^{\rm ssc}_{\rm m}<\nu < \nu^{\rm ssc}_{\rm c}$ and $\propto n^{\frac{6-8p-6\epsilon + p\epsilon}{4(8-\epsilon)}}\,(n^{\frac{8 - 2\epsilon - p(4+\epsilon)}{4(8-\epsilon)}})$ for $\nu^{\rm ssc}_{\rm m}<\nu^{\rm ssc}_{\rm c} < \nu$. Considering the value of $p=1.9$ (2.1), the expected flux as a function of the density is $F^{\rm}_{\nu}\propto n^{\alpha_w}$ with $1.15 \leq\alpha_w \leq 1.06$ ($1.11 \leq\alpha_w \leq 1.03$) for $\nu^{\rm ssc}_{\rm m} < \nu < \nu^{\rm ssc}_{\rm c}$ and $0.29 \leq -\alpha_w \leq 0.36$ ($0.01 \leq -\alpha_w \leq 0.07$) for $\nu^{\rm ssc}_{\rm m}<\nu^{\rm ssc}_{\rm c} < \nu$.  The right-hand panels in Figures \ref{LC_wind} and \ref{LC_ism} show the SSC spectra with the CTA (Southern array, green line), MAGIC (purple line) and {\itshape Fermi}/LAT (red line) sensitivities between 75 and 250~GeV at $3\times 10^4\,{\rm s}$ for a zenith angle of 20$^\circ$ \citep{2019ICRC...36..673F}. These panels display that whereas all the expected fluxes are below the {\itshape Fermi}/LAT sensitivity, only some of them are above of the Cherenkov Telescope Array (CTA) or the  Major Atmospheric Gamma-ray Imaging Cherenkov Telescop (MAGIC) Telescopes; depending on the set of parameter values. For example, the expected flux could be detected in both MAGIC and CTA for [$A_{\rm W}=0.1$ ($n=0.1\,{\rm cm^{-3}}$); $E=10^{52}\,{\rm erg}$], and not be detected by MAGIC or CTA for [$10^{-3}$ ($10^{-3}\,{\rm cm^{-3}}$); $10^{52}\,{\rm erg}$]. Given the values of [$10^{-3}$ ($10^{-3}\,{\rm cm^{-3}}$); $10^{53}\,{\rm erg}$], we can see that the expected flux could be detected by CTA for ${\rm p}=2.1$ and 2.3, but not for ${\rm p}=1.7$ or $1.9$. In the former case, MAGIC could not detect the expected flux for any parameter values.\\

\section{Application: 2FLGC}
\label{sec:app}
\subsection{Our representative sample of GRBs}\label{sub:afterglow}

In order to apply the current model, we select those GRBs from the 2FLGC \citep{2019ApJ...878...52A} with values of temporal and spectral indices with $\alpha_{\rm L}\gtrsim 1.5$ and $\Gamma_{\rm L}\approx 2$ (see Table~\ref{tab:sample}), which can hardly be described with closure relations of the standard synchrotron afterglow model, and also exhibit energetic photons above the synchrotron limit \cite[e.g., see][]{2010MNRAS.403..926G, 2019ApJ...883..134T, 2022ApJ...934..188F}. Our sample is formed by ten bursts (one short GRB and nine long GRBs), seven of which have a measured redshift. We briefly describe the multi-wavelength observations of our representative sample of GRBs.

\subsubsection{GRB~080825C}

On 25 August 2008 at 14:13:48 UT, the Gamma-ray Burst Monitor (GBM) instrument on board the {\itshape Fermi} telescope was triggered by GRB~080825C \citep{GCN8141}. The initial estimation set a duration of $T_{90}=23\ \rm s$ \citep{2008GCN..8141....1V}. Nevertheless, further spectral analysis of the data from \textit{Fermi}/GBM, revealed that the main emission lasted $T_{90}=27\ \rm s$ in the energy band $8-1000\ \rm keV$ \citep{2009ApJ...707..580A}. The fluence measured by the GBM instrument was $(0.11\pm0.04)\times 10^{-5}\,{\rm erg\,cm^{-2}}$. Moreover, this burst is the first detection of the \textit{Fermi}/LAT instrument of a GRB \citep{2008GCN..8183....1B,2016MNRAS.458.1728M}. During all the emission, the photons had energies below 1~GeV \citep{2008GCN..8183....1B}. No redshift was associated to this event.\\
The circumburst medium remains unconstrained for this burst due to the absence of X-ray and optical data.

\subsubsection{GRB~090510}

GRB~090510 was detected by the Burst Alert Telescope (BAT) instrument on board the {\it Neil Gehrels Swift Observatory} and by the {\itshape Fermi}/LAT instrument. Subsequently, the other instruments of both facilities observed the field of GRB~090510 \citep{2010ApJ...709L.146D, 2009GCN..9337....1U}. The duration of this burst was estimated to be $T_{90}=0.3\pm 0.1\,{\rm s}$ \citep{2009GCN..9337....1U, 2010ApJ...709L.146D}, with a corresponding fluence $(1.7\pm0.6)\times 10^{-5}\,{\rm erg\,cm^{-2}}$ and isotropic energy of $E_{\rm \gamma, iso}=(5.8\pm 0.5) \times 10^{53}\, {\rm erg}$ \citep{Ajello_2019}. Due to the large energy released by this event, {\itshape Fermi}/LAT reported 12 photons with energy greater than 1~GeV during the first three seconds after the trigger. Spectroscopic data from the Very Large Array (VLA) allowed to estimate a redshift of $z=0.903$ from the OII and H$\beta$ lines \citep{2009GCN..9353....1R}.\\
\cite{2010ApJ...709L.146D, 2012A&A...538L...7N} analyzed the X-ray and UV/optical/IR afterglow observations of GRB~090510. Given the SMC dust extinction with $A^{\rm host}_{V}=0.17^{+0.21}_{-0.17}$ mag, they reported early temporal and spectral indices of $\alpha_{\rm X}=0.74\pm0.03$ and $\alpha_{\rm Opt,1}\approx -0.2\pm 0.2$ evolving to $\alpha_{\rm Opt,2}= 0.80\pm 0.1$, and $\beta_{\rm X}= 0.8\pm 0.1$ and $\beta_{\rm Opt}=0.85\pm 0.05$ for X-ray and UV-optical-IR observations, respectively. The closure relations of the temporal and spectral indices of late X-ray and optical observations are {\small $F_{\rm \nu,X}\propto t^{-0.74.30\pm 0.03}\,\nu^{-0.8\pm 0.1}$ and $F_{\rm \nu,Opt}\propto t^{-0.80\pm 0.1}\,\nu^{-0.85\pm 0.05}$}, respectively. Although, the closure relations are similar to each other, the evolution in a constant-density medium under the condition $\nu^{\rm syn}_{\rm m}<\nu_{\rm Opt} < \nu_{\rm X} < \nu^{\rm syn}_{\rm c} $ is more favorable for $ p\approx 2.4\pm0.2$ or $ p\approx 2.6\pm0.2$, respectively. The evolution in stellar-wind environment in the same cooling condition leads to an atypical value of the spectral index with $p<1.4$.

\subsubsection{GRB~090902B}
GRB~090902B was detected by the {\itshape Fermi}/GBM instrument on 2 September 2009 at 11:05:08.31 UTC. The duration of this burst was estimated to be $T_{90}=19.33\,{\rm s}$, with a corresponding fluence of $(7.0\pm1.0)\times 10^{-5}\,{\rm erg\,cm^{-2}}$ and isotropic energy of $E_{\rm \gamma, iso}=(3.7\pm 0.3) \times 10^{53}\, {\rm erg}$ \citep{Ajello_2019}. This bright event was within the {\itshape Fermi}/LAT field of view initially at an angle of $51^\circ$ from the line of sight, and therefore this instrument showed an increment correlated with the {\itshape Fermi}/GBM trigger. Later, GRB~090902B was detected by the X-ray Telescope (XRT) \citep{2009GCN..9868....1K}, and by the Ultraviolet/Optical Telescope (UVOT) instruments \citep{2009GCN..9869....1S} on board the {\it Neil Gehrels Swift Observatory}, as well as by several other ground-based telescopes. The spectrum obtained with the Gemini-North telescope \citep{2009GCN..9873....1C} showed a series of metal absorption features corresponding to a redshift of $z=1.822$ \citep{2009GCN..9873....1C}.\\
\cite{2010ApJ...714..799P} conducted an analysis of the X-ray and UV-optical-IR afterglow data associated with GRB~090902B. Given the SMC-like dust extinction with $A^{\rm host}_{V}=0.20\pm 0.06$, the authors performed a temporal and spectral analysis resulting in early temporal indices of $\alpha_{\rm X}=1.30\pm0.04$ and $\alpha_{\rm Opt}\approx 1.60$, and $\beta_{\rm X}= 0.9\pm 0.1$ and $\beta_{\rm Opt}=0.68\pm 0.11$ for X-ray and UV-optical-IR observations, respectively. The closure relations of the temporal and spectral indices of late X-ray and optical observations are {\small $F_{\rm \nu,X}\propto t^{-1.30\pm 0.04}\,\nu^{-0.9\pm 0.1}$ and $F_{\rm \nu,Opt}\propto t^{-1.60}\,\nu^{-0.68\pm0.11}$}, respectively. The fact that the temporal (spectral) index for the optical observations is larger (lower) than the X-ray observations indicates that the closure relations of the synchrotron FS model evolve in a slow cooling regime through a wind-like medium ($\nu^{\rm syn}_{\rm m}<\nu_{\rm Opt}<\nu^{\rm syn}_{\rm c}< \nu_{\rm X}$) for $ p\approx 2.3\pm0.3$.

\subsubsection{GRB~090926A}

The {\itshape Fermi}/GBM triggered on GRB~090926A at 04:20:26.99 UTC on 26 September 2009 \citep{2011ApJ...729..114A}. The duration of this burst was estimated to be around $T_{90}=20\,{\rm s}$ \citep{2009GCN..9933....1B}, whereas the isotropic energy was measured as $E_{\rm \gamma, iso}=(3.7\pm 0.3) \times 10^{53}\, {\rm erg}$ \citep{2009GCN..9959....1G}. \cite{2011ApJ...729..114A} analyzed data from the {\itshape Fermi}/LAT and {\itshape Fermi}/GBM instruments for GRB~090926A, concluding the presence of a characteristic high-energy power law component. The photometry data set includes observations from {\itshape Swift}/XRT, {\itshape INTEGRAL}/SPI-ACS \citep{2009GCN..9933....1B}, {\itshape Suzaku}/WAM \citep{2009GCN..9951....1N}, {\itshape CORONAS}/Photon \citep{2009GCN.10009....1C}, the Konus-wind experiment \citep{2009GCN..9959....1G} and the {\itshape Swift/UVOT} \citep{2009GCN..9942....1M} instruments. Using data from the {\itshape VLT}/X-shooter \cite{2009GCN..9942....1M} estimated a redshift of $z=2.11$.\\ 
\cite{2010ApJ...718L..14S} and \cite{2011ApJ...732...29C} performed a temporal and spectral analysis of X-ray and optical data including UVOT observations from GRB~090926A. They reported temporal indices of $\alpha_{\rm X}=1.43\pm0.03$ and $\alpha_{\rm Opt}\approx 1.01^{+0.07}_{-0.03}$, and spectral indices of $\beta_{\rm X}= 1.12\pm 0.13$ and $\beta_{\rm Opt}=0.88\pm 0.07$ for X-ray and UV-optical-IR observations, respectively. The closure relations of the temporal and spectral indices of early X-ray and optical observations are {\small $F_{\rm \nu,X}\propto t^{-1.43\pm0.03}\,\nu^{-1.12\pm 0.13}$ and $F_{\rm \nu,Opt}\propto t^{-1.01^{+0.07}_{-0.03}}\,\nu^{-0.88\pm 0.07}$}, respectively. The fact that the temporal (spectral) index for the X-ray observations is greater than the optical observations suggests that the closure relations of the synchrotron FS model evolve in a slow cooling regime through a homogeneous medium ($\nu^{\rm syn}_{\rm m}<\nu_{\rm Opt}<\nu^{\rm syn}_{\rm c}< \nu_{\rm X}$) for $ p\approx 2.5\pm0.2$.

\subsubsection{GRB~110731A}

The {\itshape Fermi}/GBM instrument detected GRB~110731A on 31 July 2011 \citep{2009GCN..9942....1M}, estimating a duration of $T_{90}=7.49\,{\rm s}$ \citep{2011GCN.12221....1G}. Independently, {\itshape Swift}/BAT triggered on this event about 30~s after the initial {\itshape Fermi}/GBM trigger. \cite{2013ApJ...763...71A} calculated the isotropic energy measured using a power law and a band function model, obtaining $E_{\rm \gamma, iso}=(7.6\pm 0.2)\times 10^{53}\,{\rm erg}$. The {\itshape Swift}/XRT and the {\itshape Swift}/UVOT instruments began to observe the field of GRB~110731A a time $T+66.4$~s and $T+75$~s after the {\itshape Swift}/BAT trigger, respectively \citep{2011GCN.12215....1O}. Due to the energy of this event, several ground facilities followed up this GRB, such as the Faulkes Telescopes North and South \citep{2011GCN.12216....1B}, the Nordic Optical Telescope equipped with ALFOSC \citep{2011GCN.12220....1M}, Konus-Wind \citep{2011GCN.12223....1G}, the EVLA \citep{2011GCN.12227....1Z}, the Suzaku Wide-band All-sky Monitor (WAM) \citep{2011GCN.12244....1H} and the SAO RAS and Terskol observatories \citep{2011GCN.12333....1M}. The redshift value was determined to be $z=2.83$ using spectroscopic observations with the GMOS-N instrument on Gemini-North \citep{2011GCN.12225....1T}.\\
\cite{2013ApJ...763...71A} conducted an analysis of the X-ray and UV/optical afterglow data associated with GRB~110731A. After performing an analysis of the broadband spectral energy distribution (SED), the authors presented the spectral indices of $\beta_{\rm X}= 0.95^{+0.07}_{-0.09}$ and $\beta_{\rm Opt}=0.45^{+0.07}_{-0.09}$ for X-ray and UV/optical observations, at a time interval of 550 s. Similarly, the temporal analysis led to X-ray and optical indices of $\alpha_{\rm X}=1.10\pm0.02$ and $\alpha_{\rm Opt}=1.37\pm0.03$, respectively. The closure relations of the temporal and spectral indices of late X-ray and optical observations are {\small $F_{\rm \nu,X}\propto t^{-1.10\pm0.02}\,\nu^{-0.95^{+0.07}_{-0.09}}$ and $F_{\rm \nu,Opt}\propto t^{-1.37\pm0.03}\,\nu^{-0.45^{+0.07}_{-0.09}}$}, respectively. The fact that the temporal (spectral) index for the optical observations is greater (smaller) than the one for the X-ray observations indicates that the closure relations of the synchrotron FS model evolve in a slow cooling regime going through a wind-like medium ($\nu^{\rm syn}_{\rm m}<\nu_{\rm Opt}<\nu^{\rm syn}_{\rm c}< \nu_{\rm X}$) for $ p\approx 2.15\pm0.15$.

\subsubsection{GRB~130502B}

On 2 May 2013 at 07:51:11.76 UT, the \textit{Fermi}/GBM instrument was triggered by GRB~130502B. The estimated duration in the $50-300\ \rm keV$ energy band was measured to be $T_{90}=24\ \rm s$ \citep{2013GCN.14530....1V}. The GBM fluence was $(0.5\pm0.2)\times 10^{-5}\,{\rm erg\,cm^{-2}}$ \citep{Ajello_2019}. In the follow-up campaign after the initial GBM trigger, some of the instruments involved which observed the field of GRB~130502B include the \textit{Fermi}/LAT,\textit{Swift}/XRT, \textit{Swift}/UVOT, P60 and Konus-Wind \citep{2013GCN.14541....1C,2013GCN.14544....1B,2013GCN.14540....1M,2013GCN.14532....1K}.\\
The best-fit values of temporal and spectral indices derived by the Swift team and shown in the {\itshape Swift}/XRT repository $\alpha_{\rm X}= 1.62^{+0.27}_{-0.26}$ and $\beta_{\rm X}= 0.76^{+0.20}_{-0.19}$ are used.\footnote{https://www.swift.ac.uk/xrt\_live\_cat/00020266/, https://www.swift.ac.uk/xrt\_spectra/00020266/} Due to a lack of optical observations, the circumburst media for this burst, can not be restricted.

\subsubsection{GRB~141207A}

The \textit{Fermi}/GBM and \textit{Fermi}/LAT instruments were simultaneously triggered by GRB~141207A on 07 December 2017 \citep{2014GCN.17150....1B,2014GCN.17146....1A}. The \textit{Fermi}/GBM light curve exhibited a duration $T_{90}$ of approximately $20$~s in the $50-300$~keV band. The GBM fluence was $(2.0\pm0.8)\times 10^{-5}\,{\rm erg\,cm^{-2}}$ \citep{Ajello_2019}. The \textit{Swift}/XRT instrument began to observe the field of GRB~141207A about $T+13$~ hours, finding an uncatalogued X-ray source corresponding to the afterglow of this burst \citep{2014GCN.17149....1A}.\\
For this burst, the circumburst environment remains unconstrained as a result of the absence of X-ray and optical data.

\subsubsection{GRB~170214A}
On 14 February 2017 at 15:34:26.92 UT, the \textit{Fermi}/GBM was triggered by GRB~170214A. The GBM light curve showed multiple overlapping peaks with a duration $T_{90}\sim123\ \rm s$ in the $50$-$300$~keV energy band \citep{2017GCN.20675....1M}. The estimated isotropic energy was $E_{\gamma, \rm iso}=(32\pm5)\times10^{52}\ \rm erg$ and the GBM fluence was $(0.9\pm0.1)\times 10^{-5}\,{\rm erg\,cm^{-2}}$ \citep{Ajello_2019}. The \textit{Fermi}/LAT instrument was simultaneously triggered by this burst, and it observed more than 160 photons above 100~MeV and more than 13 photons above 1~GeV. The highest-energy event was a photon with an energy of 7.8~GeV \citep{2017GCN.20676....1R}. Approximately 41 hours after the initial GBM trigger, \cite{2017GCN.20686....1K} observed the optical counterpart of the burst with the ESO Very Large Telescope UT 2 equipped with the X-shooter spectrograph. The authors claimed a redshift of $z=2.53$ due to various absorption features in the low-energy optical observations.\\
The best-fit values of temporal and spectral indices derived by the Swift team and shown in the {\itshape Swift}/XRT repository $\alpha_{\rm X}= 1.3^{+0.5}_{-0.4}$ and $\beta_{\rm X}= 0.9\pm 0.5$ are used.\footnote{https://www.swift.ac.uk/xrt\_live\_cat/00020740/, https://www.swift.ac.uk/xrt\_spectra/00020740/} The value of the temporal index that best fits the observations collected with the optical instruments is $\alpha_{\rm Opt}= 1.38\pm 0.09$. The closure relations of the temporal and spectral indices of late X-ray and optical observations are {\small $F_{\rm \nu,X}\propto t^{-1.3^{+0.5}_{-0.4}}\,\nu^{-0.9\pm0.5}$ and $F_{\rm \nu,Opt}\propto t^{-1.38\pm0.09}$}, respectively. Similar to what happened in the GRB~131108A, the spectral index does not include optical frequencies.
Although, the closure relations are similar to each other, the evolution in homogeneous medium under the condition $\nu^{\rm syn}_{\rm m}< \nu^{\rm syn}_{\rm c}< \nu_{\rm Opt}< \nu_{\rm X}$ for $ p\approx 2.4\pm0.2$ is more favorable due to the absence of temporal breaks in both X-ray and optical observations.

\vspace{2cm}
\subsubsection{GRB~180720B}

The {\itshape Fermi} and {\itshape Swift} satellites triggered on GRB~180720B at 14:21:39.65 UT on 20 July 2018 \citep{2018GCN.22981....1R, 2018GCN.22980....1B}. The duration of the burst was confirmed by a posterior analysis to be $T_{90}=48.90\,{\rm s}$. The preliminary multipeaked structure lasted beyond the available event data range of the {\itshape Swift}/BAT instrument \citep{2018GCN.22998....1B}. The estimated isotropic energy was $E_{\rm \gamma, iso}=(0.39\pm 0.09)\times 10^{52}\, {\rm erg}$ and the GBM fluence was $(0.19\pm0.05)\times 10^{-5}\,{\rm erg\,cm^{-2}}$ \citep{Ajello_2019}. Optical and near-infrared (NIR) follow-up of GRB~180720B began observations about $T+73$~s \citep{2018GCN.22977....1S}. \cite{2018GCN.22996....1V} monitored the field of GRB~180720B with the {\itshape VLT}/X-shooter spectrograph, estimating a redshift of $z=0.654$ from the match of several absorption features revealed in the spectrum.\\
\cite{2019ApJ...885...29F} performed a temporal and spectral analysis of the X-ray and R-band optical observations of GRB~180720B. Based on early observations, the authors reported spectral indices of $\beta_{\rm X}= 0.697^{+0.010}_{-0.010}$ and $\beta_{\rm Opt}= 0.68\pm0.06$, and temporal indices of $\alpha_{\rm X}=1.26\pm 0.06$ and $\alpha_{\rm Opt}=1.22\pm0.02$ for X-ray and optical observations, respectively.
The closure relations of the temporal and spectral indices of X-ray and optical observations are {\small $F_{\rm \nu,X}\propto t^{-1.26\pm 0.06}\,\nu^{-0.697^{+0.010}_{-0.010}}$ and $F_{\rm \nu,Opt}\propto t^{-1.22\pm0.02}\,\nu^{-0.68\pm 0.06}$}, respectively. Although, the closure relations are similar to one another, the evolution in stellar-wind or constant-density medium under the condition $\nu^{\rm syn}_{\rm m}<\nu_{\rm Opt} < \nu_{\rm X} < \nu^{\rm syn}_{\rm c} $ is more favorable for $ p\approx 2.0\pm0.2$ or $ p\approx 2.6\pm0.2$, respectively. The evolution in the condition ${\rm max} \{\nu^{\rm syn}_{\rm m}, \nu^{\rm syn}_{\rm c}\} <\nu_{\rm Opt} < \nu_{\rm X} $ leads to an atypical value of the spectral index with $p<1.5$. The temporal break exhibited at $2.6\times 10^{5}\,{\rm s}$ with a temporal index $1.70\pm0.19$ in the X-ray light curve \citep{2019ApJ...885...29F} is consistent with the post-jet break phase in stellar wind for $ p\approx 2.0\pm0.2$.

\subsection{Data Reduction}

\subsubsection{{\itshape Fermi}/LAT Data}

The {\itshape Fermi}/LAT data files were retrieved from the science data repository.\footnote{https://fermi.gsfc.nasa.gov/cgi-bin/ssc/LAT/LATDataQuery.cgi} The {\itshape Fermi}/LAT data set was analyzed in the 0.1-100~GeV energy range using time-resolved likelihood analysis and the {\itshape Fermi} Science tools \texttt{ScienceTools 2.2.0}.\footnote{https://fermi.gsfc.nasa.gov/ssc/data/analysis/software/} Following the unbinned likelihood analysis presented by the {\itshape Fermi}/LAT team,\footnote{https://fermi.gsfc.nasa.gov/ssc/data/analysis/scitools/likelihood\_tutorial.html} we use the responses provided by \cite{2019ApJ...878...52A} for each burst. We use the \texttt{gtselect} tool to select a region of interest (ROI) within a radius of 15$^{\circ}$ around the point of the burst and impose a cut on the zenith angle greater than 100$^{\circ}$. Furthermore, before evaluating the ROI cut, we acquire the most relevant time intervals in the data using the \texttt{gtmktime} tool. We use diffuse components and point sources from 4FGL-DR3 \citep[e.g., see \texttt{make4FGLxml}][]{2022ApJS..260...53A} to define the model required to characterize the source. Using GALPROP \texttt{gll\_iem\_v07} and a PL spectrum, we establish a point source at the location of this burst and a diffuse galactic component. In addition, the extragalactic background \texttt{iso\_P8R3\_SOURCE\_V3\_v1} was used.\footnote{https://fermi.gsfc.nasa.gov/ssc/data/access/lat/BackgroundModels.html} The spectral index for each burst is set at the value stated by \cite{2019ApJ...878...52A} in Table 4. We use the tool \texttt{gtltcube} with a step $\delta \theta=0.025$, a bin size of 0.5 and a maximum zenith angle of 100$^{\circ}$ to create a lifetime cube. We consider a region of 30$^{\circ}$ around the GRB position and define 100 spatial bins in longitude/latitude and 50 energy bins to create the exposure map with \texttt{gtexpmap}. Furthermore, we carry out the likelihood analysis using \texttt{pyLikelihood}.\footnote{https://fermi.gsfc.nasa.gov/ssc/data/analysis/scitools/python\_tutorial.html} Finally, using the \texttt{gtsrcprob} tool, we retrieve photons with a probability greater than 90$\%$ to be correlated with each burst. As follows, we describe the relevant features of the energetic photons associated with each burst.

\paragraph{GRB~080825C}~ At 3.06~s after the trigger time, the first high-energy photon was detected with measured energy of 153.4~MeV. In this burst, there were 14 photons with energy over 100~MeV. The highest energy photon detected in the LAT data was 682.9~MeV, 28.3~s after the trigger time.

\paragraph{GRB~090510}~ At 0.18~s after the GBM trigger, the first high-energy photon was detected with measured energy of 526.4~MeV. The energy range of the photons in this burst was extensive, with 261 photons over 100~MeV and 33 exceeding 1~GeV. At 0.82~s after the GBM trigger, the highest-energy photon in the LAT data had a measured energy of 19.9~GeV.

\paragraph{GRB~090902B}~ The first detection of a high-energy photon occurred around 1.86~s after the trigger of the GBM, with a measured energy of 284.4~MeV. The photon energy spectrum in this burst had a wide range, with 469 photons with energies over 100~MeV, 67 photons surpassing 1~GeV, and seven photons surpassing 10~GeV. The highest-energy photon in the LAT data was measured to have an energy of 39.88~GeV at about 81.7~s after the GBM trigger.

\paragraph{GRB~090926A}~A photon with an energy of 130.6~MeV, the first photon in a series of high-energy photons, was seen 2.21~s after the GBM trigger. The burst under analysis exhibited a diverse spectrum of photon energies, whereby 339 photons had energies above 100~MeV, 31 possessed energies surpassing 1~GeV, and two possessed energies exceeding 10~GeV. Approximately 24.84~s after the GBM trigger, the LAT instrument detected a photon with a maximum energy of 19.46~GeV.

\paragraph{GRB~110731A}~The first detection of a high-energy photon, with an energy measurement of 817.1~MeV, occurred at a time interval of 3.19~s after the trigger time. A diverse range of photon energies was generated during this burst, with 40 photons over 100~MeV and an additional four photons surpassing 1~GeV. The most energetic photon seen by the LAT instrument was observed to occur 1.93~s after the trigger time, with measured energy of 8.27~GeV.

\paragraph{GRB~130502B}~There was a wide variety of energies present in this burst, with 68 photons having energy more than 100~MeV and 2 having energy more than 10~GeV. At 222.1~s after the GBM trigger, the highest energetic photon detected in the LAT data had an energy of 31.1~GeV, followed by a photon with an energy of 17.3~MeV detected at 48.2~s after the GBM trigger.

%\paragraph{GRB~131108A}~The first detection of a high-energy photon occurred 0.02~s after the BAT trigger, with an energy of 168.5 MeV. The energy spectrum of the photons seen in this burst had a significant degree of variability, spanning a wide range. Specifically, 182 photons were detected with energies over 100 MeV and three photons surpassing 1~GeV. The LAT instrument detected the photon at 66.3~s after the BAT trigger, measuring its energy to be 1.5~GeV.

\paragraph{GRB~141207A}~The first energetic photon, measured at 765.3.5~MeV, was detected 3.9~s after the GBM trigger. In this burst, there were 19 photons with energy over 100~MeV and 11 with energies above 1~GeV. At 734.3~s after the GBM trigger, the highest energetic photon detected in the LAT data had an energy of 5.5~GeV.

\paragraph{GRB~170214A}~The first high-energy photon, with a measured energy of 152.5~MeV, was detected 39.5~s after the GBM trigger. A wide variety of energies was present in this burst, with 217 photons having energy more than 100~MeV and 13 having energies greater than 1~GeV. At 103.6~s after the GBM trigger, the highest energetic photon detected in the LAT data had an energy of 7.8~GeV.

\paragraph{GRB~180720B}~The first detection of a high-energy photon occurred 12.5~s after the BAT trigger, and its energy was measured to be 175.2~MeV. The energy spectrum of the photons in this burst exhibited a wide range, with 129 photons with energies over 100~MeV and eight photons above the threshold of 1~GeV. The LAT instrument detected the photon with a maximum energy of 142.4~s after the BAT trigger, and its energy was measured to be 4.9~GeV.

\subsubsection{Swift/XRT Data}

The Swift/XRT followed-up GRB~090510, 090902B, 090926A, 110731A, 130502B, 131108A, 170214A and 180720B in different series of observations \citep{2010ApJ...709L.146D, 2009GCN..9868....1K, 2011ApJ...729..114A, 2011GCN.12215....1O, 2013GCN.14541....1C, 2013GCN.15468....1S, 2017GCN.20691....1B, 2018GCN.22986....1E}. This instrument monitored these bursts in the photon counting (PC) and windowed-timing (WT) modes with spectrum exposures from hundreds to hundreds of thousands of seconds. The best-fitting absorption columns (intrinsic) ranges from $2.1^{+1.4}_{-1.3}\times 10^{21}$ to $4.4^{+3.1}_{-3.0}\times 10^{21}\,{\rm cm^{-2}}$, and from $10^{+11.2}_{-10.0}\times 10^{20}$ to $2.3^{+0.8}_{-0.6}\times 10^{22}\,{\rm cm^{-2}}$ for WT and PC modes, respectively. Data sets from the Swift/XRT instrument were obtained from the publicly accessible database of the Swift website.\footnote{{\rm https://www.swift.ac.uk/burst\_analyser/00922968/}} The flux density at 10~keV is converted to 1~keV using the conversion factor determined in \cite{2010A&A...519A.102E}.

\subsubsection{Optical Data}

Optical data for GRB~090510 (White-band), GRB~090902B (R-band), GRB~090926A (V-band), GRB~110731A (White- and V-band), GRB~131108A (White-, B-, U- and W1-band), GRB~170214A (White- and R-band) and GRB~180720B (R-band) were taken from \cite{2016ApJ...831...22F}, \cite{2010ApJ...714..799P}, \cite{2010ApJ...720..862R}, \cite{2015ApJ...804..105F}, \cite{2019ApJ...886L..33A, 2014arXiv1407.0238G, 2013GCN.15502....1C, 2013GCN.15485....1V, 2013GCN.15484....1V}, \cite{2017ApJ...844...56T, 2017GCN.20691....1B, 2017GCN.20687....1M, 2017GCN.20686....1K} and \cite{2019ApJ...885...29F}, respectively.

\subsection{Results and Discussion}

\begin{figure*}
\centering{
\subfloat[][\centering{GRB~080825C}]{\includegraphics[scale=0.45]{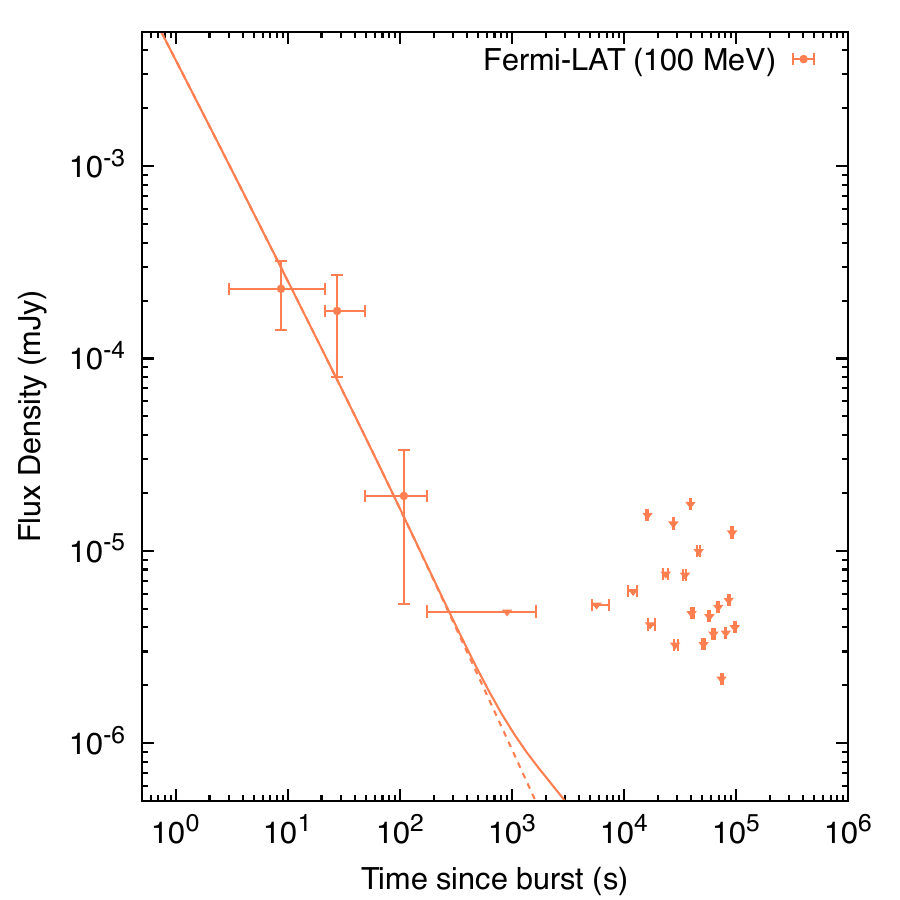} \label{fig:ISM_GRB080825C}}
\subfloat[][\centering{GRB~080825C}]{\includegraphics[scale=0.45]{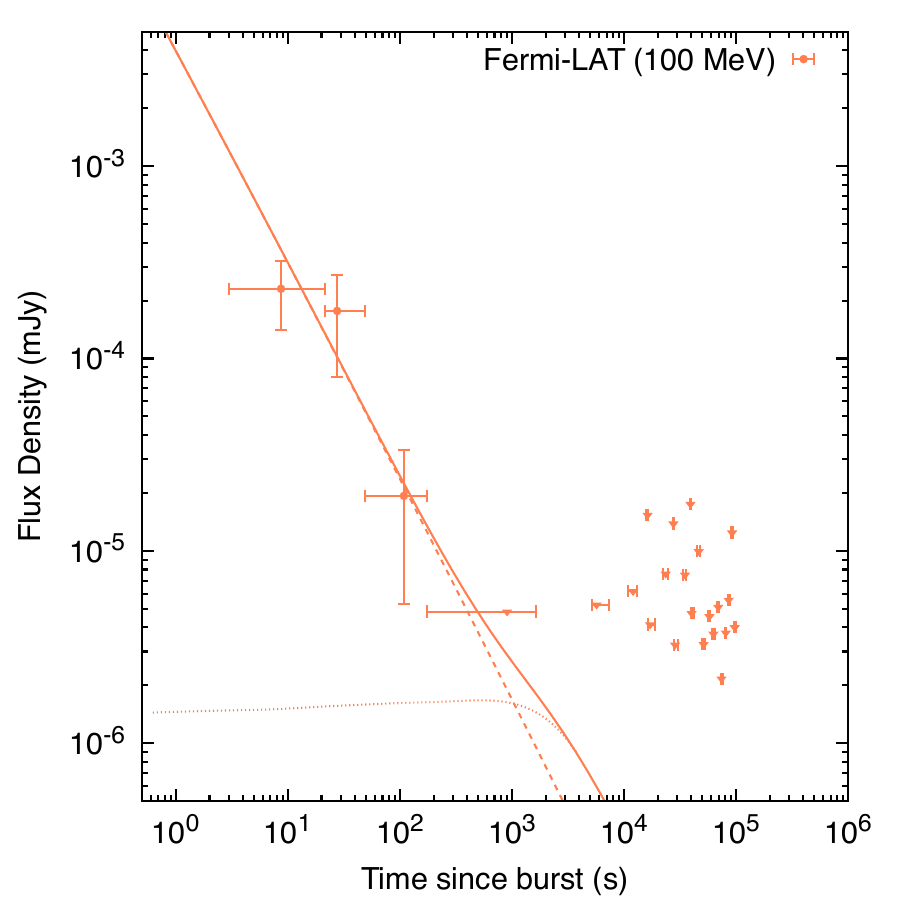} \label{fig:Wind_GRB080825C}}

\subfloat[][\centering{GRB~130502B}]{\includegraphics[scale=0.45]{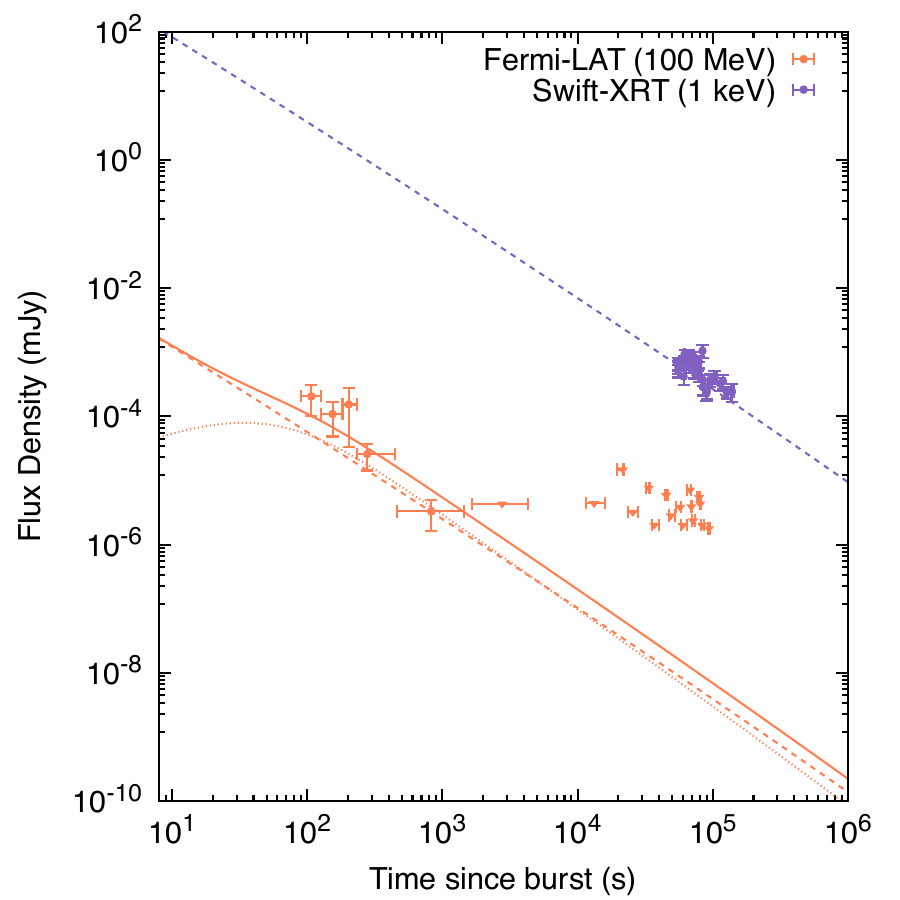} \label{fig:ISM_GRB130502B}}
\subfloat[][\centering{GRB~130502B}]{\includegraphics[scale=0.45]{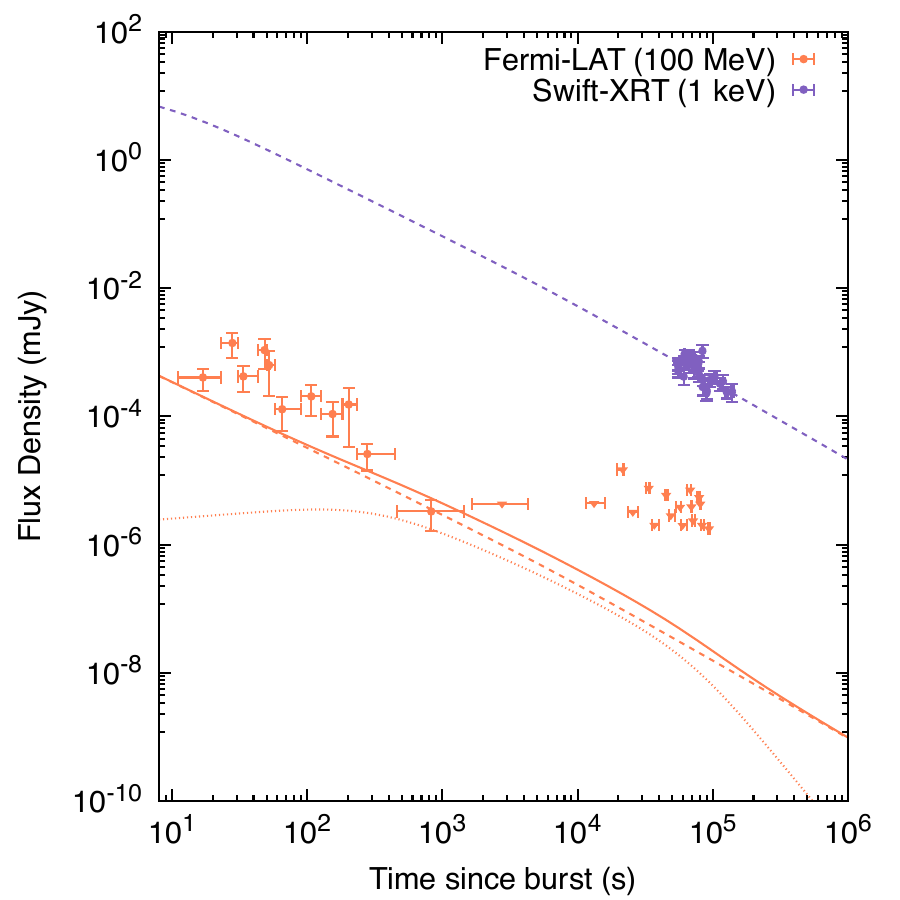} \label{fig:Wind_GRB130502B}}

\subfloat[][\centering{GRB~141207A}]{\includegraphics[scale=0.45]{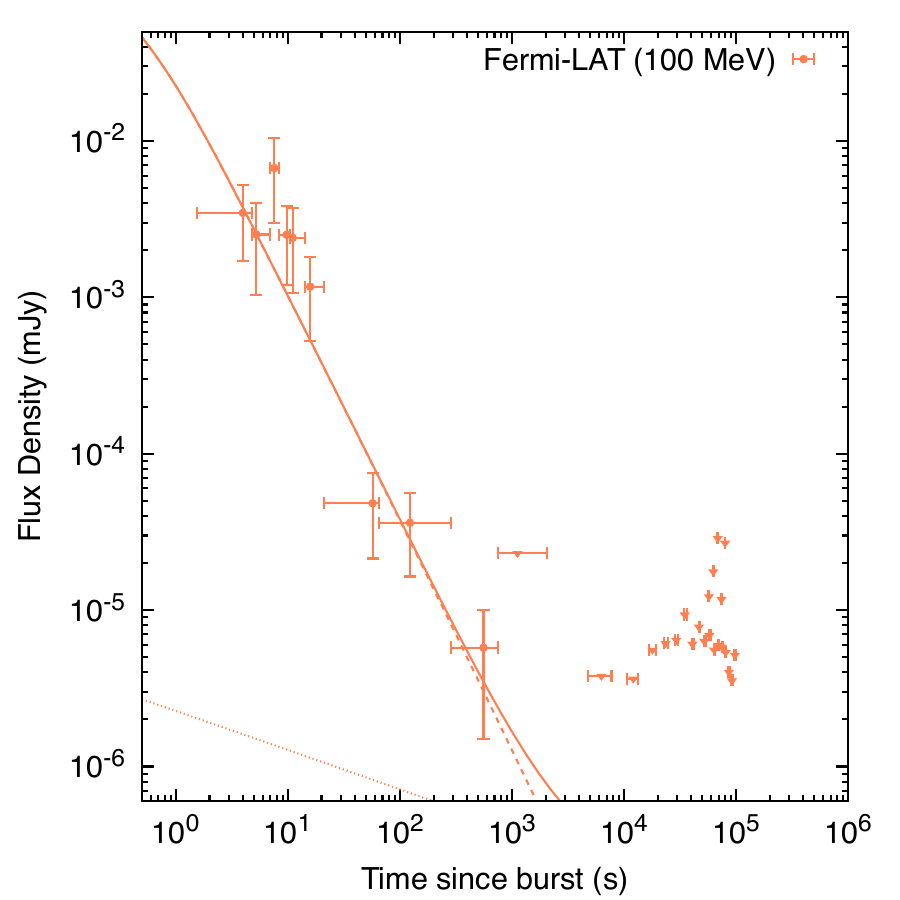} \label{fig:ISM_GRB141207A}}
\subfloat[][\centering{GRB~141207A}]{\includegraphics[scale=0.45]{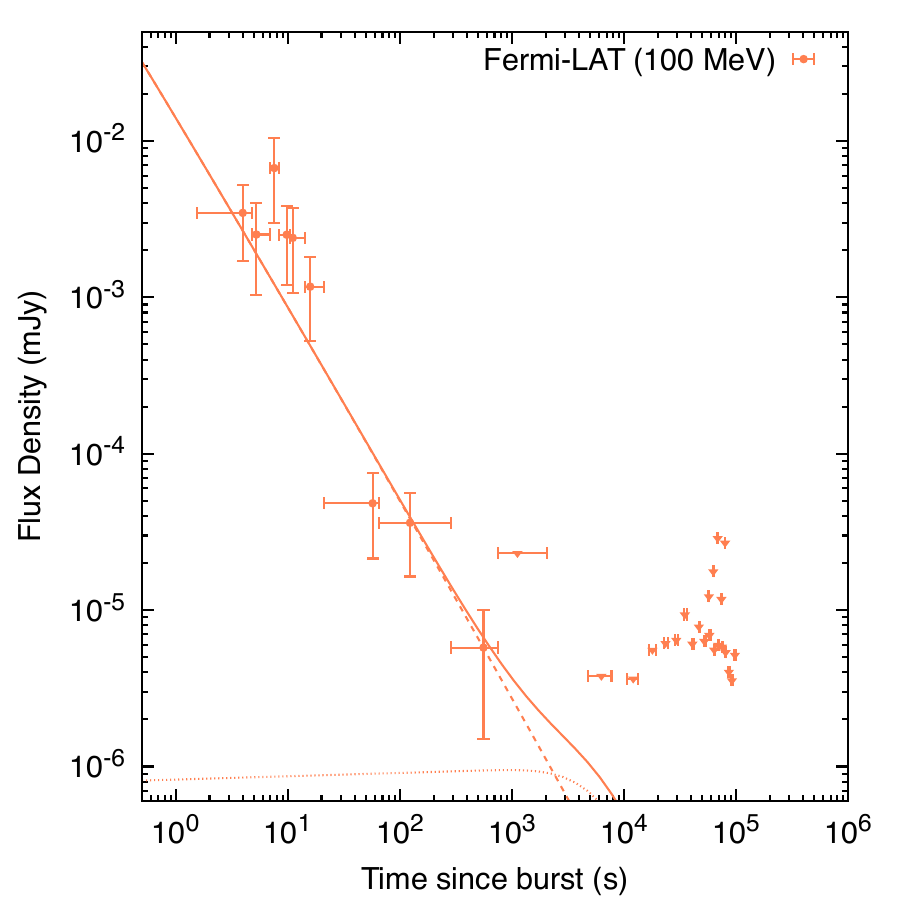} \label{fig:Wind_GRB141207A}}
}
\caption{The LAT and X-ray observations with the best-fit curves using the FS model evolving in the stellar-wind (right) and homogeneous (left) environment. The dashed and dotted lines correspond to synchrotron and SSC models, respectively.}\label{fig:LC1}
\end{figure*}

\begin{figure*}
\centering{
\subfloat[][\centering{GRB~090510}]{\includegraphics[scale=0.6]{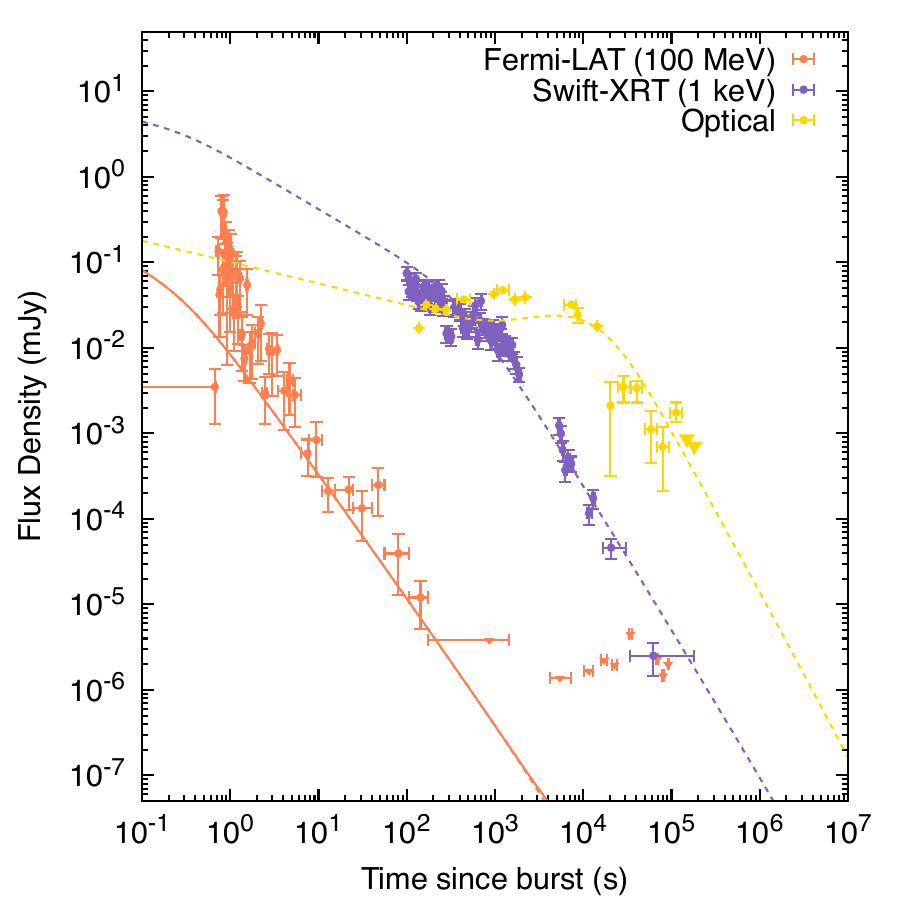} \label{fig:ISM_GRB90510.0}}
\subfloat[][\centering{GRB~090926A}]{\includegraphics[scale=0.6]{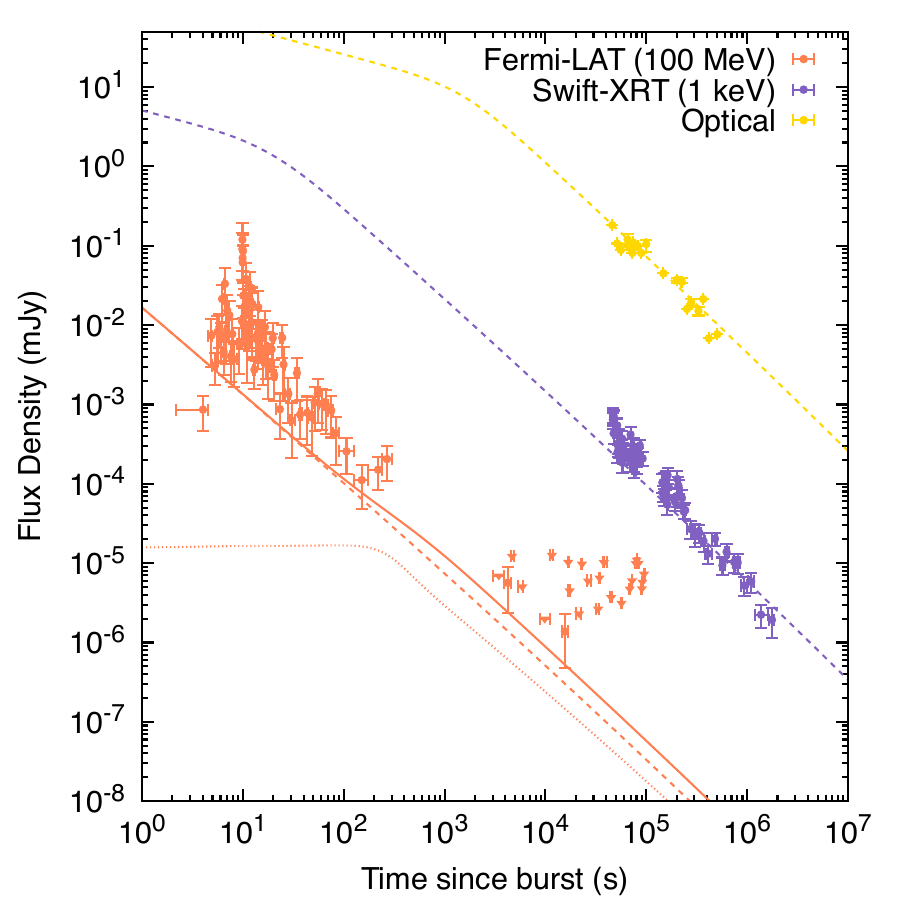} \label{fig:ISM_GRB090926A}}

%\subfloat[][\centering{GRB~131108A}]{\includegraphics[scale=0.6]{ISM_GRB131108ALightCurve.pdf} \label{fig:ISM_GRB131108A}}
\subfloat[][\centering{GRB~170214A}]{\includegraphics[scale=0.6]{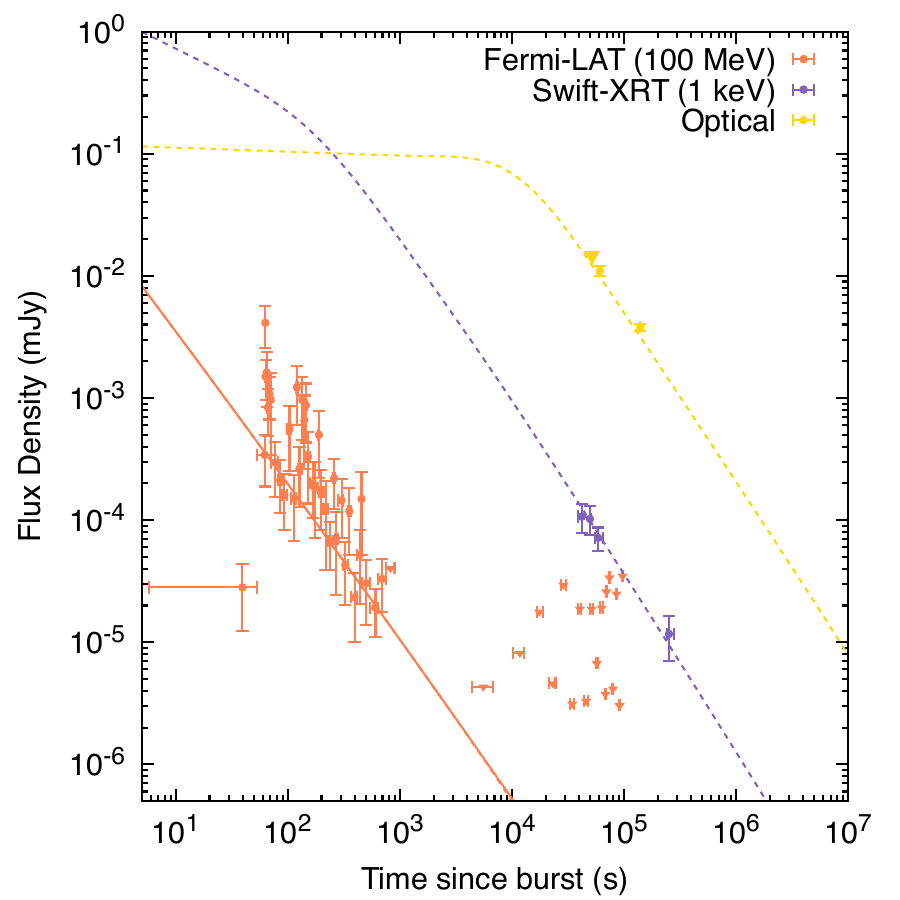} \label{fig:ISM_GRB170214A}}
}
\caption{The LAT (peach), X-ray (purple) and optical (yellow) observations with the best-fit curves using the forward-shock model evolving in the homogeneous medium. The dashed and dotted lines correspond to synchrotron and SSC models, respectively.}\label{fig:LC2}
\end{figure*}

\begin{figure*}
\centering{
\subfloat[][\centering{GRB~090902B}]{\includegraphics[scale=0.6]{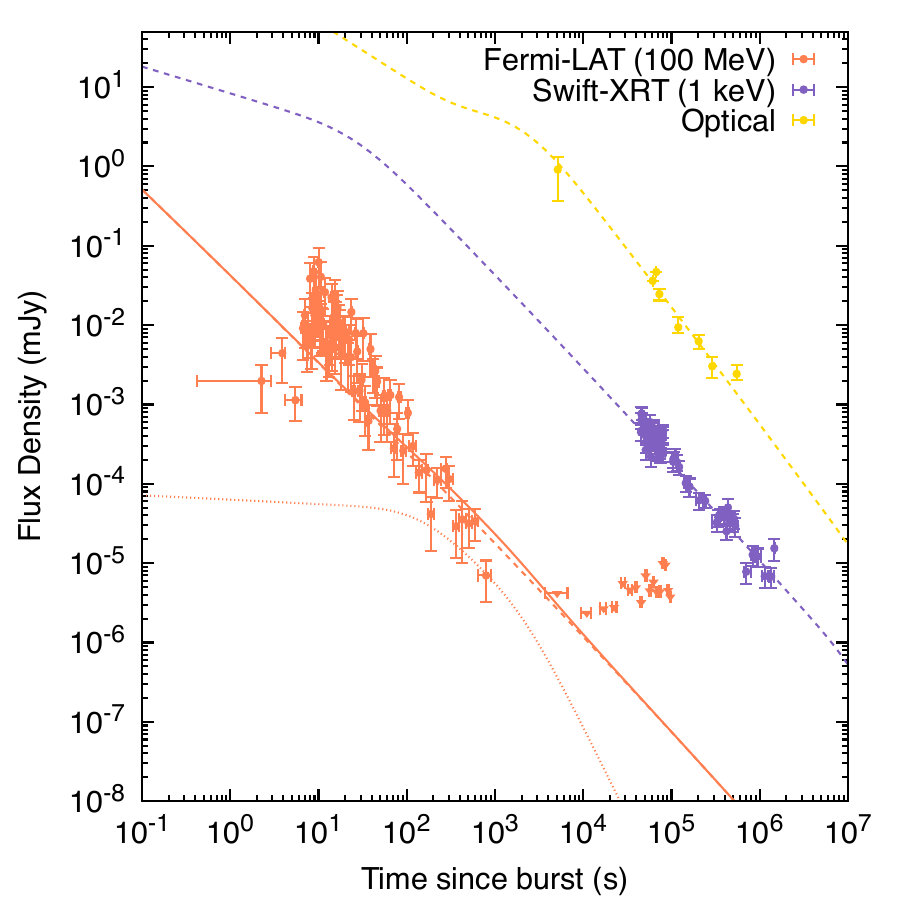} \label{fig:Wind_GRB090902B}}
\subfloat[][\centering{GRB~110731A}]{\includegraphics[scale=0.6]{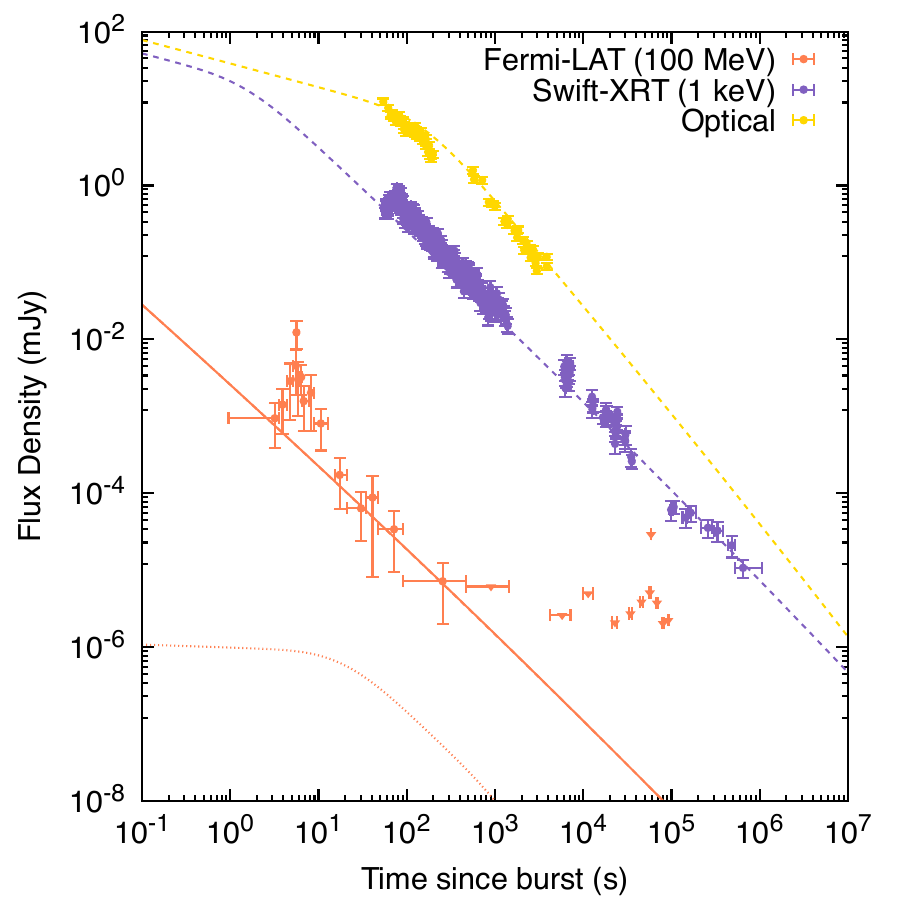} \label{fig:Wind_GRB110731A}}

\subfloat[][\centering{GRB~180720B}]{\includegraphics[scale=0.6]{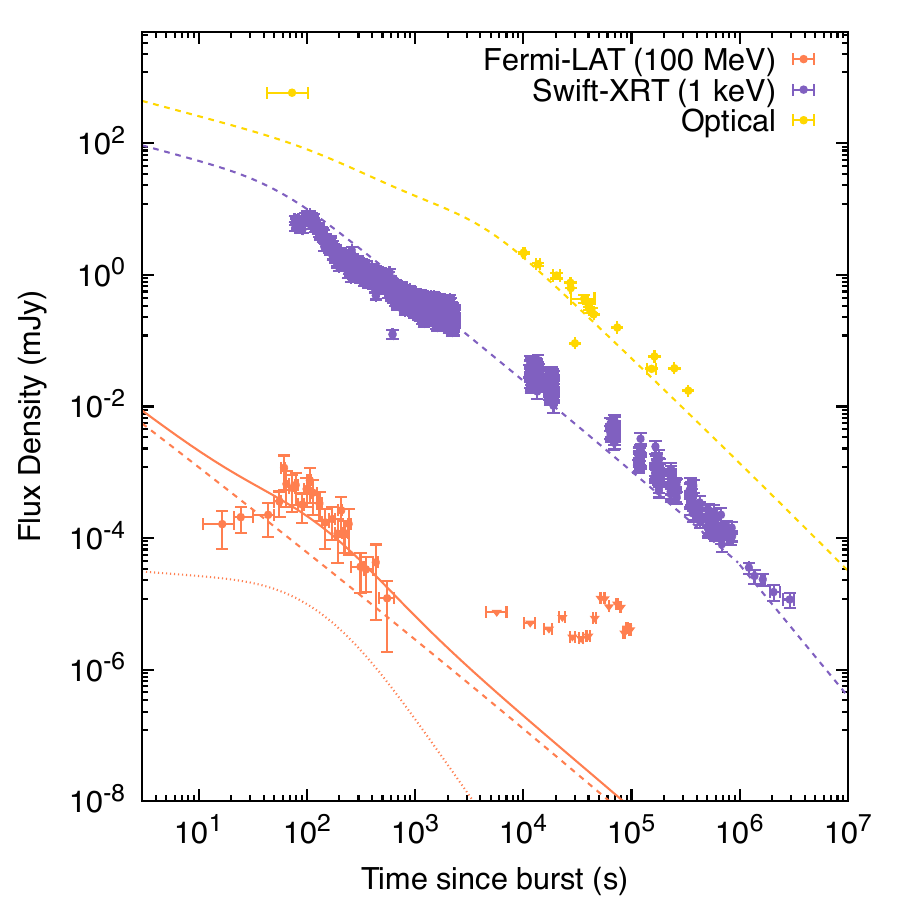} \label{fig:Wind_GRB180720B}}

}
\caption{The LAT, X-ray and optical observations with the best-fit curves using the FS model evolving in stellar-wind environment.  The dashed and dotted lines correspond to synchrotron and SSC models, respectively.}\label{fig:LC3}
\end{figure*}

\begin{figure*}
 \centering
 \includegraphics[scale=0.5]{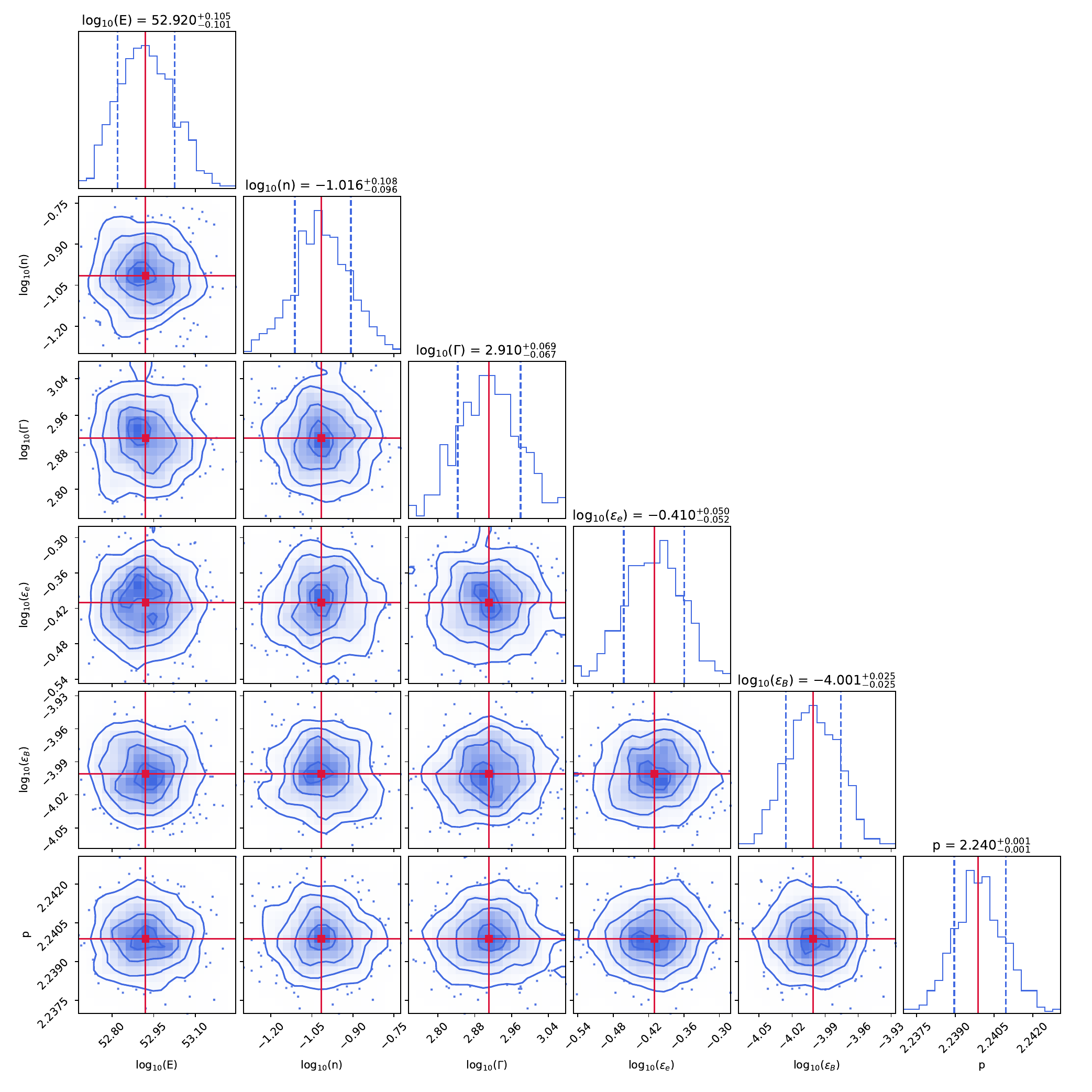}
 \caption{Corner plot of the parameters obtained from modelling the multiwavelegth afterglow observations of GRB~080825C with the constant-density model shown in Section \ref{sec:model}. The statistics for all parameters involved in the MCMC simulations are reported in Table \ref{Table:ISM_Fit}.}
 \label{fig:MCMC-ISM_GRB080825C}
\end{figure*}

We use the analysis of the closure relations shown in the subsection \ref{sub:afterglow} to describe our GRB sample with the current model evolving in the stellar wind or homogeneous environment. The panels in Figure~\ref{fig:LC1} display the LAT observations of GRB~080825C, GRB~130502B, and GRB~141207A with the best-fit curve generated by the FS model evolving in the stellar wind (right) and constant density (left) environment. Due to these three bursts having unknown redshifts, we assume a value of $z=1.0$ to estimate the total radiated energy and the luminosity distance. We show the afterglow evolution in both the stellar wind and homogeneous environment because the circumburst environments cannot be constrained as a result of the absence of optical data for GRB~130502B and X-ray and optical data for GRB~080825C and GRB~141207A. The panels in Figure~\ref{fig:LC2} show the LAT, X-ray, and optical observations of GRB~090510, GRB~090926A and GRB~170214 with the best-fit curve generated by the FS model evolving in the constant-density medium, and the panels in Figure~\ref{fig:LC3} show the LAT, X-ray and optical observations of GRB~090902B, GRB~110731A and GRB~180720B with the best-fit curve generated by the FS model evolving in the stellar-wind environment. We use Markov-Chain Monte Carlo (MCMC) simulations with the eight parameters used for the complete sample of GRBs to find the best-fit values that describe the multi-wavelength afterglow observations with the SSC and synchrotron FS models. To represent all the data in this case, a total of 15900 samples and 4400 tuning steps are used. The effect of EBL absorption as proposed in \cite{2017A&A...603A..34F} was adopted. 
We only display Figure~\ref{fig:MCMC-ISM_GRB080825C}, which corresponds to GRB 080825C, for showing the best-fit values and the median of the parameter posterior distributions. Tables~\ref{Table:ISM_Fit} and \ref{Table:Wind_Fit} list the best-fit values found with MCMC simulations after describing the multiwavelength afterglow observations with a synchrotron and SSC model evolving in both types of considered media. Tables~\ref{Table:ISM_der} and \ref{Table:Wind_der} display the synchrotron and SSC spectral breaks in a constant-density and a stellar-wind medium, respectively, which are calculated with the best-fit values reported in Tables~\ref{Table:ISM_Fit} and \ref{Table:Wind_Fit}.   We note that, while it may appear that the early LAT lightcurves are better fitted by the pure SSC model, we must also simultaneously explain the X-ray and optical observations. These are well fitted with the synchrotron model, so the synchrotron component is required and we are not able to consider a pure SSC model just for the LAT curves. Furthermore, with the parameters found, the SSC flux decreases very slowly and gives a small contribution of the early LAT data, which is less compared to the synchrotron radiation.

%Table~\ref{Table3} shows the synchrotron and SSC spectral breaks from reverse and forward shocks, which are estimated with the best-fit values reported in Table~\ref{Table6}.

\subsubsection{Microphysical parameters}

The best-fit values of the microphysical parameter given to accelerate electrons lie in the range of $0.3\leq \varepsilon_{\rm e}\leq 0.9$. In the constant-density scenario, the synchrotron afterglow model used for modeling the LAT light curves of GRB~090926A, GRB~141207A and GRB~170214A shows that they lie in the fast-cooling regime, while the light curves of GRB~080825C, GRB~090510 and GRB~130520B lie in the slow-cooling regime; see Table \ref{Table:ISM_der}. In the stellar-wind scenario, the synchrotron afterglow model used for modeling our sample shows that the light curves lie in the fast-cooling regime; see Table \ref{Table:Wind_der}. The results indicate that although the fraction of the total energy density given to accelerate electrons is much greater than $\varepsilon_{\rm e} \gg 0.1$, the shock-accelerated electrons are not in the fast-cooling regime during the entire LAT light curve. This indicates that for some GRBs a transition from radiative to adiabatic regime occurs at the beginning of the LAT observations.\\

The best-fit values of the magnetic microphysical parameter lie in the interval $10^{-5}\leq \varepsilon_{\rm B}\leq 10^{-1}$. As such, they are in the range of values required to model the multiwavelength afterglow observations in a large sample of GRBs; $10^{-5} \lesssim \epsilon_{B} \lesssim 10^{-1}$ \citep{1999ApJ...523..177W, 2002ApJ...571..779P, 2003ApJ...597..459Y, 2005MNRAS.362..921P, 2014ApJ...785...29S}. \cite{2019ApJ...883..134T} conducted a comprehensive analysis of temporal and spectral indices, meticulously examining the closure relations within a sample of 59 LAT-detected bursts that were carefully chosen. They showed that although the traditional synchrotron emission model adequately explains the spectrum and temporal indices in the majority of instances, a significant proportion of bursts can hardly be characterized by this model. Furthermore, they reported that those satisfying the closure relations are in the slow-cooling regime ($\nu^{\rm syn}_{\rm m}<\nu_{\rm LAT}<\nu^{\rm syn}_{\rm c})$ as long as the microphysical parameter is unusually low ($\epsilon_B<10^{-7}$). Our results show that the closure relations of a fraction of bursts can be satisfied with the synchrotron afterglow model in the radiative regime and typical values of $\epsilon_B$.

%\vspace{1cm}
\subsubsection{The post jet-break decay phase}

In a time scale of days, during the post-jet break decay phase, it is expected that the afterglow lies in the adiabatic regime rather than the radiative regime \citep{Racusin2009}. Therefore, the multi-wavelength observations could be described with $F^{\rm syn}_\nu \propto t^{-p}$, for $\nu^{\rm syn}_{\rm m, f} < \nu < \nu^{\rm syn}_{\rm c, f}$ or ${\rm max}\{\nu^{\rm syn}_{\rm m, f}, \nu^{\rm syn}_{\rm c, f}\}<\nu$ \citep[see e.g.][]{Fraija2022}, which are distinct from the temporal decay indices found for our sample \citep{Pereyra2022,Becerra2019}, except GRB~090510 and GRB~110731A. This implies that, except for these bursts, they were most likely emitted from a wide outflow with a significant half-opening angle, as shown by multi-wavelength observations, which display no indication of late steep decays. Based on the best-fit values, the jet opening angles become $\gtrsim 8^\circ$, and for GRB~090510 and GRB~110731A, they are $\theta_j\approx 0.5^\circ$ and $ 2^\circ$, respectively, which lies in the usual values \citep[$\theta_j\lesssim 10^\circ$;][]{Bloom2001}.

\vspace{1cm}

\subsubsection{Efficiency of equivalent kinetic energy}

The efficiency provides crucial information on the gamma-ray emitting process. The best-fit values of the equivalent kinetic energies $5.72\times 10^{52}\leq E\leq 10^{54}\,{\rm erg}$ and the isotropic energies in gamma-rays reported by the GBM and LAT instruments in the range of $4.0 \times 10^{51}\leq E_{\rm \gamma, iso}\leq 1.3\times 10^{54}\,{\rm erg}$ \citep{Ajello_2019} lead to kinetic efficiencies in the range of $0.03 \lesssim \eta_{\rm K} \lesssim 0.32$, which are typical compared to those values reported in the literature \citep{2001ApJ...557..399G, 2007ApJ...655..989Z, 2015PhR...561....1K}, and a kinetic efficiency of $\eta\approx 0.03$ for GRB~180720B , which is very low. The atypical value of efficiency for GRB~180720B was estimated considering the isotropic energy reported in the 2FLGC. If we would have considered the isotropic energy reported in other analyses \citep[$3\times10^{53}\,{\rm erg}$][]{2019Natur.575..464A,Fraija2019}, the kinetic efficiency would have been $\eta\approx 0.24$.
%Similarly, the second-lowest efficiency, which corresponds to $\eta=0.03$ for GRB~130502B. If we had considered other redshifts, a typical value would be obtained.

% These results suggest that isotropic energy of the long-lasting emission must be considered \\

\vspace{0.7cm}
\subsubsection{The profile of the circumburst environment}
 The best-fit values of the wind parameter lie in the range of $10^{-2}\lesssim A_{\rm W} \lesssim 1$, typical for GRBs identified as powerful bursts \citep{2013ApJ...763...71A, 2014ApJ...781...37P, 2014Sci...343...38V, 2012ApJ...751...33F, 2008Natur.455..183R,Fraija2017,Becerra2017}. Similarly, the values found of the homogeneous medium in the range $4.6\times 10^{-3} \lesssim n \lesssim 1 \,{\rm cm^{-3}}$ are usual with those found for other GRBs \citep{2019ApJ...885...29F, 2021ApJ...908...90A, 2021arXiv210602510H, 2023arXiv230606372L}. Considering the fact that short bursts detonate at very low densities \citep{2006ApJ...650..261S,2014ARA&A..52...43B}, the value of $n=4.6\times 10^{-3}\,{\rm cm^{-3}}$ obtained for GRB~090510 is compatible with the observations. Furthermore, the joint detection and modeling of the gravitational and electromagnetic signatures \citep{2017ApJ...848L..13A, 2017ApJ...848L..14G}, which were associated with a fusion of two neutron stars \citep{2017PhRvL.119p1101A}, provided values of circumburst densities consistent with the ones obtained for GRB~130502B.

Similarly to other bursts identified by the LAT instrument and predicted values from numerical simulations \citep{Tchekhovskoy2008}, the best-fit values of the initial bulk Lorentz factor fall in the range $10^2\lesssim \Gamma\lesssim 10^3$. Since our GRB sample included the highest energetic photons, we expect the bulk Lorentz factor values to coincide with those of the strongest bursts seen by the {\itshape Fermi}/LAT \citep{2011ApJ...729..114A, 2012ApJ...755...12V, 2013ApJ...763...71A, 2009ApJ...706L.138A, 2010ApJ...716.1178A, 2014Sci...343...42A, 2019ApJ...879L..26F, 2019ApJ...885...29F}.

%Therefore, a distinct process needs to be invoked to interpret these photons. 
\subsubsection{The highest energy photons}
\begin{figure*}
 \centering{
 
  \subfloat[Sub a) GRB~080825C][\centering{LAT light curve for GRB~080916C.}]{\includegraphics[scale=0.35]{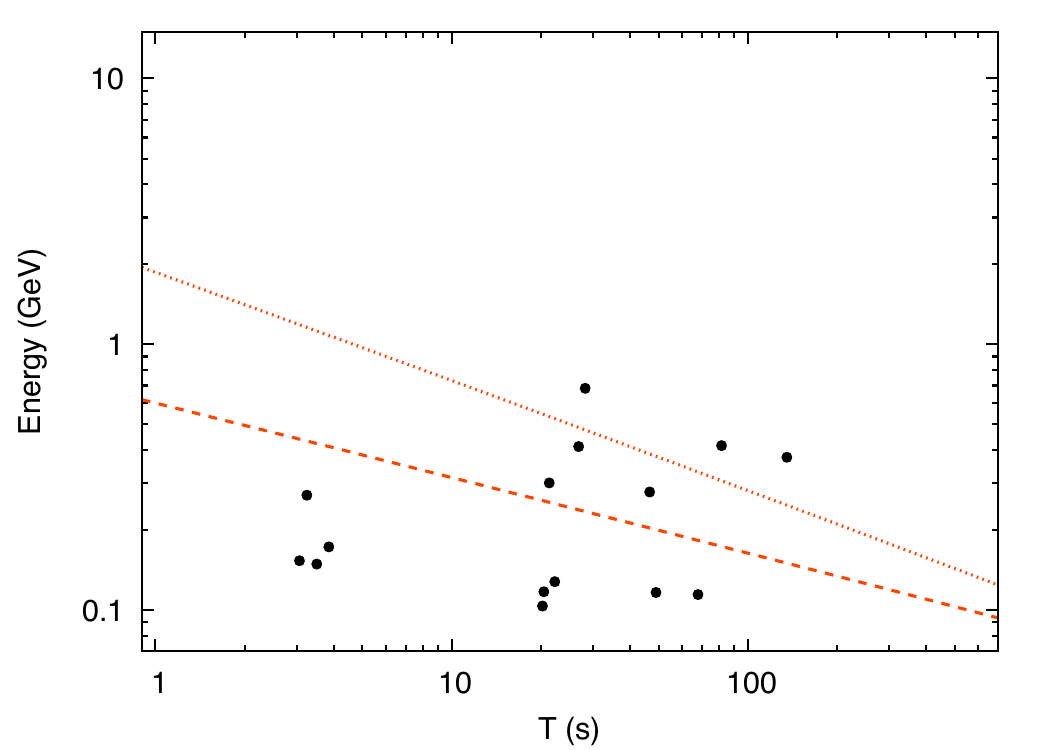}
  \label{fig4a):GRB080825C}}
  \subfloat[Sub b) GRB~090510][\centering{LAT light curve for GRB~090323.}]{\includegraphics[scale=0.35]{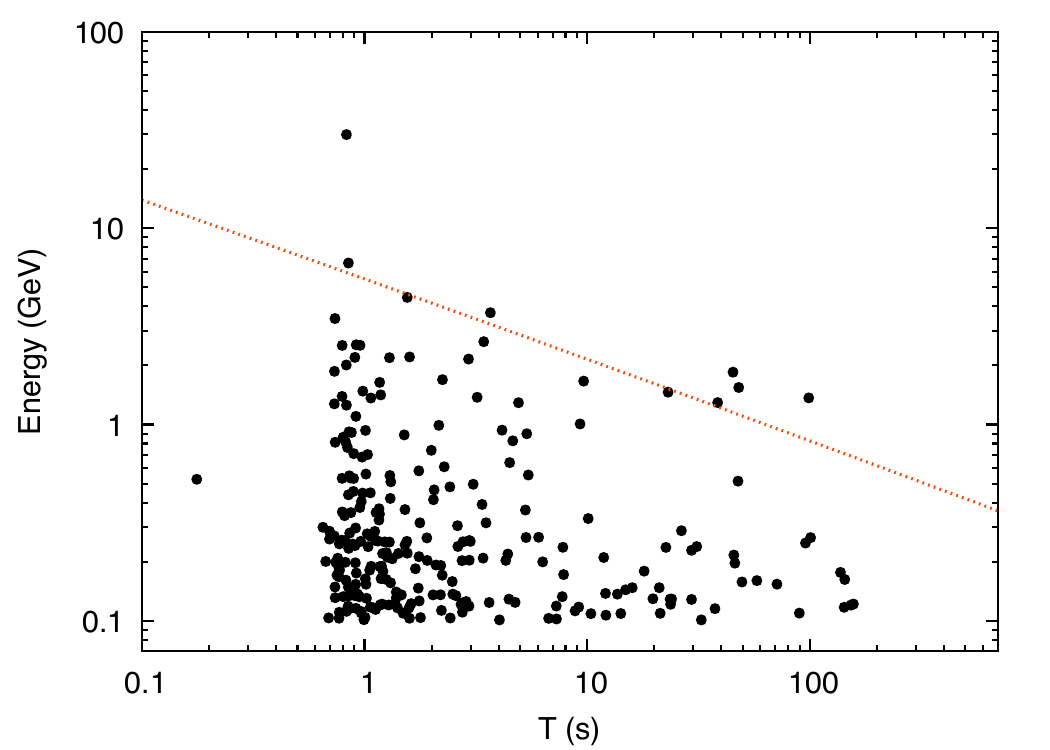}
  \label{fig4b):GRB090510}}
  \subfloat[Sub c) GRB~090902B][\centering{LAT light curve for GRB~090902B.}]{\includegraphics[scale=0.35]{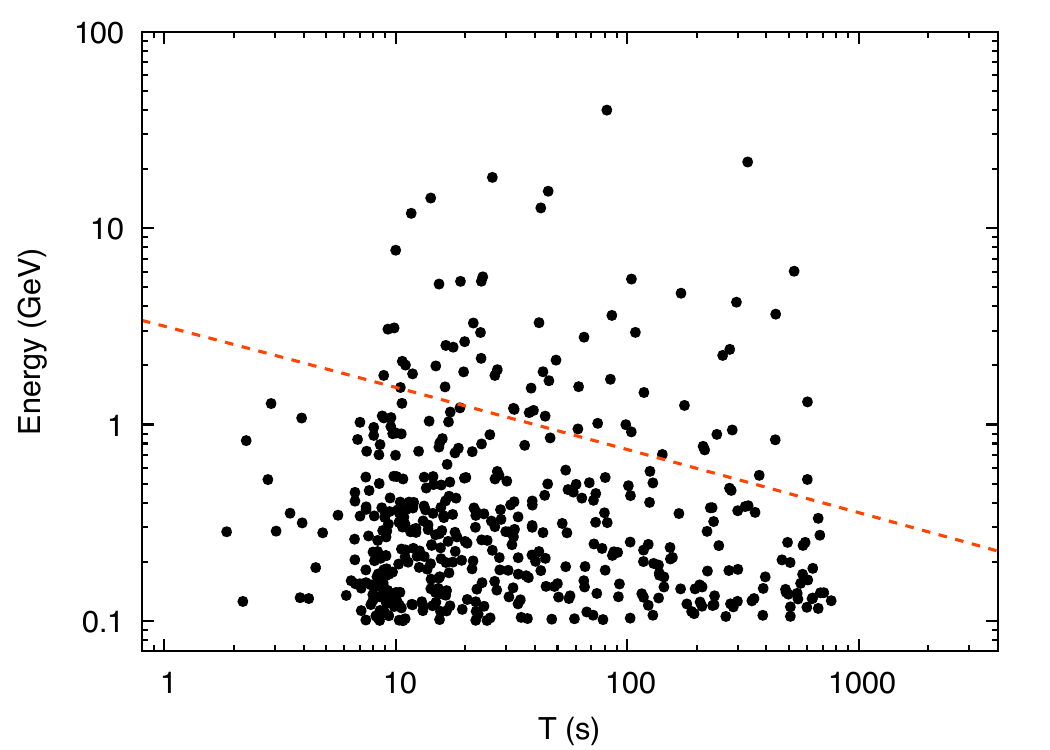}
  \label{fig4d):GRB090902B}}
  
  \subfloat[Sub d) GRB~090926A][\centering{LAT light curve for GRB~090926A.}]{\includegraphics[scale=0.35]{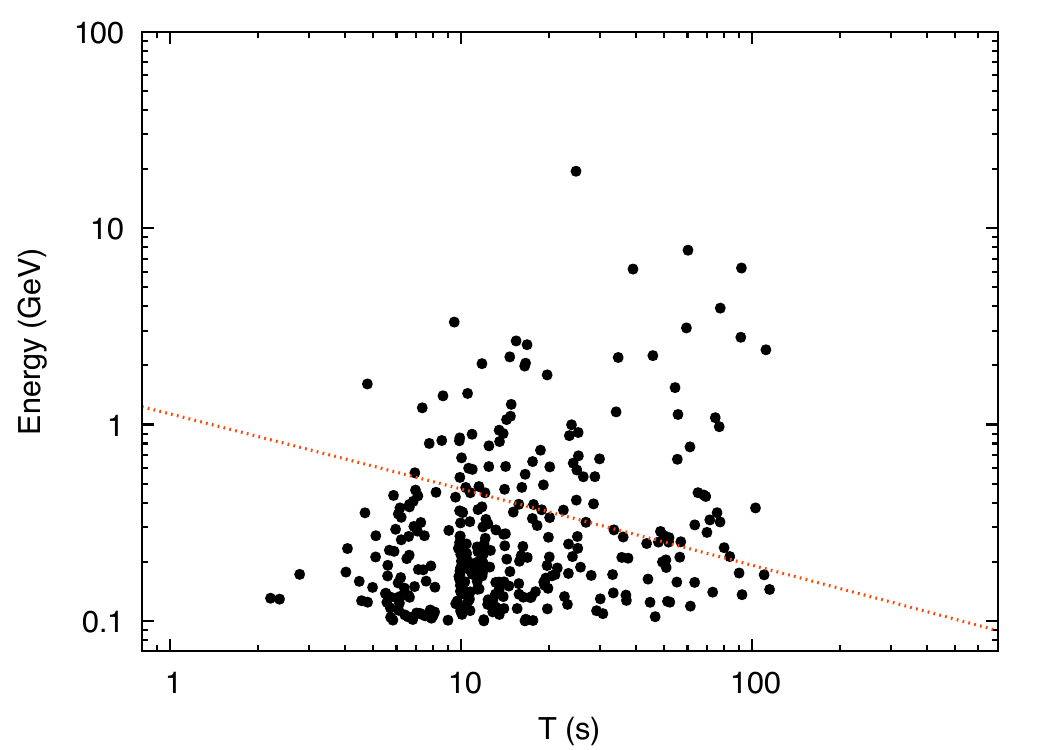}
  \label{fig4e):GRB090926A}}
  \subfloat[Sub e) GRB~110731A][\centering{LAT light curve for GRB~110731A.}]{\includegraphics[scale=0.35]{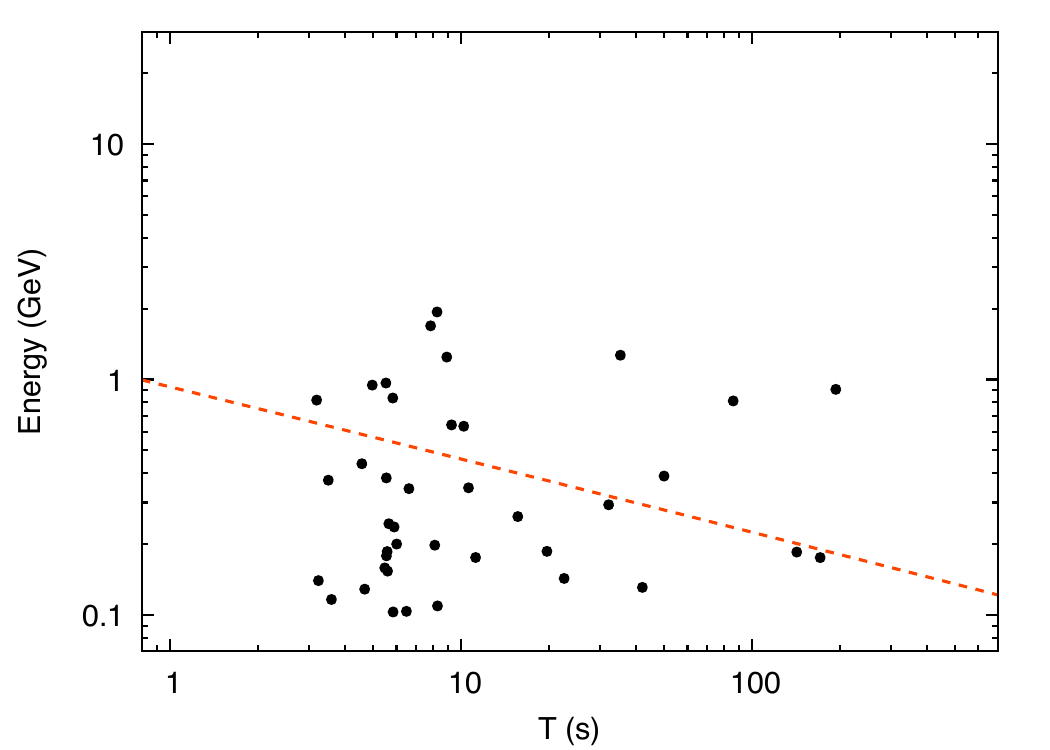}
  \label{fig4f):GRB110731A}}
  \subfloat[Sub f) GRB~130502B][\centering{LAT light curve for GRB~130502B.}]{\includegraphics[scale=0.35]{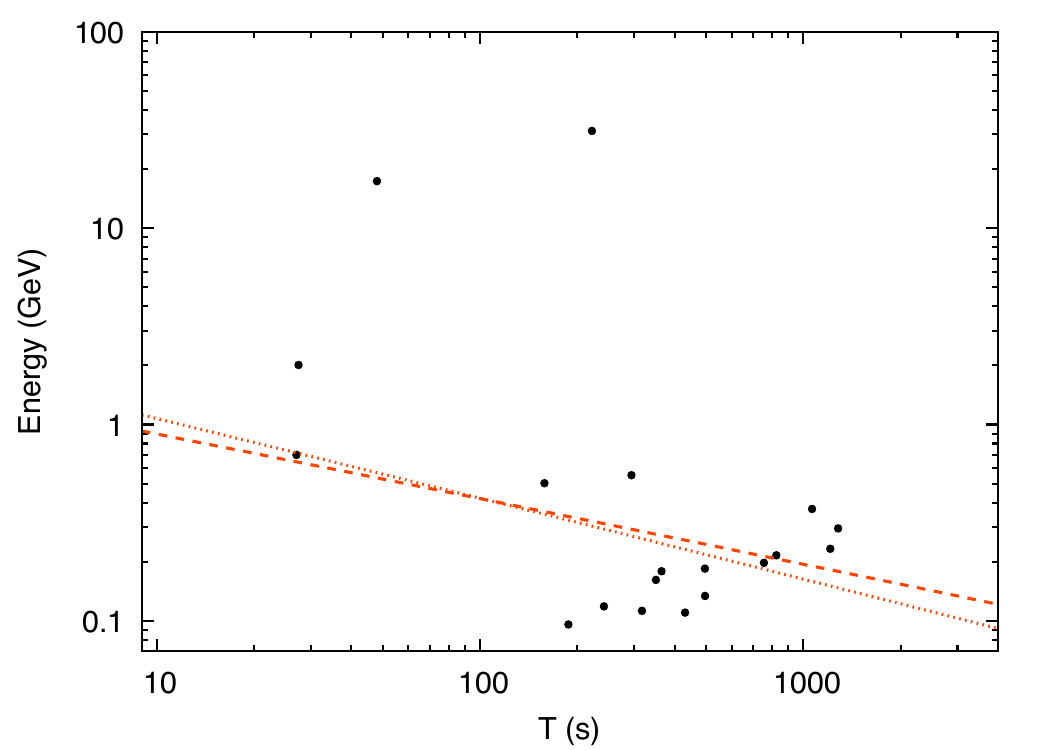}
  \label{fig4g):GRB130502B}}
  %\subfloat[Sub g) GRB~131108A][\centering{LAT light curve for GRB~131108A.}]{\includegraphics[scale=0.35]{photons_GRB131108A.pdf}
  %\label{fig4h):GRB131108A}}
  
  \subfloat[Sub h) GRB~141207A][\centering{LAT light curve for GRB~141207A.}]{\includegraphics[scale=0.35]{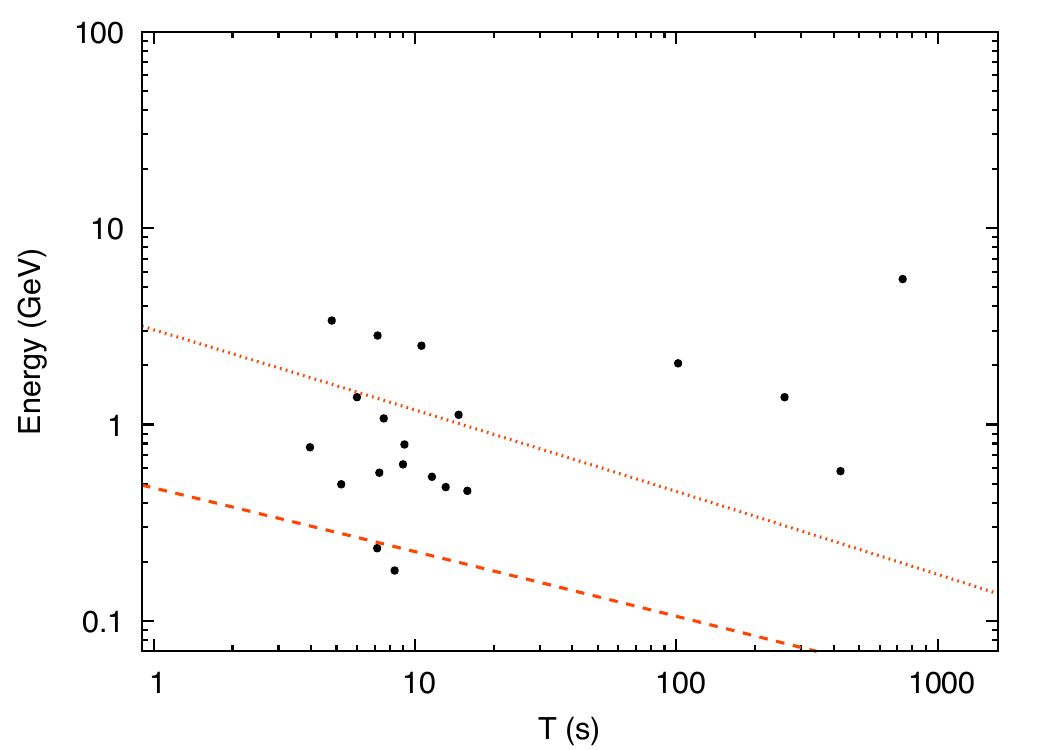}
  \label{fig4h):GRB141207A}}
  \subfloat[Sub i) GRB~170214A][\centering{LAT light curve for GRB~170214A.}]{\includegraphics[scale=0.35]{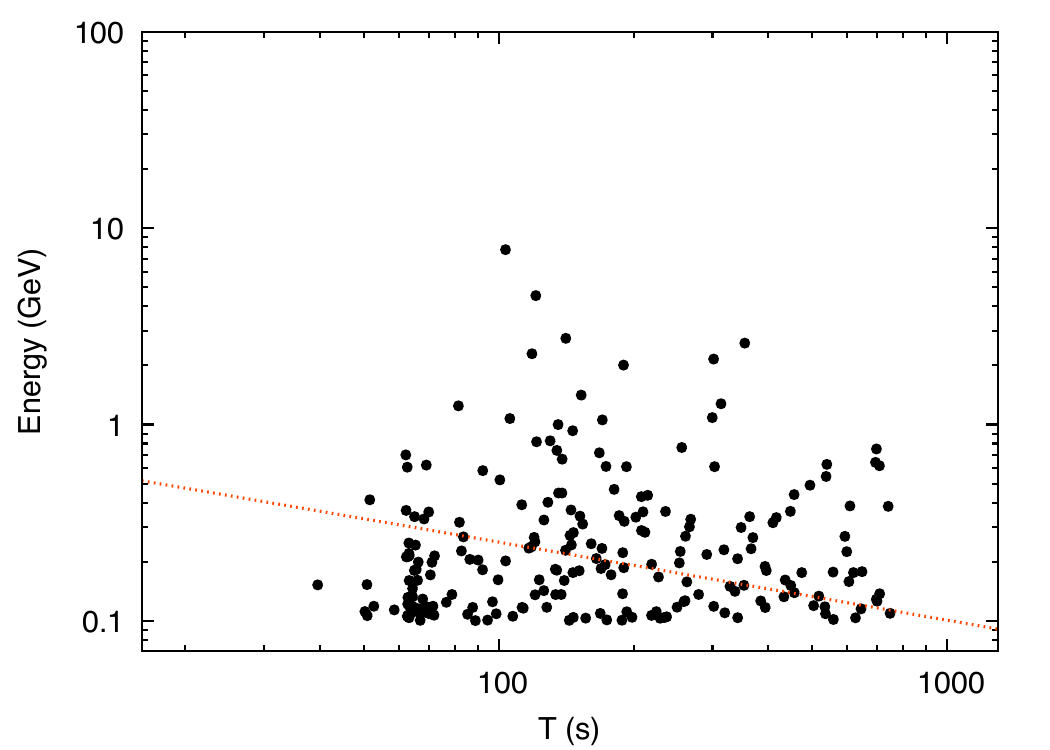}
  \label{fig4h):GRB170214A}}  
  \subfloat[Sub j) GRB~180720B][\centering{LAT light curve for GRB~180720B.}]{\includegraphics[scale=0.35]{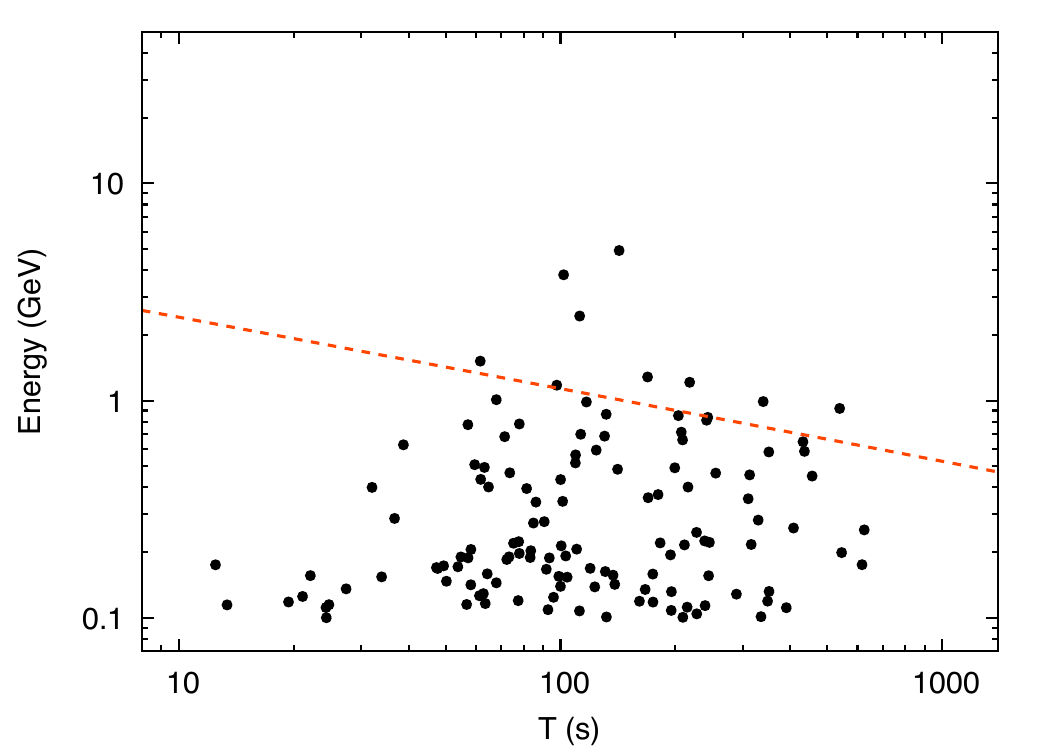}
  \label{fig4h):GRB180720B}}
 }
 \caption{All the photons with energies $> 100$~MeV and probabilities $>90$\% of being associated with each burst in our sample.  The red lines correspond to the maximum photon energies from our synchrotron afterglow model evolving in a constant-density (dotted) and stellar-wind (dashed) medium.} \label{fig4}
\end{figure*}

Figure~\ref{fig4} exhibits all photons with energies $>100$~MeV and probabilities $>90$\% of being associated with each burst of our representative sample. Additionally, we show in red lines the maximum photon energies released by the synchrotron afterglow model evolving in a constant-density (dotted) and stellar-wind (dashed) environment, estimated with the best-fit values reported in Tables~\ref{Table:ISM_Fit} and \ref{Table:Wind_Fit}. This Figure shows that the synchrotron FS model cannot explain the highest energy photons collected by {\itshape Fermi}/LAT.
Energetic photons above $>$ 10~GeV are usually explained via hadronic and SSC scenarios. In the hadronic scenarios, high-energy gamma-ray emission has been interpreted via photo-hadronic interactions; ultrarelativistic protons accelerated in the jet with internal synchrotron photons \citep{2009ApJ...705L.191A, 2000ApJ...537..255D}, inelastic proton-neutron collisions \citep{2000ApJ...541L...5M}, and relativistic neutrons with seed photons coming from the outflow \citep{2004A&A...418L...5D, 2004ApJ...604L..85A}. Even though GRBs are among the most plausible candidates to accelerate cosmic rays up to ultra-high energies \citep[$\gtrsim 10^{18}$ eV; ][]{1995PhRvL..75..386W, 1995ApJ...453..883V} and thus, potential candidates for neutrino detection, the IceCube collaboration reported no coincidences between neutrinos and GRBs after analyzing years of data \citep{2022arXiv220511410A,2012Natur.484..351A, 2016ApJ...824..115A, 2015ApJ...805L...5A}. Because of this, we rule out hadronic models as an explanation for the observed properties of GRBs and conclude that the number of hadrons is too small for hadronic interactions to efficiently generate observable gamma-ray signals in GRBs. On the other hand, a few bursts GRB~160821B, GRB~180720B, GRB~190114C, GRB~190829A, GRB~201216C and GRB~221009A have been detected, up to now, with photons above $100\, {\rm GeV}$ by the Major Atmospheric Gamma Imaging Cherenkov Telescopes \citep[MAGIC;][]{2019Natur.575..459A, 2021ApJ...908...90A}, the High Energy Stereoscopic System \citep[H.E.S.S.;][]{2019Natur.575..464A, 2021arXiv210602510H} and the Large High-Altitude Air Shower Observatory \citep[LHAASO;][]{2023arXiv230606372L}. The VHE emission in all these bursts has been successfully described via SSC FS model \citep[e.g., see][]{2021ApJ...908...90A, 2019Natur.575..464A, 2021arXiv210602510H, 2023arXiv230606372L}. Therefore, the most appropriate mechanism to explain the photons above 10~GeV in our representative sample is SSC mechanism from FSs. It should be noted that the values of the spectral breaks derived from the best-fit parameters (see Tables~\ref{Table:ISM_der} and \ref{Table:Wind_der}) indicate that the KN effects cannot be neglected in some GRBs of our sample.
%2019arXiv191108961A
Based on the maximum photon energy emitted by synchrotron radiation during the deceleration phase, several authors have claimed that some LAT light curves cannot be adequately interpreted in terms of only synchrotron FS radiation \citep{2010MNRAS.403..926G, 2011MNRAS.415...77M,2015ApJ...804..105F, 2016ApJ...818..190F, 2020ApJ...905..112F}. Therefore, although the closure relations of the synchrotron standard model could satisfy the 29 LAT-detected GRBs with VHE emission above $10$~GeV reported in the 2FLGC, this model is not the appropriate one, and a new mechanism such as SSC would be the better favored as shown in this manuscript.

\section{Summary}
\label{sec:summary}
Based on the external FS scenario in the stellar wind and homogeneous medium, we have presented a general analytical model of the synchrotron and SSC processes in the fully adiabatic ($\epsilon=0$), fully radiative ($\epsilon=1$) or partially radiative or adiabatic ($0 <\epsilon < 1$) regimes for an electron spectral index in the ranges of $1<p<2$ and $2\leq p$. Using the typical values of a GRB afterglow and assuming that all electrons are accelerated during the FS, we explicitly derived and plotted the expected synchrotron and SSC light curves and spectra in the stellar-wind and constant-density medium for each range of p. We calculated the spectral breaks in the KN regime \citep{2009ApJ...703..675N, 2010ApJ...712.1232W} and introduced the effect of EBL absorption as proposed in \cite{2017A&A...603A..34F}. We discuss the evolution of the SSC flux as a function of the equivalent kinetic energy, density parameter, electron spectral index, and radiative parameter. We compared the expected SSC fluxes with the CTA, MAGIC, and {\itshape Fermi}/LAT sensitivities \citep{2019ICRC...36..673F}. We showed that all the expected fluxes are below the {\itshape Fermi}/LAT sensitivity, and depending on the parameter values, they could be detected by the CTA or MAGIC Telescopes. In particular, when $\epsilon=0$, the standard synchrotron and SSC light curves derived in the standard stellar wind and homogeneous medium afterglow models for $1<p<2$ and $2\leq p$ are recovered \citep{1998ApJ...497L..17S, 1998ApJ...501..772P, 2000ApJ...536..195C, 2001BASI...29..107B, 2001ApJ...548..787S, 2013NewAR..57..141G, 2019ApJ...883..162F}. 

%Deviations around $\epsilon=0$ will modify the evolution of spectral breaks and synchrotron and SSC light curves, producing significant variations of the spectral and temporal features of the afterglow emission. 

Adiabatic breaks around the transition time between fast- and slow-cooling regimes are expected in the light curves. However, if the value of $p$ does not deviate from $2$, adiabatic breaks are not observed. We have derived the closure relations between the temporal and spectral indices that describe the evolution of the synchrotron and SSC flux as a function of $\epsilon$ and $p$. Significant variations of the spectral and temporal features of the afterglow emission are introduced by radiative losses only if $\epsilon$ is large and approaches to unity. Otherwise, Deviations around $\epsilon=0$ will produce small variations of the spectral and temporal features. In the fully adiabatic regime, the temporal evolution of the synchrotron flux in the stellar wind and constant-density medium is identical. On the contrary, in the radiative regime, they evolve differently in both density profiles. Therefore, an afterglow transition between stellar wind and homogeneous medium would be easily identifiable in the radiative regime.

 Given the closure relations (see Table~\ref{ClosureRelations}), we notice that synchrotron and SSC fluxes could have a similar evolution in time and energy when condition $\alpha_{\rm syn}(\beta) \approx \alpha_{\rm ssc}(\beta)$ is satisfied. For a stellar wind and constant density medium, the condition is reached when $\beta\to 1$ and the observed frequency evolves under the cooling condition of $\{\nu^{\rm j}_{\rm m}, \nu^{\rm j}_{\rm c}\}<\nu$ with ${\rm j=ssc}$ and ${\rm syn}$. Irrespective of the value of $p$, the temporal decay indices become $\alpha\approx\frac{4}{4-\epsilon}$ and $\approx\frac{2(4+\epsilon)}{8-\epsilon}$ for a wind and homogeneous medium, respectively. Temporal and spectral similarities observed in two different bands of afterglow emission (e.g., TeV gamma-rays, X-rays and optical bands) could be explained through the synchrotron and SSC FS model with a hard electron spectral index $p\approx 2$. If $\epsilon\approx 0$, the expected flux in the stellar wind or constant-density medium evolves with a temporal index of $\alpha\approx 1$, the parameter $\epsilon$ could be used to discriminate between the afterglow models. It can be seen in Table \ref{ClosureRelations} that the closure relations of the synchrotron and SSC models with $\epsilon\approx 1$, and the cooling condition $\{\nu^{\rm k}_{\rm m}, \nu^{\rm k}_{\rm c}\} < \nu$ (with ${\rm k=syn}$ and {\rm ssc}) favor the powerful LAT-detected GRBs described with $\beta\sim 1$ and $\alpha \sim 1.5$ as reported by \cite{2010MNRAS.403..926G}. GRBs described with $\alpha >2$ and $\beta<1$ satisfied the closure relations of the SSC model under the cooling condition $\nu^{\rm ssc}_{\rm m}<\nu<\nu^{\rm ssc}_{\rm c}$ with $0\leq\epsilon\leq 1$, which is difficult to explain with the standard synchrotron model. GRBs satisfy the synchrotron closure relations of a homogeneous medium of a cooling condition $\nu^{\rm syn}_{\rm m}<\nu<\nu^{\rm syn}_{\rm c}$ for $\epsilon=0$ or a cooling condition of $\{\nu^{\rm syn}_{\rm m}, \nu^{\rm syn}_{\rm c}\}<\nu$ for $\epsilon=1$. For example, given a value of $p=2.8$ and a cooling condition of $\nu^{\rm syn}_{\rm m}<\nu<\nu^{\rm syn}_{\rm c}$, the corresponding spectral and temporal indices are $\beta=0.9$ and $\alpha=1.35$, respectively. A similar result could be obtained considering a value of $p=1.8$ and a cooling condition of $\{\nu^{\rm syn}_{\rm m}, \nu^{\rm syn}_{\rm c}\}<\nu$ with $\epsilon=0.92$. It is worth noting that if $h\nu\approx 100\,{\rm MeV}$, the first condition $\nu^{\rm syn}_{\rm m}<\nu<\nu^{\rm syn}_{\rm c}$ might require (due to $\nu^{\rm syn}_{\rm c}\propto \varepsilon_B^{-\frac32}$) an atypical value of the magnetic microphysical parameter \citep[e.g. $\epsilon_B\lesssim 10^{-6}$;][]{2009MNRAS.400L..75K,2010MNRAS.409..226K} whereas the second one leads to an usual value of the microphysical parameter \citep[e.g., $3\times10^{-5}\lesssim\epsilon_B\lesssim 0.3$;][]{2014ApJ...785...29S}.\\

%%%%%%%%%%%%%%%%%%%%%%%%%%%%%%%%%%%%%%%%%%%%%%%%%%%%%%%%%%%%%%%%%%%%%%%%%%%%%%%%%%%%%%%%%%%%%%%%%%%%%%%%%%%%%%%%%%%%%%%%%%%%%%%%%%%%%%%%%%%%%%%%%%%%%%%%%%%%
%%%%%%%%%%%%%%%%%%%%%%%%%%%%%%%%%%%%%%%%%%%%%%%%%%%%%%%%%%%%%%%%%%%%%%%%%%%%%%%%%%%%%%%%%%%%%%%%%%%%%%%%%%%%%%%%%%%%%%%%%%%%%%%%%%%%%%%%%%%%%%%%%%%%%%%%%%%%
%%%%%%%%%%%%%%%%%%%%%%%%%%%%%%%%%%%%%%%%%%%%%%%%%%%%%%%%%%%%%%%%%%%%%%%%%%%%%%%%%%%%%%%%%%%%%%%%%%%%%%%%%%%%%%%%%%%%%%%%%%%%%%%%%%%%%%%%%%%%%%%%%%%%%%%%%%%%
%%%%%%%%%%%%%%%%%%%%%%%%%%%%%%%%%%%%%%%%%%%%%%%%%%%%%%%%%%%%%%%%%%%%%%%%%%%%%%%%%%%%%%%%%%%%%%%%%%%%%%%%%%%%%%%%%%%%%%%%%%%%%%%%%%%%%%%%%%%%%%%%%%%%%%%%%%%%
%In order to apply our theoretical model to

As particular cases, we have derived the {\itshape Fermi}/LAT light curves together with the photons with energies $\geq 100$~ MeV associated with each burst.
We have selected those GRBs (080825C, 090510, 090902B, 090926A, 110731A, 130502B, 141207A, 170214A and 180720B) from the 2FLGC \citep{2019ApJ...878...52A} with values of temporal and spectral indices with $\alpha_{\rm L}\gtrsim 1.5$ and $\Gamma_{\rm L}\approx 2$, respectively. We have applied our adiabatic-radiative afterglow model to fit the observations of this sample. We want to highlight that the standard SSC or synchrotron afterglow model cannot describe the closure relations of GRBs with both the temporal decay index $\alpha_L\gtrsim 1.5$ and the spectral index $\Gamma_L \approx 2$. It is always possible to assume light curves from synchrotron and SSC models to superimpose them and describe some temporal evolution different from that predicted by the standard afterglow model. However, light curves in the radiative regime can be done in an evident and clean way. We have used the multiwavelength observations to constrain the parameters in the synchrotron and SSC mechanism and model the LAT light curves of the sample via MCMC simulations. We have fitted the LAT observations of GRB~080825C, GRB~130502B and GRB~141207A with the FS model evolving in the stellar-wind and constant-density environment, the LAT, X-ray and optical observations of GRB~090510, GRB~090926A and GRB~170214 using the constant-density medium, and GRB~090902B, GRB~110731A and GRB~180720B with the stellar-wind environment. The results indicate that although the fraction of the total energy density given to accelerate electrons is much greater than $\varepsilon_{\rm e} \gg 0.1$, the shock-accelerated electrons are not in the fast-cooling regime during the entire LAT light curve. This indicates that in some cases a transition from radiative to adiabatic regime occurs in the LAT observations.

\section*{Acknowledgements}

We would like to extend our appreciation to the anonymous referee for their thorough evaluation of the article and valuable suggestions, which significantly improved the quality and clarity of our manuscript. We also thank Tanmoy Laskar, Paz Beniamini and Bin Zhang for useful discussions. NF acknowledges financial support from UNAM-DGAPA-PAPIIT through grants IN106521 and IN119123. PV acknowledges financial support from NASA grants 80NSSC19K0595 and NNM11AA01A.

%%%%%%%%%%%%%%%%%%%%%%%%%%%%%%%%%%%%%%%%%%%%%%%%%%
\section*{Data Availability}

{No new data were generated or analysed in support of this research.}



\bibliographystyle{mnras}
\bibliography{main} % if your bibtex file is called example.bib

%%%%%%%%%%%%%%%%%%%%%%%%%%%%%%%%%%%%%%%%%%%%%%%%%%

%%%%%%%%%%%%%%%%% APPENDICES %%%%%%%%%%%%%%%%%%%%%

\appendix

\section{Some extra material}

\begin{table*}
\centering \renewcommand{\arraystretch}{1.85}\addtolength{\tabcolsep}{1pt}
\caption{Closure relations of the SSC and synchrotron afterglow model in stellar-wind and homogeneous medium with ${\rm j=ssc}$ and ${\rm syn}$, respectively.}
\label{ClosureRelations}
\begin{tabular}{ccccccc}
\hline
 & & & Synchrotron & & SSC & \\
 & & $\beta$ & $\alpha(\beta)$ & $\alpha(\beta)$ & $\alpha(\beta)$ & $\alpha(\beta)$ \\
 & & & $(1<p<2)$ & $(2<p)$ & $(1<p<2)$ & $(2<p)$ \\
\hline
 & & & ISM, Slow Cooling & & & \\
\hline
1 & $\nu^{\rm j}_m<\nu<\nu^{\rm j}_c$ & $\frac{p-1}{2}$ & $\frac{3(2\beta+3+2\epsilon)}{2(8-\epsilon)}$ & $\frac{3(4\beta+\epsilon)}{8-\epsilon}$ & ----- & $\frac{2(2\epsilon+9\beta-1)}{8-\epsilon}$ \\
2 & $\nu^{\rm j}_c<\nu$ & $\frac{p}{2}$ & $\frac{3\beta+5+2\epsilon}{8-\epsilon}$ & $\frac{2(6\beta+\epsilon-2)}{8-\epsilon}$ & ----- & $\frac{2(\epsilon+9\beta-5)}{8-\epsilon}$ \\
\hline
 & & & ISM, Fast Cooling & & & \\
\hline
3 & $\nu^{\rm j}_c<\nu<\nu^{\rm j}_m$ & $\frac{1}{2}$ & $\frac{4\beta(\epsilon+1)}{8-\epsilon}$ & $\frac{4\beta(\epsilon+1)}{8-\epsilon}$ & $\frac{2(2\epsilon-1)\beta}{8-\epsilon}$ & $\frac{2(2\epsilon-1)\beta}{8-\epsilon}$ \\
4 & $\nu^{\rm j}_m<\nu$ & $\frac{p}{2}$ & $\frac{3\beta+5+2\epsilon}{8-\epsilon}$ & $\frac{2(6\beta+\epsilon-2)}{8-\epsilon}$ & ----- & $\frac{2(\epsilon+9\beta-5)}{8-\epsilon}$ \\
\hline
& & & Wind, Slow Cooling & & & \\
\hline
5 & $\nu^{\rm j}_m<\nu<\nu^{\rm j}_c$ & $\frac{p-1}{2}$ & $\frac{2\beta+9-\epsilon}{2(4-\epsilon)}$ & $\frac{2+\beta(6-\epsilon)}{4-\epsilon}$ & $\frac{9-2\epsilon+\beta(\epsilon-2)}{4-\epsilon}$ & $\frac{4-\epsilon+\beta(8-\epsilon)}{4-\epsilon}$ \\
6 & $\nu^{\rm j}_c<\nu$ & $\frac{p}{2}$ & $\frac{\beta+3}{4-\epsilon}$ & $\frac{\epsilon-2+\beta(6-\epsilon)}{4-\epsilon}$ & $\frac{6-\epsilon+\beta(\epsilon-2)}{4-\epsilon}$ & $\frac{\epsilon-4+\beta(8-\epsilon)}{4-\epsilon}$ \\
\hline
 & & & Wind, Fast Cooling & & & \\
\hline
7 & $\nu^{\rm j}_c<\nu<\nu^{\rm j}_m$ & $\frac{1}{2}$ & $\frac{(\epsilon+2)\beta}{4-\epsilon}$ & $\frac{(\epsilon+2)\beta}{4-\epsilon}$ & $\frac{\epsilon\beta}{4-\epsilon}$ & $\frac{\epsilon\beta}{4-\epsilon}$ \\
8 & $\nu^{\rm j}_m<\nu$ & $\frac{p}{2}$ & $\frac{\beta+3}{4-\epsilon}$ & $\frac{\epsilon-2+\beta(6-\epsilon)}{4-\epsilon}$ & $\frac{6-\epsilon+\beta(\epsilon-2)}{4-\epsilon}$ & $\frac{\epsilon-4+\beta(8-\epsilon)}{4-\epsilon}$ \\
\hline
\end{tabular}
%Note: 
\end{table*}

\begin{table*}
%\footnotesize
\centering \renewcommand{\arraystretch}{1.5}\addtolength{\tabcolsep}{2pt}

\caption{Sample of 10 GRBs used here. Temporal and spectral indices are taken from \citet{2019ApJ...878...52A} with $\beta_{\rm L}=\Gamma_{\rm L}-1$.}
\label{tab:sample}
\begin{tabular}{lcc}
\hline
GRB & $\alpha_{\rm L} \pm \delta_{\alpha_{\rm L}}$ & $\beta_{\rm L} \pm \delta_{\beta_{\rm L}}$ \\
\hline
080825C	&	$1.45	\pm	0.03$	&	$1.40	\pm	0.40$		\\
090510	&	$1.81	\pm	0.08$	&	$1.15	\pm	0.08$			\\
090902B	&	$1.63	\pm	0.08$	&	$0.92	\pm	0.06$			\\
090626A	&	$1.39	\pm	0.08$	&	$0.86	\pm	0.07$			\\
110731A	&	$1.50	\pm	0.10$	&	$1.00	\pm	0.20$			\\
130502B	&	$1.44	\pm	0.06$	&	$1.00	\pm	0.10$			\\
%131108A	&	$1.50	\pm	0.02$	&	$1.80	\pm	0.30$			\\
141207A	&	$1.88	\pm	0.03$	&	$0.80	\pm	0.30$			\\
170214A	&	$1.70	\pm	0.30$	&	$1.30	\pm	0.10$			\\
180720B	&	$1.90	\pm 0.10$	&	$1.15	\pm	0.10$			\\

\hline
\end{tabular}
%\label{tab:PLsample}
\end{table*}

\begin{table*}
 \centering \renewcommand{\arraystretch}{1.5}\addtolength{\tabcolsep}{2pt}
 \caption{The best-fit values found with MCMC simulations after describing the multiwavelength afterglow observations with a synchrotron model evolving in a constant circumstellar medium.}
 \label{Table:ISM_Fit}
 \begin{tabular}{l c c c c c c}
 \hline
  GRB Name & $\mathrm{log_{10}(E/erg)}$ & $\mathrm{log_{10}(n/cm^{-3})}$ & $\mathrm{log_{10}(\Gamma)}$ & $\mathrm{log_{10}(\varepsilon_{e})}$ & $\mathrm{log_{10}(\varepsilon_{B})}$ & $\mathrm{p}$ \\ \hline

 080825C & $52.92_{-0.10}^{+0.10}$ & $-1.01_{-0.09}^{+0.10}$ & $2.91_{-0.06}^{+0.06}$ & $-0.41_{-0.05}^{+0.05}$ & $-4.00_{-0.02}^{+0.02}$ & $2.24_{-0.01}^{+0.01}$ \\
 090510 & $53.98_{-0.10}^{+0.10}$ & $-1.70_{-0.09}^{+0.09}$ & $2.92_{-0.07}^{+0.06}$ & $-0.09_{-0.05}^{+0.05}$ & $-4.59_{-0.03}^{+0.02}$ & $2.60_{-0.03}^{+0.04}$ \\
 090926A & $53.85_{-0.08}^{+0.12}$ & $-0.02_{-0.10}^{+0.10}$ & $2.00_{-0.06}^{+0.07}$ & $-0.51_{-0.05}^{+0.05}$ & $-1.02_{-0.03}^{+0.02}$ & $2.00_{-0.01}^{+0.02}$ \\
 130502B & $52.60_{-0.69}^{+0.67}$ & $-1.43_{-0.09}^{+0.12}$ & $2.01_{-0.64}^{+0.63}$ & $-0.56_{-0.02}^{+0.01}$ & $-3.19_{-1.20}^{+1.42}$ & $2.00_{-0.01}^{+0.02}$ \\
 %131108A & $52.703_{-0.100}^{+0.102}$ & $-1.712_{-0.105}^{+0.095}$ & $1.994_{-0.068}^{+0.070}$ & $-0.525_{-0.046}^{+0.053}$ & $-1.728_{-0.024}^{+0.025}$ & $2.600_{-0.010}^{+0.009}$ \\
 141207A & $52.95_{-0.10}^{+0.07}$ & $-1.45_{-0.10}^{+0.11}$ & $2.45_{-0.06}^{+0.07}$ & $-0.07_{-0.04}^{+0.05}$ & $-2.82_{-0.02}^{+0.03}$ & $2.55_{-0.11}^{+0.10}$ \\
 170214A & $53.97_{-0.08}^{+0.10}$ & $-1.00_{-0.11}^{+0.10}$ & $2.99_{-0.06}^{+0.05}$ & $-0.39_{-0.05}^{+0.04}$ & $-4.41_{-0.02}^{+0.02}$ & $2.39_{-0.02}^{+0.01}$ \\ \hline
 \end{tabular}
\end{table*}

\begin{table*}
 \centering \renewcommand{\arraystretch}{1.5}\addtolength{\tabcolsep}{2pt}
 \caption{The best-fit values found with MCMC simulations after describing the multiwavelength afterglow observations with a synchrotron model evolving in a stellar-wind environment.}
 \label{Table:Wind_Fit}
 \begin{tabular}{l c c c c c c}
 \hline
  GRB Name & $\mathrm{log_{10}(E/erg)}$ & $\mathrm{log_{10}(A_{w})}$ & $\mathrm{log_{10}(\Gamma)}$ & $\mathrm{log_{10}(\varepsilon_{e})}$ & $\mathrm{log_{10}(\varepsilon_{B})}$ & $\mathrm{p}$ \\ \hline

 080825C & $52.64_{-1.55}^{+0.92}$ & $-1.06_{-0.93}^{+0.83}$ & $2.51_{-0.34}^{+0.33}$ & $-0.26_{-0.53}^{+0.13}$ & $-2.13_{-2.24}^{+1.05}$ & $2.53_{-0.51}^{+0.31}$ \\
 090902B & $54.12_{-0.01}^{+0.01}$ & $-2.05_{-0.01}^{+0.01}$ & $2.10_{-0.02}^{+0.01}$ & $-0.34_{-0.01}^{+0.01}$ & $-2.20_{-0.01}^{+0.02}$ & $2.02_{-0.01}^{+0.02}$ \\
 110731A & $53.95_{-0.50}^{+0.47}$ & $-1.47_{-0.52}^{+0.43}$ & $2.03_{-0.55}^{+0.39}$ & $-0.52_{-0.42}^{+0.45}$ & $-2.98_{-0.55}^{+0.38}$ & $2.02_{-0.07}^{+0.11}$ \\
 130502B & $52.69_{-0.51}^{+0.52}$ & $ 0.01_{-0.51}^{+0.53}$ & $2.99_{-0.48}^{+0.50}$ & $-0.55_{-0.45}^{+0.49}$ & $-4.98_{-0.51}^{+0.52}$ & $2.21_{-0.10}^{+0.09}$ \\
 141207A & $52.91_{-0.51}^{+0.54}$ & $-1.33_{-0.48}^{+0.55}$ & $2.51_{-0.50}^{+0.50}$ & $-0.06_{-0.46}^{+0.56}$ & $-3.27_{-0.51}^{+0.47}$ & $2.18_{-0.08}^{+0.11}$ \\
 180720B & $53.59_{-0.52}^{+0.43}$ & $-3.01_{-0.49}^{+0.56}$ & $2.40_{-0.05}^{+0.05}$ & $-0.07_{-0.05}^{+0.05}$ & $-1.30_{-0.05}^{+0.04}$ & $2.03_{-0.01}^{+0.02}$ \\ \hline
 
 \end{tabular}
\end{table*}

%4.000e+01	1.624e+00	4.187e+01	1.400e-01	1.825e+08	1.213e+11	2.709e-08	8.260e+02	4.195e+03	4.304e-01	4.899e-08 #ism GRB080825C
%4.000e+01	5.590e+02	4.976e+03	1.039e-03	1.178e+14	9.339e+15	4.867e-13	8.135e+04	2.427e+05	9.788e-01	2.325e-13 #ism GRB090510
%4.000e+01	2.813e+04	1.555e-01	5.505e-02	1.010e+15	3.085e+04	3.314e-09	1.960e+04	4.609e+01	5.715e-01	8.175e-11 #ism GRB090926A
%4.000e+01	4.651e-02	1.668e+02	2.420e-02	3.566e+05	4.585e+12	1.014e-08	2.032e+02	1.217e+04	4.055e-01	9.448e-9 #ism GRB130502B
%4.000e+01	2.057e+02	7.633e-02	6.550e-02	6.565e+10	9.038e+03	8.387e-08	1.410e+03	2.717e+01	4.361e-01	1.308e-9 #ism GRB131108A
%4.000e+01	7.006e+02	2.834e+01	2.358e-02	2.420e+13	3.961e+10	2.587e-10	1.374e+04	2.764e+03	4.084e-01	1.196e-9 #ism GRB141207A
%4.000e+01	5.401e+00	2.559e-01	1.908e-01	8.571e+08	1.924e+06	1.628e-06	7.360e+02	1.602e+02	3.227e-01	1.073e-07 #ism GRB170214A

\begin{table*}
\centering \renewcommand{\arraystretch}{1.6}\addtolength{\tabcolsep}{-1pt}
\caption{Derived quantities at $40\,{\rm s}$ from the best-fit parameter values found with MCMC simulations with an afterglow model evolving in a homogeneous medium.}
\label{Table:ISM_der}
\begin{tabular}{ l c c c c c c c}
\hline
\hline
{\large GRB }	& {080825C} & {090510} & {090926A} & {130502B} & {141207A} & {170214A} 		 \\ 
%{\large Parameters}	& {\large Median} & {\large Median} & {\large Median} 		 \\ 
%
\hline \hline

$B\,({\rm G})$ 	\hspace{1.5cm}&  \small{$2.9\times10^{-1}$} & \small{$1.4\times10^{-1}$} & \small{$4.6\times 10$} &  \small{$5.1\times10^{-1}$}& \small{$8.6\times 10^{-1}$} &  \small{$3.9\times 10^{-1}$} \\
$\gamma_{\rm m}$ 	\hspace{1.5cm}&  \small{$3.3\times 10^4$} & \small{$2.8\times 10^5$} & \small{$2.1\times10^{3}$} &  \small{$4.6\times10^2$}& \small{$1.7\times 10^5$} &  \small{$1.1\times 10^5$} \\
$\gamma_{\rm c}$ 	\hspace{1.5cm}&  \small{$1.7\times 10^4$} & \small{$2.1\times 10^4$} & \small{$3.3\times10^{1}$} &  \small{$1.5\times 10^4$}& \small{$3.9\times 10^3$} &  \small{$8.5\times 10^3$} \\

$h\nu^{\rm syn}_{\rm m}\,({\rm eV})$ 	\hspace{1.5cm}&  \small{$6.1\times 10$} & \small{$2.8\times 10^3$} & \small{$2.3\times10^{1}$} &  \small{$2.4 \times 10^{-2}$}& \small{$6.1 \times 10^3$} &  \small{$5.9\times 10^2$} \\
$h\nu^{\rm syn}_{\rm c}\,({\rm eV})$	\hspace{1.5cm}&  \small{$1.7\times 10$}& \small{$1.5\times 10$}&  \small{$5.5 \times 10^{-3}$}&  \small{$2.4 \times 10^1$}&  \small{$3.4$}&  \small{$3.8 $}	 \\
$F^{\rm syn}_{\rm max}\,({\rm mJy})$	\hspace{1.5cm}&  \small{$1.7\times 10^{-1}$} & \small{$2.5\times 10^{-2}$} & \small{$5.6\times 10^{-2}$} &  \small{$7.9 \times 10^{-2}$}& \small{$2.1\times 10^{-1} $} &  \small{$9.8\times 10^{-3}$} \\

$h\nu^{\rm ssc}_{\rm m}\,({\rm~GeV})$	\hspace{1.5cm}&  \small{$6.6\times 10$} & \small{$2.2 \times 10^5$} & \small{$1.0 \times 10^{-1}$} &  \small{$4.9 \times 10^{-6}$}& \small{$1.7 \times 10^{5}$} &  \small{$6.6 \times 10^{3}$} \\
$h\nu^{\rm ssc}_{\rm c}\,({\rm~GeV})$	\hspace{1.5cm}&  \small{$5.1$}& \small{$6.2$}& \small{$5.9 \times 10^{-9}$}& \small{$5.2 $}& \small{$5.2 \times 10^{-2}$}&  \small{$2.8 \times 10^{-1}$} \\
$F^{\rm ssc}_{\rm max}\,({\rm mJy})$	\hspace{1.5cm}&  \small{$3.2\times 10^{-10}$} & \small{$2.6\times 10^{-11}$} & \small{$1.3 \times 10^{-9}$} &  \small{$9.4 \times 10^{-11}$}& \small{$3.2 \times 10^{-10}$} &  \small{$1.9 \times 10^{-11}$} \\
$h\nu^{\rm ssc}_{\rm KN, m}\,({\rm~GeV})$	\hspace{1.5cm}&  \small{$3.6\times 10^3$} & \small{$3.8 \times 10^4$} & \small{$1.3\times 10^2 $} &  \small{$5.6 \times 10^1$}& \small{$2.3\times 10^4$} &  \small{$8.0 \times 10^3$} \\
$h\nu^{\rm ssc}_{\rm KN, c}\,({\rm~GeV})$	\hspace{1.5cm}&  \small{$1.9\times 10^3$}& \small{$2.8\times 10^3$}&  \small{$2.1$}&  \small{$1.8\times 10^3$}& \small{$5.5\times 10^2$}&  \small{$6.4 \times 10^2$} \\

%\cdashline{1-2}
\cline{1-1}
%\hhline{:=:~~~~~~~~}

\hline
\end{tabular}
\end{table*}
%

%4.000e+01	1.151e+04	1.070e-04	3.098e+01	2.944e+13	2.546e-03	5.647e-04	1.706e+03	1.646e-01	1.864e-01	2.111e-9 #wind GRB080825C
%4.000e+01	1.435e+03	5.786e+00	7.303e-02	2.380e+13	3.872e+08	2.862e-09	2.073e+04	1.317e+03	8.891e-01	8.688e-9 #wind GRB090902B
%4.000e+01	3.737e+02	8.613e-04	2.851e-01	4.095e+10	2.175e-01	5.230e-06	4.879e+02	7.406e-01	2.575e-01	4.008e-10 #wind GRB110731A
%4.000e+01	9.596e-03	8.028e-04	1.026e+02	9.027e+01	6.316e-01	8.042e-01	3.935e+00	1.138e+00	2.241e-01	1.004e-08 #wind GRB130502B
%4.000e+01	1.010e+04	1.312e-05	1.942e+01	1.437e+13	2.427e-05	1.571e-03	8.276e+02	2.983e-02	1.212e-01	7.720e-10 #wind GRB141207A
%4.000e+01	9.887e+01	3.969e-01	2.221e-01	1.992e+10	3.210e+05	1.655e-07	2.011e+03	1.274e+02	7.827e-01	1.541e-08 #wind GRB180720B

\begin{table*}
\centering \renewcommand{\arraystretch}{1.6}\addtolength{\tabcolsep}{-1pt}
\caption{The same as Table \ref{Table:ISM_der}, but for an afterglow model evolving in a stellar-wind medium.}
\label{Table:Wind_der}
\begin{tabular}{ l c c c c c c}
\hline
\hline
{\large GRB }	& {080825C} & {090902B} & {110731A} & {130502B} & {141207A} & {180720B} 		 \\ 
%{\large Parameters}	& {\large Median} & {\large Median} & {\large Median} 		 \\ 
%
\hline \hline
$B\,({\rm G})$ 	\hspace{1.5cm}&  \small{$1.9\times 10$} & \small{$2.0$} & \small{$3.2$} &  \small{$4.7$}& \small{$2.8$} &  \small{$8.5\times10^{-1}$} \\
$\gamma_{\rm m}$ 	\hspace{1.5cm}&  \small{$4.6\times10^4$} & \small{$8.1\times10^3$} & \small{$1.0\times10^3$} &  \small{$6.1\times10^3$}& \small{$4.4\times 10^4$} &  \small{$5.5\times10^4$} \\
$\gamma_{\rm c}$ 	\hspace{1.5cm}&  \small{$4.9\times10^1$} & \small{$1.6\times10^3$} & \small{$6.5\times10^2$} &  \small{$8.6\times10^1$}& \small{$3.8\times 10^2$} &  \small{$7.9\times10^3$} \\

$h\nu^{\rm syn}_{\rm m}\,({\rm eV})$ 	\hspace{1.5cm}&  \small{$4.3\times 10^{3}$} & \small{$4.0\times 10^{1}$} & \small{$5.0\times 10^{-1}$} &  \small{$9.7 $}&  \small{$7.9 \times 10^{2}$}&  \small{$1.7 \times 10^3$} \\
$h\nu^{\rm syn}_{\rm c}\,({\rm eV})$	\hspace{1.5cm}&  \small{$4.9\times 10^{-3}$}& \small{$1.7$}&  \small{$2.0 \times 10^{-1}$}&  \small{$2.0 \times 10^{-3}$}&  \small{$6.0 \times 10^{-2}$}&  \small{$3.5\times 10$} \\
$F^{\rm syn}_{\rm max}\,({\rm mJy})$	\hspace{1.5cm}&  \small{$2.9$} & \small{$4.3\times 10^{-2}$} & \small{$8.2\times 10^{-2}$} &  \small{$3.2 \times 10$}&  \small{$2.1 $}&  \small{$1.1\times 10^{-2}$} \\

$h\nu^{\rm ssc}_{\rm m}\,({\rm~GeV})$	\hspace{1.5cm}&  \small{$9.0\times 10^3$} & \small{$2.6$} & \small{$5.2 \times 10^{-4}$} &  \small{$3.6 \times 10^{-1}$}&  \small{$1.5 \times 10^{3}$}&  \small{$5.3 \times 10^3$} \\
$h\nu^{\rm ssc}_{\rm c}\,({\rm~GeV})$	\hspace{1.5cm}&  \small{$1.2\times 10^{-8}$}& \small{$4.5\times10^{-3}$}& \small{$8.5 \times 10^{-5}$}& \small{$1.4 \times 10^{-8}$}&  \small{$8.8\times 10^{-6}$}&  \small{$2.2$} \\
$F^{\rm ssc}_{\rm max}\,({\rm mJy})$	\hspace{1.5cm}&  \small{$4.6\times 10^{-7}$} & \small{$3.2\times 10^{-11}$} & \small{$7.7 \times 10^{-10}$} &  \small{$2.2 \times 10^{-4}$}&  \small{$9.7 \times 10^{-8}$}&  \small{$3.2 \times 10^{-13}$} \\
$h\nu^{\rm ssc}_{\rm KN, m}\,({\rm~GeV})$	\hspace{1.5cm}&  \small{$2.7\times 10^{3}$} & \small{$1.4 \times 10^3$} & \small{$8.4 \times 10$} &  \small{$1.9\times 10^2$}&  \small{$3.6\times 10^3$}&  \small{$2.1\times10^{4}$} \\
$h\nu^{\rm ssc}_{\rm KN, c}\,({\rm~GeV})$	\hspace{1.5cm}&  \small{$2.9$}& \small{$2.9\times 10^2$}&  \small{$5.3 \times 10$}&  \small{$2.6$}&  \small{$3.1 $}&  \small{$3.0\times 10^3$} \\

%\cdashline{1-2}
\cline{1-1}

\hline
\end{tabular}
\end{table*}
%

%\begin{figure*}
%\centering
%\includegraphics[width=0.48\textwidth]{EpsilonEvolution_WIND.pdf}
%\includegraphics[width=0.48\textwidth]{EpsilonEvolution_ISM.pdf}\\
%\includegraphics[width=0.48\textwidth]{Energy_Ratio.pdf}
%\includegraphics[width=0.48\textwidth]{find_k_eps.pdf}\\
%\caption{Upper panels: Evolution of the radiative parameter in the stellar-wind (left) and uniform-density (right) medium considering the synchrotron and expansion timescales (solid lines), or the ratio of Compton scattering to synchrotron luminosities (dotted lines). Lower panels show the evolution of energy radiated away (left) and relation between the radiative parameters $\epsilon$ and $s$ (right). The uniform-density and stellar-wind environments are shown in red solid and green dashed lines, respectively. The term $E_0$ corresponds to the initial equivalent kinetic energy. We use the values of $\Gamma_0=60$, $E_0=10^{52}\,{\rm erg}$, $\varepsilon_e=0.5$, $\varepsilon_B=5\times10^{-2}$, $\chi_{\rm e}=1$, $A_{W}=0.1$, $n=1\,{\rm cm^{-3}}$.}
%\label{k_eps}
%\end{figure*}

%%%%%%%%%%%%%%%%%%%%%%%%%%%%%%%%%%%%%%%%%%%%%%%%%%%%%%%%%%%%%%%%%%%%%%%%%%%%%%%%

%%%%%%%%%%%%%%%%%%%%%%%%%%%%%%%%%%%%%%%%%%%%%%%%%%

% Don't change these lines
\bsp	% typesetting comment
\label{lastpage}
\end{document}